\documentclass[preprint,11pt, authoryear,dvipsnames]{elsarticle}

\makeatletter
\def\ps@pprintTitle{%
	\let\@oddhead\@empty
	\let\@evenhead\@empty
	\def\@oddfoot{\reset@font\hfil\thepage\hfil}
	\let\@evenfoot\@oddfoot
}
\makeatother

\usepackage[utf8]{inputenc}
\usepackage{amsmath}
\usepackage{amssymb}
\usepackage{bbm}
\usepackage{todonotes}
\usepackage{algpseudocode}
\usepackage{algorithm} 
\usepackage{float}
\usepackage[english]{babel}
\usepackage{tabularx}
\usepackage{longtable}
\usepackage{comment}
\usepackage{hyperref}
\usepackage{adjustbox}
\usepackage{lscape}
\usepackage{amsthm}
\usepackage{makecell} 
\usepackage{bm}
\usepackage{comment}
\usepackage{physics}
\usepackage{tablefootnote}
\usepackage{array}
\newcolumntype{H}{>{\setbox0=\hbox\bgroup}c<{\egroup}@{}} 
\usepackage{graphicx}
\usepackage{subcaption}
\usepackage{mwe}
\usepackage{multirow}
\usepackage{booktabs}
\usepackage{setspace}
\usepackage{geometry}
\newtheorem{definition}{Definition}

\usepackage{xcolor}
\usepackage{soul}

\newboolean{includeHidden}
\setboolean{includeHidden}{false}
\newcommand{\hide}[1]{\ifthenelse{\boolean{includeHidden}}{{\tiny\textbf{HIDDEN:~}#1}}{}}

\definecolor{darkviolet}{rgb}{0.58, 0.0, 0.83}

\journal{Energy Economics}

\begin{document}
	
\begin{frontmatter}
	
	\title{Zonal vs. Nodal Pricing: An Analysis of Different Pricing Rules in the German Day-Ahead Market}

    \author[1]{Johannes Kn\"orr}
	\ead{knoerr@cit.tum.de}
    \author[1]{Martin Bichler}
	\ead{bichler@cit.tum.de}
    \author[1]{Teodora Dobos\corref{cor1}}
	\ead{dobos@cit.tum.de}

	\cortext[cor1]{Corresponding author}

	\affiliation[1]{organization={School of Computation, Information and Technology, Technical University of Munich},Department and Organization
		addressline={Boltzmannstr.~3}, 
		city={Garching},
		postcode={85748}, 
		country={Germany}}
        
\begin{abstract}
\color{black}
The European electricity market is based on large pricing zones with a uniform day-ahead price. The energy transition leads to changes in supply and demand and increasing redispatch costs. In an attempt to ensure efficient market clearing and congestion management, the EU Commission has mandated the Bidding Zone Review (BZR) to reevaluate the configuration of European bidding zones. 
Based on a unique data set published in the context of the BZR for the target year 2025, we analyze the short-run effects of various pricing rules for the German-Luxembourgish bidding zone. 
We compare market clearing and pricing for different zonal models, including their generation and redispatch costs. In numerical experiments with this dataset, the differences in the average prices in different zones are low. 
The total costs across different configurations are similar and the reduction of standard deviations in prices is also small.
This might be different with other load and generation scenarios, but the BZR data is important as it was created to make a decision about splits of the existing bidding zones. We can replicate several results from the BZR study, except the large cost savings when moving from one to two price zones in Germany and Luxembourg. 
In addition to the four zonal configurations analyzed in the BZR study, we compare these against a nodal pricing system.
While the total cost savings after introducing zonal splits were less than 1\%, nodal pricing led to savings of 5-6\%. We also evaluate differences of nodal pricing rules with respect to the necessary uplift payments, which is relevant in the context of the discussion on non-uniform pricing in the EU. While the study focuses on Germany, the analysis is relevant beyond and feeds into the broader discussion about pricing rules in non-convex markets.
\color{black}
\end{abstract}

	
\end{frontmatter}

\section{Introduction}
A central discussion on electricity market pricing revolves around zonal and nodal pricing (aka. locational marginal pricing). Among the electricity wholesale markets implementing nodal pricing are Argentina, Mexico, New Zealand, Russia, Singapore, and several regions in the United States (U.S.). In contrast, a zonal pricing approach has been employed in the large and coupled EU market. Large groups of nodes are aggregated into zones, and the market clearing problem only considers flow constraints between zones. It is supposed that no or only limited congestion occurs inside a zone.  
With some exceptions (e.g., Italy, Norway), there is a single price zone within a country, and the day-ahead market provides a uniform national electricity price. Ignoring transmission constraints can lead to dispatch decisions that are not physically feasible. 
Thus, after the market clearing, transmission operators conduct redispatch to ensure that the final \textcolor{black}{dispatch solution (or, allocation) satisfies the power flow equations}. In the USA, all Independent System Operators (ISOs) moved from a zonal to a nodal model, partly due to steeply increasing redispatch costs. \citet{aravena2021transmission} provide an excellent summary of historical developments in different jurisdictions. 

\subsection{Zonal vs. Nodal Pricing}

Due to the ongoing energy transition, the zonal pricing paradigm has further come under scrutiny \citep{eicke2022fighting, Bertsch.2016, Trepper.2015}. \textcolor{black}{Volatile renewable energy sources such as wind and solar, combined with the limitations of the electricity grid can lead to system overloads -- commonly referred to as congestions or bottlenecks} \citep{Neuhoff.2013}. \textcolor{black}{Structural congestion refers to predictable and recurring bottlenecks.} We will focus on Germany as a case in point. 
In this region, energy is cleared as if there were no transmission constraints, but the traded energy can often not be delivered due to transmission constraints. 
As a result, redispatch volumes and costs occur, which have increased substantially in recent years to 34 TWh and EUR 3.1 billion in 2023 \citep{Bundesnetzagentur.2024}. 
In an attempt to ensure efficient congestion management, the EU Commission -- as part of the Clean Energy Package -- has mandated a Bidding Zone Review (BZR) to reevaluate the configuration of European bidding zones. An integral part of this process is a locational marginal pricing (LMP) study conducted by the European Network of Transmission System Operators for Electricity (ENTSO-E) as a basis to identify structural congestion and modified bidding zones. Having accurate data and grid models is central for the quality of the BZR. The LMP study constitutes the results of a multi-year effort of the European TSOs to provide a representative data set, and it is unparalleled in terms of the quality and degree of detail of the data. 
This data set includes an electricity grid model derived from the Ten-Year Network Development Plan (TYNDP) 2020\footnote{Ongoing network expansion projects such as the development of the Südlink connection (expected to be finished in 2028) are not modeled in the data set.}, and supply and demand projections for 2025 under three climate years.
In the LMP study, ENTSO-E solved linearized unit commitment models with marginal pricing to obtain nodal prices \citep{ENTSOE.2022}. 
Based on these prices, in August 2022, the European Union Agency for the Cooperation of Energy Regulators (ACER) decided on a set of alternative bidding zone configurations \citep{ACER.2022.Decision}. 
Following Article 16 of the BZR methodology, ENTSO-E has published non-confidential data related to the LMP study \citep{ENTSOE.2023.BZRData}. In what follows, we refer to this data set as the BZR data and leverage it to expand on the results of the LMP study, focusing on the German BZ.\footnote{Although Germany and Luxembourg together form a single BZ, we will refer to it simply as the German BZ.} \textcolor{black}{This bidding zone is particularly interesting, because ACER suggested four alternative configurations \citep{ACER.2022.BZRConfig} that define a split of the large zone into 2, 3 or 4 smaller zones.\footnote{The configurations were identified by solving a clustering problem, where the features represent the time series of prices computed in the LMP study \citep{acer-clustering-algo}.}. Subsequently, the German TSOs proposed a new configuration comprising 5 zones, with the Schleswig-Holstein region represented as a separate zone.}

\subsection{Challenges in Congestion Management}
A key driver of the BZR is ensuring efficient congestion management. Congestion occurs when transmission capacity is insufficient to accommodate the flow of electricity from generation units to demand centers, leading to bottlenecks that require operational adjustments like redispatch or curtailment. Over the last years, redispatch costs and curtailment rates have increased considerably in Europe due to the increasing integration of renewable energy sources (RES), limited storage capabilities and insufficient transmission capacity.  
In 2024, the European Commission proposed an emissions reduction of 90\% by 2040 compared to the 1990 levels \citep{EuropeanCommission2024Climate}. This objective depends on the deployment of RES. Current efforts primarily focus on installing renewable capacity in areas with the highest resource potential and overlook the grid topology and the local demand. 
Regions rich in renewable generation potential may not align with areas of high electricity demand. Thus, the need to transmit power within zones could frequently surpass the capacity of the grid, leading to congestion and increased curtailment of RES. 
\cite{newbery2023high} and \cite{newbery2024marginal} discuss the inefficiency of current European market designs from the perspective of the RES merchants, emphasizing that investments are driven by average curtailment rates, which fail to account for significantly higher marginal curtailment rates.\footnote{Marginal curtailment represents the extra curtailment caused by the last MW of capacity installed. Average curtailment is defined as the total curtailed volume divided by the total installed capacity.} This misalignment risks promoting overinvestment in poorly connected areas, since, according to the authors, marginal curtailment is more than three times higher than average curtailment.
A recent study by \cite{thomassen2024redispatch} suggests that in a grid expansion scenario that extrapolates current trends, up to 310 TWh of renewable generation could be curtailed in 2040 due to the limitations in the grid. 
Moreover, even in an extreme grid expansion scenario that involves expanding the length of the European grid by more than a third, the total redispatch volumes would increase almost sixfold compared to 2022.
While the increase in curtailment rates and redispatch volumes is predictable, addressing these challenges requires a thorough evaluation of alternative market designs and pricing techniques.

\subsection{{Related Literature}}

A few recent studies aimed to analyze the impact of a bidding zone split on Germany or other countries in Europe. 
These studies are typically based on proprietary data sets and models, and assume different target years. 
The consulting company Aurora published results of such a study where they predict 5 EUR/MWh price difference between the north and the south of Germany in the year 2030, increasing to 9 EUR in 2045 \citep{AuroraStudy2023}. Models and data are proprietary. 
Another study by EWI and THEMA consulting \citep{EWITHEMAStudy2023} predicts differences in prices in a southern and a northern zone of around 10-15 EUR/MWh between 2024 and 2033, with a decreasing trend. The study is based on two separate models and data by EWI and THEMA consulting that are not publicly available. The size of the spread differs in THEMA's and EWI's results according to the authors. For 2030, the study makes a number of assumptions regarding electrolysers and new lines being built, batteries, and a shift of renewables from north to south. 
\citet{tiedemann2024gebotszonenteilung} discuss the market value of renewable energy sources after a zonal split for 2030. Their study is based on a data set and grid model by Fraunhofer IEE and they predict around 10 EUR/MWh difference between the north and south of Germany in 2030. They do not comment on reductions in redispatch.  
\textcolor{black}{Moreover, \cite{AgoraFraunhofer2025} show that a a locally organized electricity market increases the cost efficiency of a climate-neutral electricity system.}

It is worth mentioning that the data sets used in these studies and the underlying grid models differ, and so do the target years analyzed. For example, there are differences in the number of lines in the BZR data set compared to the widely used JAO Static Grid model.\footnote{\textcolor{black}{JAO Static Grid model contains 837 lines $\geq$ 220kV (see \url{https://www.jao.eu/static-grid-model}, accessed in May 2024). The dataset we use consists of 2407 lines spanning various voltage levels, both below and above 220kV.}} Also, the assumptions made for the future supply and demand and the redispatch methods used in these studies differ.\footnote{There are also earlier studies on a zonal split, such as those by \citet{Trepper.2015}, but the data sets and assumptions are even harder to compare due to the significant changes in the past ten years.} The study by EWI and THEMA, while not disclosing the implementation details, calculated redispatch volumes ``\textit{using a welfare optimal redispatch procedure with limited countertrading}''. The study by Aurora does not disclose any details regarding how redispatch is computed.

In November 2024, the average market prices for the alternative configurations were published \citep{ENTSOE.2024}. 
Importantly, the BZR study evaluates the alternative configurations at the Central European (CE) level, whereas our study evaluates them at the German level. We highlight the key modelling differences between our study and the BZR study in \ref{app:modelling-differences}.
The BZR results reveal that for the scenario year 2009 (see Section \ref{sec:data}), a two-zone split leads to a zonal price difference of 4.77 EUR/MWh. Averaging over all climate years considered in the BZR, the average prices observed in the north and in the south differ by $\approx$ 7 EUR/MWh. The results published by ENTSO-E also suggest that a split of Germany into 2-5 zones leads to price increases higher than 1 EUR/MWh in Czechia, Austria, Hungary, Slovakia, Slovenia, Croatia and Romania.
The final results of the BZR study were published in April 2025 \citep{ENTSOE.2025MainReport} and show that the welfare gains associated with any alternative configuration are less than 1\% of the simulated system costs in the CE region. These configurations also lead to redispatch savings of $\approx$ 50\% compared to the status quo European BZ layout.
We analyze these results in detail in \ref{sec:comparison}. 

All of the  previously mentioned studies analyze \textit{static} models where demand and supply are given as we do. 
\citet{ambrosius2020endogenous} use a multi-level mixed-integer non-linear model based on \citet{grimm2019optimal} to analyze investment incentives over time. Such models cannot accommodate the level of detail in static models due to their computational complexity, such that the paper is based on an aggregation of the German electricity network into 28 nodes. However, the approach complements static models taking into account the long-run effects of prices on investment. Several articles discuss the impact of price splits on transmission and generation investment incentives and emission reductions \citep{grimm2016transmission, grimm2021impact, grimm2022emissions}. For example, \citet{grimm2021impact} predict that a split of the German market into two price zones has only a small impact on efficiency.
Considering 3 grid expansion scenarios to be implemented by 2040, \cite{thomassen2024redispatch} analyze the evolution of redispatch costs for a large part of the EU based on a model consisting of 1024 nodes. Their study is built upon the PyPSA-Eur framework \citep{horsch2018pypsa} and shows that the redispatch volumes and costs will increase substantially regardless of the grid expansion scenario. 
\textcolor{black}{Several studies, focusing on various jurisdictions, estimate the static dispatch inefficiencies of a zonal market design to range between 0.5\% and 3\% \citep{Simshauser2025competition, Holmberg2023survey, Green.2007, Leuthold2008efficient, aravena2016renewable}.}

Another stream in the literature asks how zones should be split. Most studies use heuristics to identify price zones configurations by clustering power transfer distribution factors \citep{kumar2004, klos2014} or LMPs \citep{stoft1997, burstedde2012, breuer2013, weber2018consistent}. 
In contrast, \cite{grimm2019optimal} propose a mixed-integer nonlinear model to divide a market area into a specific number of price zones, ensuring the resulting configuration is welfare-optimal. \cite{ambrosius2020endogenous} extend this model by including capacity investment decisions, aiming to determine the available transfer capacities between zones.
\citet{Dobos.2024} use the BZR data and the published nodal prices to compute zones with low price variance. They show that the clusterings they found are not robust across time or clustering method used.

Finally, there is an extensive literature on nodal pricing and different pricing rules to deal with the non-convexities of power markets \citet{liberopoulos2016critical}. The most prominent examples of nodal pricing rules are Integer Programming (IP) pricing \citep{oneill2005efficient} and Convex Hull (CH) pricing \citep{gribik2007market, hogan2003minimum, Bichler.2021}. Both are used in a number of U.S. markets \citep{MISO.2019,PJM.2018} and require side-payments by the market operator to compensate losses of market participants (make-whole payments) or even to compensate all incentives to deviate from the efficient allocation (lost opportunity costs). These pricing rules are now being discussed in Europe in the context of non-uniform pricing rules. In particular, \cite{Pollitt2023} analyzes whether and how nodal prices should be implemented in Europe. CH pricing is known to provide poor congestion signals. We also draw on a recent proposal by \citet{Ahunbay.2024OR} that trades off make-whole payments and the quality of congestion signals, and evaluate these three pricing rules in the context of our study, which provides input for the discussion on non-uniform pricing in the EU day-ahead market \citep{AllNEMOCommittee.2023.Nonuniform}. 

\subsection{Contributions}

\color{black}
In this study, we conduct an independent analysis of the BZR dataset, which is based on supply and load data collected in 2019 and models the target year 2025. This dataset is arguably the most comprehensive and recent dataset available for the European power grid and serves as a foundation for decisions by the European Commission.
Unlike the BZR study, which simulates the entire CE continent, we focus exclusively on modeling Germany and its surrounding nodes, which allows for exact optimization of the dispatch and price computation and a focus on the effects in Germany. 

Our analysis considers not only zonal but also nodal pricing, for which we evaluate the average costs and day-ahead prices.
For the nodal market design, we formulate the unit commitment and economic dispatch problem, or welfare maximization problem, as a mixed-integer linear program (MILP). 
This differs from the LMP study, which used a linear programming relaxation of the problem for CE due to its large size \citep{ENTSOE.2022}. 
We then compute prices using three distinct pricing approaches: IP (Integer Programming) pricing (as used in U.S. power markets) \citep{oneill2005efficient}, CH (Convex Hull) pricing \citep{hogan2003minimum,gribik2007market}, and Join pricing \citep{Ahunbay.2022}. These methods compute prices based on the optimal allocation solution, minimizing different classes of lost opportunity costs. We note the difference to the LMP study, in which nodal prices were computed as shadow prices for the relaxed allocation problem. While the nodal prices computed in our study are close to the prices computed in the LMP study on average, they can differ on individual nodes. For the climate year 2009, the average value of the prices in the LMP study is 46.83 EUR/MWh, while the average value of the prices computed in this study with IP pricing is 46.63 EUR/MWh. 

For the zonal market design, we evaluate Germany's current single-zone configuration and the alternative configurations proposed by ACER, which include 2, 3, or 4 zones \citep{ACER.2022.BZRConfig}. We formulate the zonal allocation problem as an MILP and model the cross-zonal capacities using a net transfer capacity approach with security limits. In a subsequent step, we compute zonal prices using the three methods mentioned previously and analyze the redispatch costs associated with each configuration.
These costs are identified based on a minimum-cost redispatch procedure, which assumes that all generators are redispatchable and that redispatch is determined through a system-wide optimization, i.e., there is a cross-border cooperation between German TSOs, but not across all states in CE. Importantly, while the implementation details of the redispatch algorithm used in the BZR study are not published, the algorithm assumes full coordination in CE. Such redispatch process would mean that the most efficient remedial action within the whole CE region is selected to relieve a bottleneck in the German grid, which is against current practice.

\subsection{Key Results} 

 
Let us now summarize our main findings with respect to prices, price differences between zones, total cost savings due to the zonal configurations and savings in redispatch costs.

As indicated earlier, the LMP prices that we computed are very close to the prices published in the BZR study (the difference between the average prices in both studies is less than 1 EUR/MWh), which confirms their findings. 
In contrast to earlier studies, the average price differences across zonal configurations between 1 and 4 zones for Germany are low in our study, less than 3.5 EUR/MWh.  
There are also individual days where the price difference between zones is more than 10 EUR/MWh, but on average the differences are small.\footnote{In 2018, Austria was split from the joint price zone with Germany. Demand and supply is significantly different between both countries, yet prices (after the gas crises in 2021 and 2022) are often remarkably close as well (see \url{https://www.energy-charts.info/}).} The differences in the BZR study between the zones are higher (4-8 EUR/MWh), but these differences are still small considering average prices of 45-50 EUR/MWh, and compared to some earlier studies based on different data sets. 

The BZR study finds significant 50\% savings in redispatch costs in the CE region after switching from 1 to 2 zones, but no significant savings when moving from 2 to 3, 4, or 5 zones. We do not find these savings in our study when introducing 2 zones. However, similar to the BZR study, the savings in redispatch costs when moving from 2 to 5 zones are negligible if they exist. The BZR does not reveal all the details of their redispatch computation, but they allow for EU-wide redispatch, which is not current practice and one important difference to our study.  

The total cost savings (including redispatch costs) of a zonal split (2-5 zones) are small for Germany in our study, and small for the CE region in the BZR study ($<$1\%) . However, we find that the total cost savings of nodal pricing in Germany on the same data set are around 7-8\%. This is remarkable, as we do not consider demand response, which is a likely consequence of nodal pricing. The savings in total costs are due to the fact that redispatch is largely avoided in nodal pricing systems. Redispatch costs were at 2.7 billion EUR in 2024 and they are expected to increase six-fold in the EU by 2040 \citep{thomassen2024redispatch}.

Our study suggests that very small price zones or nodal prices should be considered for Germany in the future, a pricing system that has not been considered in the EU so far. A split in 2-5 larger price zones seems not to have much effect based on the BZR data. Larger effects on welfare and redispatch costs only emerge with much smaller or nodal zones. Similar findings are made in a recent study by \cite{AgoraFraunhofer2025} based on a different proprietary data set. 
Another aspect to consider is that it is very difficult to determine zones that are robust over time \citep{Dobos.2024}. Bottlenecks arise at different locations at different times, and it is hardly possible to determine zones that are robust even with historical data, when different time frames are considered. 

One might argue that the BZR data set is already several years old and the energy transition has progressed since then. The share of renewables in Germany's electricity generation has increased from 42\% in 2019 to 50\% in 2023 \citep{Bundesnetzagentur.2024}. The grid is being expanded, but also the demand grows significantly. 
Recent forecasts predict, in Germany, a growth from 457 TWh electricity demand in 2022\footnote{\url{https://www.energy-charts.info/index.html?l=en&c=DE}} to 726 TWh by 2030.\footnote{\url{https://www.renewable-ei.org/pdfdownload/activities/02_DimitriPescia_231128.pdf?}} This makes it even harder to assess the impact of a zonal split. Yet, the BZR data set is one of the most accurate descriptions of the EU power system and it allows us to assess at least the changes that a split would have had if it was initiated a few years ago. The impact on total cost savings based on this data set is smaller than what one might have expected. The costs and benefits of a zonal split deserve a thorough analysis, and this study aims to contribute to this discussion.



\color{black}

\subsection{Limitations} \label{sec:limitations}

Like any \textcolor{black}{model-based analysis}, our study makes assumptions and faces some limitations. 
\textit{First}, a central assumption that all such studies need to make for a BZ split are the cross-zonal line capacities. 
Transmission capacity allocation is a complex task with several decision made by TSOs, which cannot easily be predicted and modeled (see Section \ref{sec:zonal-market-clearing}). In line with earlier studies (see Section \ref{sec:zonal-market-clearing}), we compute market clearing solutions in which only 80\% of the transmission line limits are considered to take into account security considerations. With higher security margins there might be higher differences in the prices between zones, and also lower redispatch costs.
\textit{Second}, as in any simulation, the way how we compute redispatch can only be an approximation of real-world practices, which are complex (see decision BK8-18-0007-A by the German Bundesnetzagentur). For example, we assume that every generator can be redispatched, but this is not the case in reality. Moreover, we do not consider countertrade measures and restrict our analysis to day-ahead redispatch. So, the computed redispatch costs serve only as an approximation of the redispatch costs encountered by TSOs, but we argue that they provide a fair comparison.
\textit{Third}, our examination focuses on the German day-ahead market without incorporating neighboring countries, cross-border trades, or loop flows through other countries. We do incorporate 100 neighboring nodes not in Germany though to mitigate the effect. \textcolor{black}{However, flows from neighboring countries could enhance congestion and, thus, lead to higher redispatch volumes and costs in Germany.}
\textit{Fourth}, \textcolor{black}{we do not consider the long-run effects of the prices and congestion management regime corresponding to specific market clearing models. While this paper focuses on short-run inefficiencies, long-run inefficiencies -- such as suboptimal location choices and technology decisions for generation and consumption -- are also crucial factors to consider when evaluating a market clearing model.} 
\textit{Fifth}, we assume fixed, \textcolor{black}{inelastic} demand from the ENTSO-E dataset\footnote{\textcolor{black}{According to \cite{nemo2020euphemia}, demand is inelastic and must be satisfied in any feasible solution.}}, neglecting the potential for demand response under nodal prices. Incentives for demand response would be in favor of nodal pricing rules. \textcolor{black}{Moreover, we assume that the demand cannot be redispatched, as it is current practice in Germany.}
\textit{Sixth}, the current European market is characterized by portfolio bidding, differing from the unit-commitment bids underlying both the study by ENTSO-E and our analysis. Portfolio bids allow optimization of trades across assets and differ from those feasible in nodal market designs. Incorporating such considerations into our analysis would require making strong and potentially unwarranted assumptions regarding market participant behavior. Neither the LMP study nor we make such assumptions. 
\textit{Seventh}, as any other study of this sort, we do not model the intraday or forward markets.
\textit{Finally}, the data set provided by ENTSO-E was constructed for the target year 2025 based on input data and assumptions collected in 2019, and might not reflect the state of the electricity market in future years. It is also important to acknowledge that analyses can vary based on differing assumptions regarding supply, demand and electricity grid, as well as the employed methods. The data assumptions made for our analysis are described in \ref{app:data}. We also highlight the modelling differences compared to the BZR study in \ref{app:modelling-differences}.

We also want to emphasize that the discussion surrounding zonal and nodal pricing extends beyond dispatch and prices. In the BZR, a comprehensive approach is adopted which considers factors such as liquidity, security of supply, transition costs, and investments in low-carbon technologies, which we can not replicate in our analysis.

\subsection{Organization}
The remainder of this article is structured as follows. In Section \ref{sec:clearing_and_pricing}, we introduce our unit commitment model and formalize nodal and zonal market clearing. We also review the IP, CH, and Join pricing rules. Section \ref{sec:data} outlines our experimental design. Section \ref{sec:results} presents the key results of our analysis. Lastly, Section \ref{sec:conclusion} provides a summary and conclusions.

\section{Methods}

\subsection{Clearing and Pricing on Electricity Markets}\label{sec:clearing_and_pricing}

\subsubsection{Preliminaries}
We consider an electricity market consisting of a set of buyers $B$ and a set of sellers $S$, each located at nodes $N$ in an interconnected electricity network. The set of power lines $L$ is encoded as pairs of nodes $(n,m)$, and we consider multiple periods $T$. An item in this market corresponds to a unit of electricity at a particular node $n \in N$ at a specific time $t \in T$. 
A buyer $b \in B$ possesses a valuation function $v_b: \mathbb{R}^{N \times T} \rightarrow \mathbb{R}$ that quantifies their preferences. 
Similarly, a seller $s \in S$ has a cost function  $c_s: \mathbb{R}^{N \times T} \rightarrow \mathbb{R}_{\geq 0}$. 

The market operator collects these buy and sell bids to calculate a feasible allocation. As electricity is transmitted over the network, the allocation is subject to physical power flows, encoded as a constraint set $\Psi$. An accurate representation of the transmission network would require $\Psi$ to be equivalent to AC optimal power flow (ACOPF) constraints \citep{Molzahn.2019}. However, it is well known that the ACOPF is intractable for realistic problem sizes, and market operators need to employ computationally scalable approximations of the ACOPF.

\subsubsection{Market Clearing}
In practice, market operators employ different bidding languages to encode $v_b$ and $c_s$, as well as different approximations $\Psi$ of the power flow constraints. 
In full generality, the market operator seeks an allocation $(x,y)$, where $x = (x_b)_{b \in B}$, $x_b \in \mathbb{R}^{N \times T}$ is the allocation vector of buyers and $y = (y_s)_{s \in S}$, $y_s \in \mathbb{R}^{N \times T}$ is the allocation vector of sellers. The market operator identifies an allocation that maximizes welfare by solving the following optimization model:
\begin{align}
	\max_{x,y} \quad & \sum_{b \in B} v_b(x_b) - \sum_{s \in S} c_s(y_s) \label{model:general} \\
	\text{subject to \quad} & x,y \in \Psi. \nonumber
\end{align}

On European day-ahead markets, buyers and sellers submit hourly bids and block orders \citep{NEMOCommittee.2019}. The BZR data, however, does not contain bids of this structure. Instead, it provides generator and load characteristics similar to unit commitment problems in U.S. markets. Specifically, for a buyer $b$ located in a node $n_b$ there is only information on a fixed demand profile $P_b \in \mathbb{R}^{N \times T}$, while price-elastic bids are unavailable. Thus, the valuation function of $b$ simplifies to strict demand satisfaction, i.e., $v_b(x_b) = 0$ if $x_b=P_b$ and $v_b = -\infty$ otherwise.

For sellers/generators $s \in S$, the BZR data release includes more detailed information. Each generator $s$ has a minimum / maximum production quantity $\underline{P}_s$ / $\overline{P}_s \in \mathbb{R}_{\geq 0}$ at their node $n_s$. Moreover, once a generator has been started, it must run for at least $\underline{R}_s \in \mathbb{Z}_0^+$ periods. A generator incurs variable costs $g_s$ and fixed costs $h_s$, and its cost function is $c_s(y_s) = \sum_{t \in T} g_s y_{s,n_st} + \sum_{t \in T} h_s u_{st}$.
Given that only price-inelastic demand bids are available, the objective of problem (\ref{model:general}) can be interpreted as minimizing generators' costs required to fulfill the demand.

During the 2000s and 2010s, many U.S. ISO markets moved from zonal to nodal pricing. Recognizing that an accurate representation of the transmission network in the form of the ACOPF is computationally infeasible, market operators resort to linearized versions of power flow. 
The most common implementation is the Direct Current OPF (DCOPF), which is based on three simplifying assumptions \citep{Stott.2009, Molzahn.2019}, i.e., (1) uniform voltage magnitudes, (2) negligibly small voltage angle differences, and (3) neglecting line resistances and reactive power. Under these assumptions, the constraint set $\Psi$ can be expressed as simple linear constraints:
\begin{equation} \label{eq:dcopf}
	\Psi^{DC} = \{(x,y, \textcolor{black}{\theta}): \sum_{s \in S} y_{s,nt} - \sum_{b \in B} x_{b,nt} = \sum_{m \in N} -B_{nm} (\theta_{nt} - \theta_{mt}) \; \forall n \in N, t \in T \}.
\end{equation}

\color{black}
\noindent
The following MILP encodes the nodal clearing problem:
\begin{align}
	\min\limits_{x, y, u, \theta} \quad & \sum_{s \in S} c_s(y_s)  \tag{Nodal Clearing} \label{model:nodal} \\
	\text{subject to} &  \sum_{s \in S} y_{s,nt} - \sum_{b \in B} x_{b,nt} = \sum_{m \in N} -B_{nm} (\theta_{nt} - \theta_{mt}) & \forall n \in N, t \in T \label{first-nodal}\\
	& \underline{F}_{nm} \leq B_{nm} (\theta_{nt} - \theta_{mt}) \leq \overline{F}_{nm} & \forall n, m \in N, t \in T \\
	& \underline{P}_{st}u_{st} \leq y_{s,n_st} \leq \overline{P}_{st}u_{st} & \; \forall s \in S, t \in T \label{c1} \\
	& y_{s,nt} = 0 & \; \forall s \in S, n \in N \setminus \{n_s\}, t \in T \\
	& \phi_{st} \geq u_{st} - u_{s(t-1)} & \; \forall  s \in S, t \in T \setminus \{T_0\} \\
	& \sum_{i=t-\underline{R}_s+1}^{t} \phi_{si} \leq u_{st} & \; \forall  s \in S, t \in T \setminus \{T_0\} \\
	& c_s(y_s) = \sum_{t \in T} g_s y_{s,n_st} + \sum_{t \in T} h_s u_{st} & \forall s \in S \\
	& x_{b,n_bt} = P_{bt} & \forall b \in B, t \in T \\
	& x_{b,nt} = 0 & \forall b \in B, n \in N \setminus \{n_b\}, t \in T \label{c2}\\
	& \theta_{n^*t} = 0 & \forall t \in T \label{last-nodal}
\end{align}
\color{black}
\ref{app:dcopf} summarizes the notation.
We use a standard DCOPF model in our analysis, similar to what was used in the ENTSO-E study. 
Note that power system operators need to consider contingencies for a secure operation of the network. In particular, the system should satisfy the ($N-1$) criterion to guarantee that the power system is operated safely in the normal condition or in any contigency case \citep{christie2000transmission, marien2013importance, weinhold2023uncertainty}. Rather than a full security-constrained optimal power flow (SCOPF) model \citep{hinojosa2020comparing}, we consider a safety margin of 20\% on all transmission lines. 

\noindent
\textit{Zonal Market Clearing}\label{sec:zonal-market-clearing}

The key idea of zonal pricing is to aggregate nodes into zones and only consider transmission flows across zones, which simplifies the constraint set $\Psi$ and enhances computational scalability. 
An important parameter of a zonal system is represented by the capacities of the cross-zonal lines. 

In the European market, two approaches have been implemented: the net transfer capacity (NTC) model and the flow-based market coupling (FBMC) methodology \citep{NEMOCommittee.2019}. 
FBMC was introduced in Central Western Europe in 2015 and has since been extended to neighboring countries. 
In FMBC, the net export possible from one zone to another is limited by the remaining available margin (RAM) on the critical network elements (CNEs) connecting the zones. The RAM is determined by the maximum allowable flow (e.g., thermal capacity) subtracting the physical flow before the optimization (the base case), and a flow reliability margin to cover uncertainty in the capacity and allocation computations \citep{marien2013importance, weinhold2023uncertainty}.
FBMC with Generation Shift Keys (FBMC-GSK) is the specific methodology currently being used in the EU. The way how cross-zonal capacity is calculated and allocated in FBMC-GSK depends on how various parameters are set and the CNEs considered by the TSOs \citep{aravena2021transmission}, and cannot be predicted precisely for an eventual split of the German BZ. 

In our zonal market clearing implementation, we consider all interconnectors as critical network elements and model their physical characteristics via the DCOPF constraints (\ref{eq:dcopf}).
In line with earlier literature \citep{tiedemann2024gebotszonenteilung, wyrwoll2018, voswinkel2019}, we compute a base case that only takes 80\% of the capacities of all interconnectors into account. 
Similar to the nodal model, this should take into account the $(N - 1)$ security criterion \citep{christie2000transmission}.
\textcolor{black}{The zonal flow constraints can be written as:}
\begin{equation} \label{eq:dcopfZonal}
	\Psi^{Zonal} = \{(x,y, \textcolor{black}{\theta}): \sum_{s \in S} y_{s,zt} - \sum_{b \in B} x_{b,zt} = \sum_{n \in z, m \notin z}-B_{nm} (\theta_{nt} - \theta_{mt}) \; \forall z \in Z, t \in T\}
\end{equation}
\color{black}
\noindent
The zonal clearing problem is formulated as follows:
\begin{align}
	\min\limits_{x, y, u, \theta} \quad & \sum_{s \in S} c_s(y_s)  \tag{Zonal Clearing}\\
	\text{subject to} & \sum_{s \in S} y_{s,zt} - \sum_{b \in B} x_{b,zt} = \sum_{n \in z, m \notin z}-B_{nm} (\theta_{nt} - \theta_{mt}) & \forall z \in Z, t \in T \\
	& \underline{F}_{nm} \leq B_{nm} (\theta_{nt} - \theta_{mt}) \leq \overline{F}_{nm} & \forall n, m \in N, z(n) \neq z(m), t \in T \\
	& \theta_{z^*t} = 0 & \forall t \in T \\
	& (\ref{c1}) - (\ref{c2}) \nonumber
\end{align}
\color{black}
\textcolor{black}{We denote by $(x^{Zonal}, y^{Zonal}, u^{Zonal}, \theta^{Zonal})$ the optimal solution of the zonal clearing problem.}

\noindent
\textit{Redispatch} \label{subsec:redis}

While zonal market clearing is computationally less expensive than nodal market clearing, it substantially simplifies the underlying physical power flows. In fact, the economic outcome of zonal market clearing is usually infeasible with physical power flows. As a result, transmission operators resort to \textit{redispatch}, i.e., modifying the economic allocation to provide a physically feasible outcome. 

In the Single Day-Ahead Coupling Region (SDAC), the economic allocation is determined using the Euphemia algorithm \citep{nemo2020euphemia}. This allocation includes decisions on order acceptance and power flows between the bidding zones. The allocation results are published around 12:30 CET\footnote{The order book closes at 12:00 CET and then Euphemia computes a solution within 12 minutes.}, setting fixed day-ahead schedules. After these results are released, redispatch begins, which occurs during the period leading up to real-time delivery.

Current redispatch mechanisms consider two paradigms: Firstly, cost-based redispatch, as implemented in Germany, reimburses redispatched generators for their additional costs or lost profits and makes them indifferent to the previous market outcome. Secondly, market-based redispatch aims at a bid-based procurement of redispatch volumes. Market-based redispatch was criticized as prone to market power abuse and causing inc-dec gaming, i.e., strategic bidding in day-ahead markets to generate higher profits in subsequent redispatch markets \citep{LionHirth.2020}.


The precise redispatch methodology implemented by TSOs is not publicly disclosed. While Article 35 of the CACM Regulation states that ``\textit{Each TSO shall coordinate the use of redispatching and countertrading resources, taking into account their impact on operational security and economic efficiency}", it does not detail the coordination process or provide formal definitions of operational security and economic efficiency. The German legislation (§13 EnWG) states that ``\textit{In the case of power and voltage-related adjustments to active power generation or active power consumption [\dots], among several suitable measures [\dots], those measures are to be selected which are expected to cause the lowest overall costs}". However, the legislation does not provide the mathematically precise definition that has to be considered by TSOs. There are several studies that touch upon redispatch computation and employ different models. 
\cite{ambrosius2022risk, grimm2019optimal} and \cite{trepper2015} consider a redispatch problem that has as objective minimizing the cost difference between the redispatch and the spot (zonal) solutions.
Moreover, \cite{Kunz2015benefits} implement a model that minimizes the sum of redispatch costs, costs associated with load shedding and costs associated with renewable generation curtailment.

In this paper, we consider a minimum-cost redispatch method in line with earlier literature. We assume that demand cannot be redispatched, as it is current practice in Germany. This means that the (inelastic) demand $P_b$ for all $b \in B$ will be fully satisfied both in the zonal and in the redispatch solution.
Let $R$ denote the set of renewables, $Cur^{res}$ the quantity of renewable generation curtailment corresponding to $res \in R$ and $c_{cur}$ the curtailment cost per MW. 
Given the outcome of the zonal allocation $y^{Zonal}$, the redispatch problem is:
\begin{align}
	\min_{x,y, \theta} \quad & \sum_{s \in S} c_s(y_s - y_s^{Zonal}) + \sum_{res \in R} c_{cur} Cur^{res} \tag{RD-Min-Cost} & \label{model:redis} \\
	\text{subject to \quad} & Cur^{res} = \min(0, y^{Zonal}_{res} - y_{res}) & \forall res \in R \\
	 & (\ref{first-nodal}) - (\ref{last-nodal}) \nonumber
\end{align}
Negative (positive) redispatch occurs when $y_s < y^{Zonal}_s (y_s > y^{Zonal}_s)$. Negative redispatch implies cost savings since generation costs are avoided, which leads to a decrease in the overall redispatch costs. These costs must be reimbursed by power plants to TSOs. Positive redispatch requires additional generation costs for upward production, causing an increase in the total redispatch costs. This method ensures that the cheap (expensive) generators are selected for upward (downward) redispatch. Note that it also assumes that every generator can be redispatched, which might not be the case in practice.\footnote{Until 2021, only production units with a capacity of at least 10 MW could be redispatched. The Grid Expansion Acceleration Act (Netzausbaubeschleunigungsgesetz, NABEG 2.0) introduced the current system, Redispatch 2.0, in which units with a capacity of at least 100 kW can be redispatched. Redispatch 3.0. currently under discussion, aims to further extend it by integrating units with capacities below 100 kW. This would allow for redispatch of all production units, aligning with the method that we implement in this study.} 
We assume that the set of renewables $R$ includes offshore wind, onshore wind and photovoltaic units. The curtailment costs are set to the variable costs presented in Table \ref{tab:costs-res-var}, which reflect the variable and annual operating costs.\footnote{Since the curtailment costs are meant to cover the lost revenue plus additional expenses incurred by curtailed renewable units, the costs from Table \ref{tab:costs-res-var} act as a lower bound for the actual curtailment costs.}
\textcolor{black}{Note that such costs are not incurred in a nodal market clearing solution (\ref{model:nodal} problem).}

The objective value of problem \ref{model:redis} represents the total redispatch costs.
A redispatch solution is interpreted as the final DCOPF-feasible allocation $(x,y, \theta)$ that can be implemented as physical power flows.
\color{black}
Note that a DCOPF-feasible solution may still be ACOPF-infeasible and require further adjustments to the allocation, which we disregard in this work. Market participants subject to redispatch must be compensated for lost profits, reducing overall welfare. In our numerical experiments, we thus consider \textit{system costs} as the total cost of generation \textit{before} redispatch, \textit{redispatch costs} themselves, and \textit{total costs} as the sum of system and redispatch costs.

\subsubsection{Pricing on Non-Convex Markets} \label{sec:pricing_on_non_convex_markets}
Once an allocation has been obtained, the market operator must provide electricity prices $p \in \mathbb{R}^{N \times T}$ for each node and time period. Due to the non-convexities implied by sellers' cost functions (i.e., the binary commitment variables $u_{st}$), this is not a trivial task. 
In microeconomics, the \textit{welfare theorems} provide a foundation for pricing in markets. In their seminal paper, \citet{arrow1954existence} demonstrate that under convex preferences, demand independence, and perfect competition with divisible items, a market operator can always achieve a set of Walrasian equilibrium prices that support the welfare-maximizing allocation. 
We assume \textit{quasilinear} utilities, i.e., the utility of each market participant is defined as the difference between valuation or costs and price. 
\begin{align}\label{def:utility}
	u_b(x|p) = v_b(x) - \sum_{n \in N, t \in T} p_{nt} x_{b,nt} & \quad \forall \ b \in B \\
	u_s(y|p) = \sum_{n \in N, t \in T} p_{nt} y_{s,nt} - c_s(y) & \quad \forall \ s \in S \nonumber
\end{align} 

At given prices $p$, each market participant has some preferred bundles that maximize utility. We call these bundles the \textit{demand set} and denote the maximum possible utility as follows.
\begin{align} \label{def:indirect-utility}
	\hat{u}_b(p) = \max_x  v_b(x) - \sum_{n \in N, t \in T} p_{nt} x_{b,nt} & \quad \forall \ b \in B \\
	\hat{u}_s(p) = \max_y  \sum_{n \in N, t \in T} p_{nt} y_{s,nt} - c_s(y) & \quad \forall \ s \in S. \nonumber
\end{align}

A Walrasian equilibrium consists of a market-clearing allocation and linear prices such that no participant has an incentive to deviate. 

\begin{definition}[Walrasian Equilibrium] \label{def:WE}
	A price vector $p$ and a feasible allocation $(x,y)$ form a \textit{Walrasian equilibrium} if:
	\begin{enumerate}
		\item (Market clearing) Supply equals demand, i.e., $\sum_{s \in S, n \in N} y_{s,nt} = \sum_{b \in B, n \in N} x_{b,nt} \; \forall t \in T$.
		\item (Envy-freeness) Every market participant maximizes their utility at the prices, i.e., $u_b(x|p) = \hat{u}_b(p) \; \forall b \in B$ and $u_s(y|p) = \hat{u}_s(p) \; \forall s \in S$.
		\item (Budget balance) The payments that sellers receive equal the payments that buyers provide, i.e., $\sum_{n \in N, t \in T} p_{nt} (\sum_{s \in S} y_{s,nt} - \sum_{b \in B} x_{b,nt}) = 0$.
	\end{enumerate}
\end{definition}

Unfortunately, electricity markets do not satisfy the assumptions underlying the welfare theorems. As in many other markets, items are not perfectly divisible and market participants exhibit non-convex preferences. A large stream of literature has investigated existence conditions for Walrasian equilibria in non-convex markets \citep{Kelso82,bikhchandani2002package,baldwin2019understanding}, but none of these conditions are satisfied in electricity markets. \citet{bikhchandani1997competitive} demonstrate that Walrasian equilibria exist if and only if the linear relaxation of the allocation problem has an integer solution. When this is not the case, market participants bear \textit{lost opportunity costs} at any set of prices, which violates envy-freeness. 

Consequently, market operators have to resort to heuristic pricing rules that aim at approximating Walrasian equilibrium prices. Many such pricing rules have been proposed \citep{liberopoulos2016critical}, each compromising on the properties of Walrasian equilibria differently. In this work, we focus on \textit{Convex Hull} and \textit{Integer Programming} pricing as two established pricing rules in practice, as well as on the \textit{Join} pricing rule as a novel multi-objective approach. We also consider a simplified version of the \textit{Euphemia} pricing rule as it is applied in Europe. 

\noindent
\textit{Convex Hull Pricing}

In the absence of Walrasian equilibrium prices, there is no set of prices $p$ maximizing the profit of every market participant. Global lost opportunity costs describe this forgone payoff. 

\begin{definition}[Global Lost Opportunity Costs (GLOCs)] \label{def:gloc}
	Given an allocation $(x,y)$ and prices $p$, a market participant's global lost opportunity costs describe the difference between their individual payoff maximum, given $p$, and their actual payoff.
	\begin{align*}
		\text{GLOC}_b = \hat{u}_b(p) - u_b(x|p) \\
		\text{GLOC}_s = \hat{u}_s(p) - u_s(y|p)
	\end{align*}
\end{definition}

The key idea of CH \textcolor{black}{(Convex Hull)} pricing \citep{hogan2003minimum, gribik2007market} is to minimize total GLOCs. Formally, CH pricing replaces each non-convex cost function $c_s(y_s)$ with its convex envelope in (\ref{model:general}), and derives prices from the dual of the resulting convex problem. 

Computing convex envelopes and determining CH prices is generally computationally challenging \citep{Schiro.2016}. However, it has been shown by \citet{Hua.2017} that CH pricing becomes tractable for the cost function we consider. Specifically, by relaxing the binary constraints $u_{st} \in \{0,1\}$ to $u_{st} \in [0,1]$ and solving the dual of (\ref{model:general}), CH prices can be obtained from a single linear program. This approach, also referred to as Extended Locational Marginal Pricing (ELMP), was also used by the recent ENTSO-E study \citep{ENTSOE.2022}.

\noindent
\textit{IP Pricing} \label{subsec:IP}

In a convex market, Walrasian equilibrium prices are equivalent to \textit{marginal} prices, i.e., the costs of the last accepted bid set the prices. Following this notion, \citet{oneill2005efficient} generalized marginal pricing for non-convex markets. Their IP \textcolor{black}{(Integer Programming)} pricing rule assumes that generators cannot easily deviate from their commitment status and sets prices equal to the variable costs of the marginal \textit{committed} unit (i.e., the marginal unit among all units with $u_{st}=1$). 

IP pricing involves three steps: (i) obtaining the optimal commitment variables $u_{st}^*$ from the allocation problem in (\ref{model:general}), (ii) fixing the commitment variables of each generator to these optimal values, i.e., setting $u_{st} \in \{0,1\}$ to $u_{st} \in [0,1]$ with $u_{st} = u_{st}^*$, and (iii) solving (\ref{model:general}) with these linearized cost functions and extracting prices from its dual.

IP prices can be considered Walrasian equilibrium prices assuming that no generator can deviate from its commitment status. In other words, every participant maximizes their utility \textit{locally}, meaning under fixed commitment. We define these \textit{local lost opportunity costs} a subset of GLOCs. 

\begin{definition}[Local Lost Opportunity Costs (LLOCs)] \label{def:lloc}
	Given an allocation $(x,y)$, generator commitments $u^*$, and prices $p$, a seller's local lost opportunity costs describe the difference between their individual payoff maximum \emph{under fixed commitment}, given $p$, and their actual payoff. 
	\begin{align*}
		\text{LLOC}_s & = \hat{u}'_s(p) - u_s(y|p)  \text{ with } \hat{u}'_s(p) = \hat{u}_s(p) \text{ s.t. } u=u^* 
	\end{align*}
\end{definition}

Besides computational tractability, IP prices accurately reflect the marginal value of transmission capacity \citep{Yang.2019}, i.e., price differences among nodes arise solely when the network experiences congestion. It has been shown that this property immediately corresponds minimizing LLOCs \citep{Ahunbay.2024OR}, making them a significant indicator of good \textit{congestion signals}.

\noindent
\textit{Join Pricing}

In most electricity markets, neither GLOCs nor LLOCs are paid out to market participants to disincentivize deviations from the optimal allocation. Instead, market operators merely ensure that no participant incurs losses by participating in the market, ensuring non-negative utilities and, equivalently, \textit{individual rationality}. The payments to ensure individual rationality are known as \textit{make-whole payments}. 

\begin{definition}[Make-Whole Payments (MWPs)] \label{def:mwps}
	Given an allocation $(x,y)$ and prices $p$, a market participant's make-whole payments describe their negative payoff, if applicable. 
	\begin{align*}
		\text{MWP}_b & = \max\{-u_b(x|p),0\} \\
		\text{MWP}_s & = \max\{-u_s(y|p),0\} 
	\end{align*}
\end{definition}

Accepted bids that require MWPs are known as \textit{paradoxically accepted bids}. MWPs can be regarded as another subset of GLOCs referring to participants' incentives to not participate in the market. Both CH and IP pricing can experience high levels of MWPs. Recent practical concerns have thus motivated the development of pricing rules that ensure lower MWPs. Among them is the Join pricing rule \citep{Ahunbay.2024OR}, which proposes a dual pricing problem that combines the objective of IP pricing (minimizing LLOCs) with minimizing MWPs. Specifically, for each participant it considers the maximum of LLOCs and MWPs in the objective, leading to a computationally tractable pricing rule with adequate congestion signals and low MWPs simultaneously. 

In the U.S., MWPs are generally not covered directly by revenues obtained from the buyers through energy purchases in the day-ahead market but require additional funding mechanisms which depend on the ISO \citep{townsend2014mwps}.\footnote{\textcolor{black}{MWPs are also referred to as bid productions cost guarantees (NYISO), revenue sufficiency guarantees (MISO) and operating reserve credits (PJM).}} ISOs often impose uplift charges or ancillary service fees on market participants to fund MWPs. These charges are ultimately passed on to electricity consumers, either directly through consumption fees or indirectly through higher electricity bills.

\subsubsection{Euphemia}
CH, IP, and Join pricing all have in common that they price the welfare-maximizing allocation. Market participants incur lost opportunity costs; some are even paradoxically accepted and need to be compensated by MWPs. 

European electricity markets follow a different approach. Paradoxically accepted bids are not permitted, and instead, the market operator deviates from the welfare-maximizing outcome to avoid paying any MWPs. This is the essence of the \textit{Euphemia} day-ahead clearing algorithm \citep{NEMOCommittee.2019}. Note that avoiding paradoxically accepted bids does not satisfy the envy-freeness property from Definition \ref{def:WE}. Market participants still incur GLOCs and LLOCs, a subset of which are even rejected to participate in the market even though it would be profitable for them to do so. Such bids are referred to as \textit{paradoxically rejected bids}. 

The Euphemia algorithm combines market clearing and pricing (unlike the pricing rules above which can be applied to any pre-computed allocation). In European markets, it computes an allocation and \textit{zonal} prices, but theoretically, it can also be applied in nodal markets. Algorithmically, it starts by computing the welfare-maximizing allocation, solving (\ref{model:general}). It then advances to examining whether prices can be established for the allocation, i.e., they must prevent any paradoxically accepted bids, ensure the acceptance of all in-the-money hourly bids and, optionally, certify that cross-zonal flows occur from a lower-price to a higher-price zone. If paradoxically accepted bids exist, Euphemia adds cuts to (\ref{model:general}) that eliminate the current candidate solution, constraining the allocation problem and leading to welfare losses. These iterations between allocation and pricing continue until a solution without paradoxically accepted bids is reached. 
Subsequently, Euphemia continues by addressing some of the idiosyncrasies of the European market (PUN search, re-insertion of certain paradoxically rejected bids, resolving indeterminacy). If, at any of these steps, violations are detected, the iterative procedure continues. Eventually, Euphemia outputs allocation, zonal prices, and cross-zonal flows. 

Since we aim to investigate prices and welfare losses associated with Euphemia, we implemented a simplified version of the algorithm for our numerical experiments. With the data provided, however, we can only replicate the first two steps of the algorithm, i.e., the iterative allocation and pricing procedure. When a paradoxically accepted seller $s$ in period $t$ is detected, we add a constraint $u_{st} = 0$ to the allocation problem (\ref{model:general}). No open-source implementation of the algorithm is available, but we followed the public description as closely as possible.

A practical problem with Euphemia is computational tractability. Due to its iterative nature, multiple MILPs must be solved until allocation and prices are obtained. With the planned 15-min market time unit introduction in Europe, the scalability of Euphemia poses a significant concern, and policymakers consider non-uniform pricing rules (such as the ones introduced above) to obtain solutions faster \citep{AllNEMOCommittee.2022}.

\subsection{Experimental Design} \label{sec:data}

We leverage the data released in the context of the European bidding zone review \citep{ENTSOE.2022, ACER.2022.BZRConfig}. We focus on Germany as arguably one of the most important bidding zones regarding clearing volume and its central position in the European grid. 
The data we use in our study and the processing steps are described in \ref{app:data}.


We implemented the models in Python 3.11 using Gurobi to solve optimization problems. We conducted experiments for eight weeks in 2009 for the German BZ. To maintain computational feasibility of calculating hourly allocations and prices, we permitted a MIP gap of 5\% for mixed-integer programs. Table~\ref{tab:experiment_overview} summarizes our experimental design and the scope of our analysis. 

We first employ different allocation rules in accordance with the ENTSO-E report. These configurations range from a national single-zone model to a fully nodal model, as well as the ACER proposals for two to four bidding zones \cite{ACER.2022.BZRConfig}. Each allocation is associated with generation costs as well as redispatch costs to obtain a physically feasible outcome. After computing an allocation, different pricing rules, such as IP, CH, or Join pricing, can be applied. Each pricing rule implies different price levels as well as MWPs for market participants. \textcolor{black}{Prices were capped at 100 EUR/MWh to limit the impact of outliers. Note that prices higher than 100 EUR/MWh are regarded as outliers also in the BZR Report (Section 2.4) \citep{ENTSOE.2022}}. We also consider the Euphemia allocation and prices as the current implementation in European markets. 

\begin{table}[!htp]
	\centering
	\begin{tabular}{c|c|c|c}
		Calendar Weeks of 2009 & Allocation Rules & Pricing Rules & Focus Variables \\
		\hline 
		04 & National & IP & Generation Costs \\
		08 & 2 Zones (k-means) & CH & Redispatch Costs \\
		11 & 2 Zones (spectral) & Join & Prices \\
		15 & 3 Zones & & MWPs \\ 
		16 & 4 Zones & &  \\
		21 & Nodal & &  \\
		31 & \multicolumn{2}{c|}{    Euphemia} &  \\
		48 & & & 
	\end{tabular}
	\caption{Overview of experiments}
	\label{tab:experiment_overview}
\end{table}

\section{Results} \label{sec:results}

In this section, we summarize the main results of our analysis. We start with comparing average generation and total costs, before discussing price levels and MWPs. We report aggregate statistics for all weeks under consideration. \ref{app:repr-days} includes an analysis of representative days with low, medium, and high prices, respectively.

\subsection{Generation and Redispatch Costs} \label{subsec:results_costs}

\textcolor{black}{Table \ref{tab:costs} presents an overview of the average daily generation and redispatch costs for the allocation rules under consideration. Table \ref{tab:percentages} shows the percentage decrease of the nodal costs relative to the total costs associated with different zonal configurations.\footnote{The values are obtained by computing $\frac{x \cdot 100}{y} - 100$, where $x$ is the zonal cost for a specific configuration and $y$ is the nodal cost.}}


\begin{table}[!htp]
	\centering
	\begin{tabular}{c|ccccccc}
		in kEUR & National & 2 Zones (k) & 2 Zones (s) & 3 Zones & 4 Zones & Nodal \\
		\hline
		Generation & 32,145.08 & 32,428.55 & 32,219.17 & 32,198.95 & 32,382.74 & 36,150.7\\
		\hline
		\textcolor{black}{RD-Min-Cost} & 6,140.06 & 5,687.83 & 5,863.74 & 5,827.86 & 5,726.9 & 0 \\
		\textcolor{black}{Total RD-Min-Cost} & 38,285.14 & 38,116.38 & 38,082.91 & 38,026.81 & 38,109.64 & 36,150.7 \\
	\end{tabular}
	\caption{Average Daily Costs}
	\label{tab:costs}
\end{table}


\begin{table}[!htp]
	\centering
	\begin{tabular}{c|ccccccc}
		\% & National & 2 Zones (k) & 2 Zones (s) & 3 Zones & 4 Zones  \\
		\hline
		Total RD-Min-Cost & 5.90 & 5.44 & 5.34 & 5.19 & 5.42  \\
	\end{tabular}
	\caption{Percentage decrease of the nodal costs relative to the total costs associated with different zonal configurations}
	\label{tab:percentages}
\end{table}

The numbers reported in Table \ref{tab:costs} have realistic orders of magnitude as compared to those provided by the German Bundesnetzagentur \citep{Bundesnetzagentur.2024}. 
However, our goal is not to come up with a precise prediction of costs for 2025, but to obtain a relative comparison between different zonal configurations and nodal pricing. 
Adding network constraints cannot lead to lower objective function values in the cost minimization. 
As the national allocation rule neglects all transmission constraints, its generation costs are a lower bound for all other allocation rules. On average, the suggested zonal configurations require only slightly more generation costs, suggesting that adding few transmission constraints to the computation has little impact on the costs. In contrast, the nodal allocation rule includes the entire transmission network. Obviously, higher generation costs are required to reach an allocation with feasible power flows. On average, generation costs increase by 5.90\% after including all network constraints for the nodal model compared to the national configuration before redispatch. 

After redispatch, the national configuration has 5.90\% higher costs than the nodal model. 
Note that we assume that all generators can be redispatched in our model, such that redispatch costs in reality might be higher.
\textcolor{black}{Also, in our zonal market clearing problem, we do not consider critical elements within zones and assume that the demand is inelastic and must be fully satisfied in any feasible allocation.}
In comparison, the average daily redispatch costs in Germany between 2020 and 2022 amount to roughly kEUR 7,200 \citep{Bundesnetzagentur.2022}.\footnote{These costs include power- and voltage-related measures, countertrading and redispatch test procedures.} The average daily redispatch costs in Germany in 2022 were even at around kEUR 11,600, having increased by almost 100\% compared to 2021. 
\textcolor{black}{We assume zero redispatch for a nodal system that satisfies $\Psi^{DC}$, although in practice a post-processing step is required to ensure that the DCOPF solution is ACOPF-feasible \citep{taheri2024acfeasibility}. In the US, this post-processing depends on the ISO and involves using an ACOPF model to adjust the DCOPF solution, accounting for real power losses, voltage magnitudes, and reactive power needs \citep{caiso2024operating}.}

\subsubsection{Welfare losses with Euphemia - Impact of Non-Uniform Pricing}

Euphemia maximizes welfare subject to linear and \textcolor{black}{uniform} prices, which is different from the allocation problem (\ref{model:general}). In our analysis, the average daily welfare loss associated with Euphemia compared to the national configuration (without any network constraints) is 0.34\% -- or EUR 109,801.24 in absolute terms. This is comparable with previous estimate of 0.05\% as a relative loss \citep{Meeus.2009}.\footnote{\textcolor{black}{Note that the 0.05\% relative welfare loss does not correspond to the official Euphemia implementation, but to a simulation that considers block order restrictions \citep{Meeus.2009}.}} Disregarding market coupling, the cuts introduced by the Euphemia algorithm to get to linear prices do not deteriorate the outcome by a lot. 
After redispatch, Euphemia suffers an average daily loss of 7.9\% (or kEUR 2,859.80) in total costs compared to the nodal allocation rule. We argue that the national, zonal, or Euphemia models differ only marginally in terms of allocation, and only a transition to a nodal allocation rule would substantially affect the average generation and redispatch costs. 

\subsection{Price Levels}

Table \ref{tab:prices} provides average prices over all hours of the weeks under consideration, while Table \ref{tab:volas} price standard deviations for all computed prices under the respective allocation and pricing rule. The median prices are presented in \ref{app:median-prices}.

\begin{table}[!htp]
	\centering
	\begin{tabular}{c|cccccc}
		in EUR/MWh & National & 2 Zones (k) & 2 Zones (s) & 3 Zones & 4 Zones &  Nodal \\
		\hline
		IP & 45.92 & 45.45 & 46.56 & 44.90 & 44.37 & 46.63 \\
		CH & 40.13 & 40.53 & 40.81 & 40.21 & 41.35 & 44.79 \\
		Join & 46.31 & 48.04 & 46.0 & 45.51 & 48.41 & 47.65 \\
		Euphemia & 47.08 & & & & &
	\end{tabular}
	\caption{Average Prices}
	\label{tab:prices}
\end{table}

\begin{table}[!htp]
	\centering
	\begin{tabular}{c|cccccc}
		in EUR/MWh & National & 2 Zones (k) & 2 Zones (s) & 3 Zones & 4 Zones &  Nodal \\
		\hline
		IP & 19.76 & 19.55 & 19.79 & 19.45 & 19.57 & 19.54 \\
		CH & 14.43 & 14.42 & 14.74 & 14.39 & 14.32 & 16.78 \\
		Join & 19.97 & 20.45 & 20.12 & 20.34 & 20.64 & 20.06  \\
		Euphemia & 20.61 & & & & &
	\end{tabular}
	\caption{Price Standard Deviation}
	\label{tab:volas}
\end{table}

Note that the mean BZR price for the weeks under consideration was EUR/MWh 40.17 and the standard deviation EUR/MWh 18.51, computed bi-hourly based on a linear relaxation. 

Average CH \textcolor{black}{(Convex Hull)} prices are slightly lower compared to the other pricing rules under consideration, with lower standard deviation and irrespective of the allocation rule. In contrast, median prices are at comparable levels across all pricing rules. Since CH prices are unaffected by pre-determined generator commitments, they tend to be lower and more stable. In particular, prices can be set by a cheaper -- yet uncommitted -- generator \citep{Schiro.2016}. In contrast, average Euphemia prices tend to be slightly higher than zonal prices, while median prices are in the same ranges. This finding implies that the price-setting generator is usually identical for Euphemia and zonal prices, yet sometimes the Euphemia algorithm introduces a cut that disallows a generator that would otherwise be dispatched. As a result, a higher-price generator may set the price.

\begin{figure*}[!htp]
	\centering
	\begin{subfigure}[b]{0.23\textwidth}
		\centering
		\includegraphics[width=\textwidth]{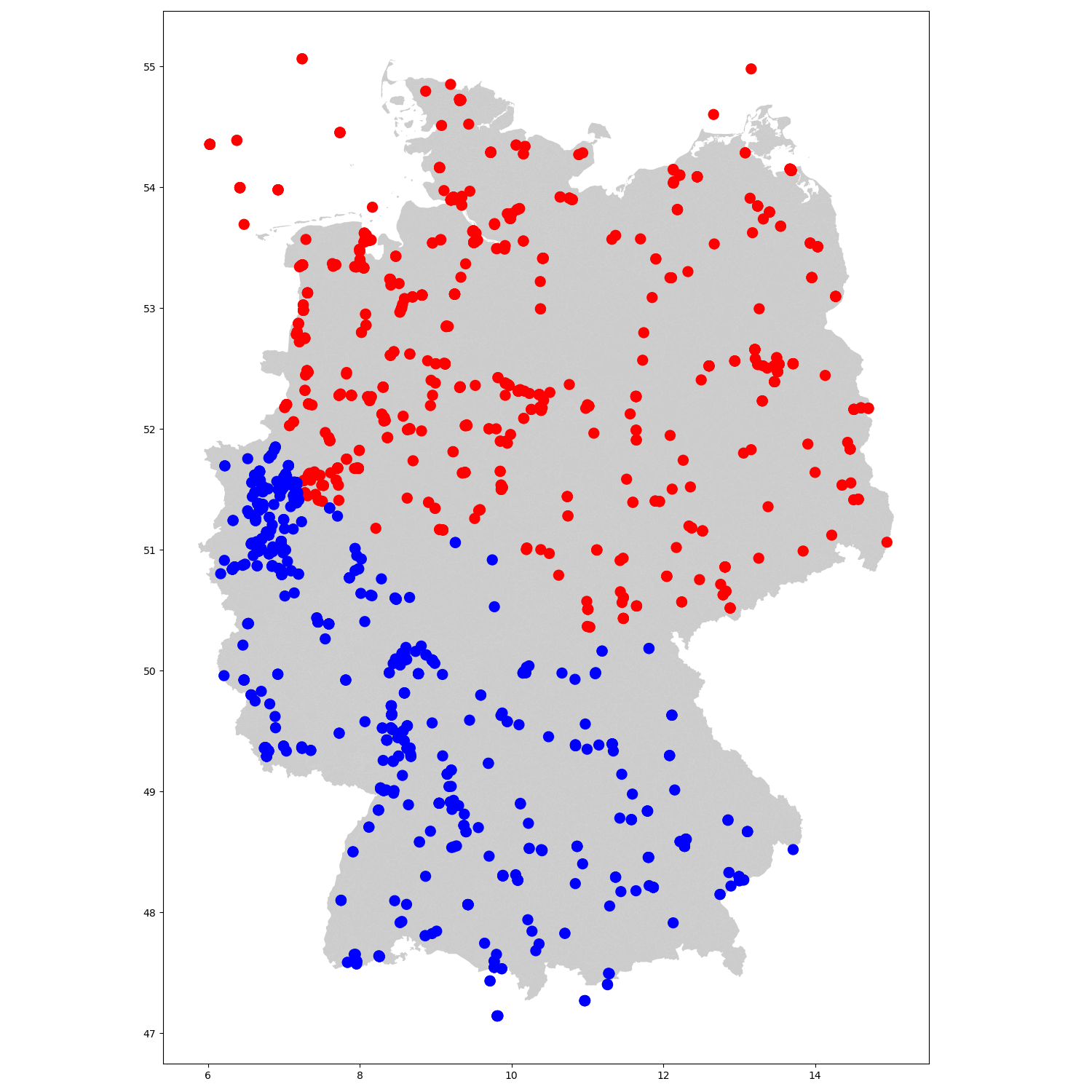}
		\caption{\small 2 Zones (k)}   
		\label{fig:DE2k}
	\end{subfigure}
	\begin{subfigure}[b]{0.23\textwidth}  
		\centering 
		\includegraphics[width=\textwidth]{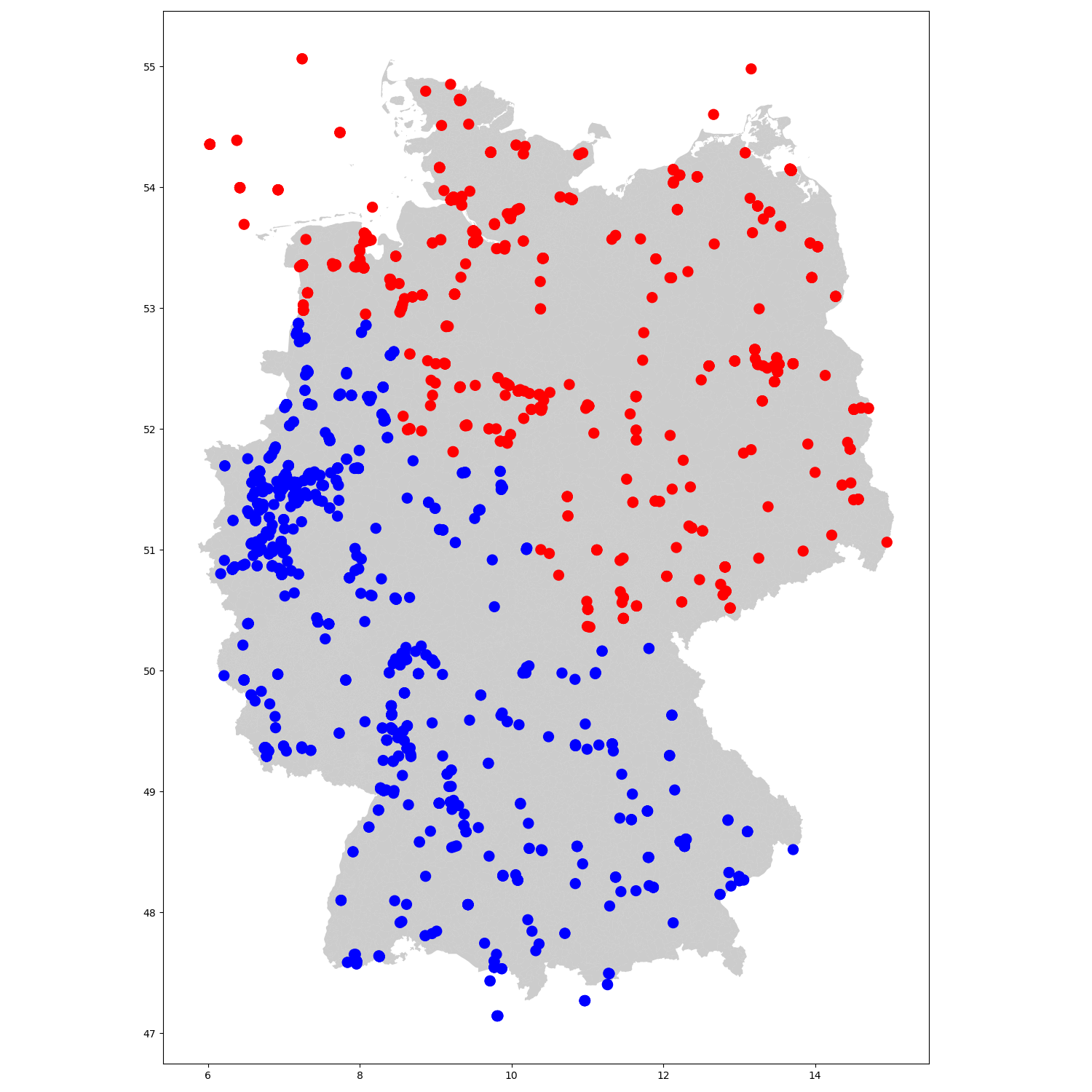}
		\caption{\small 2 Zones (s)}
		\label{fig:DE2s}
	\end{subfigure}
	\begin{subfigure}[b]{0.23\textwidth}   
		\centering 
		\includegraphics[width=\textwidth]{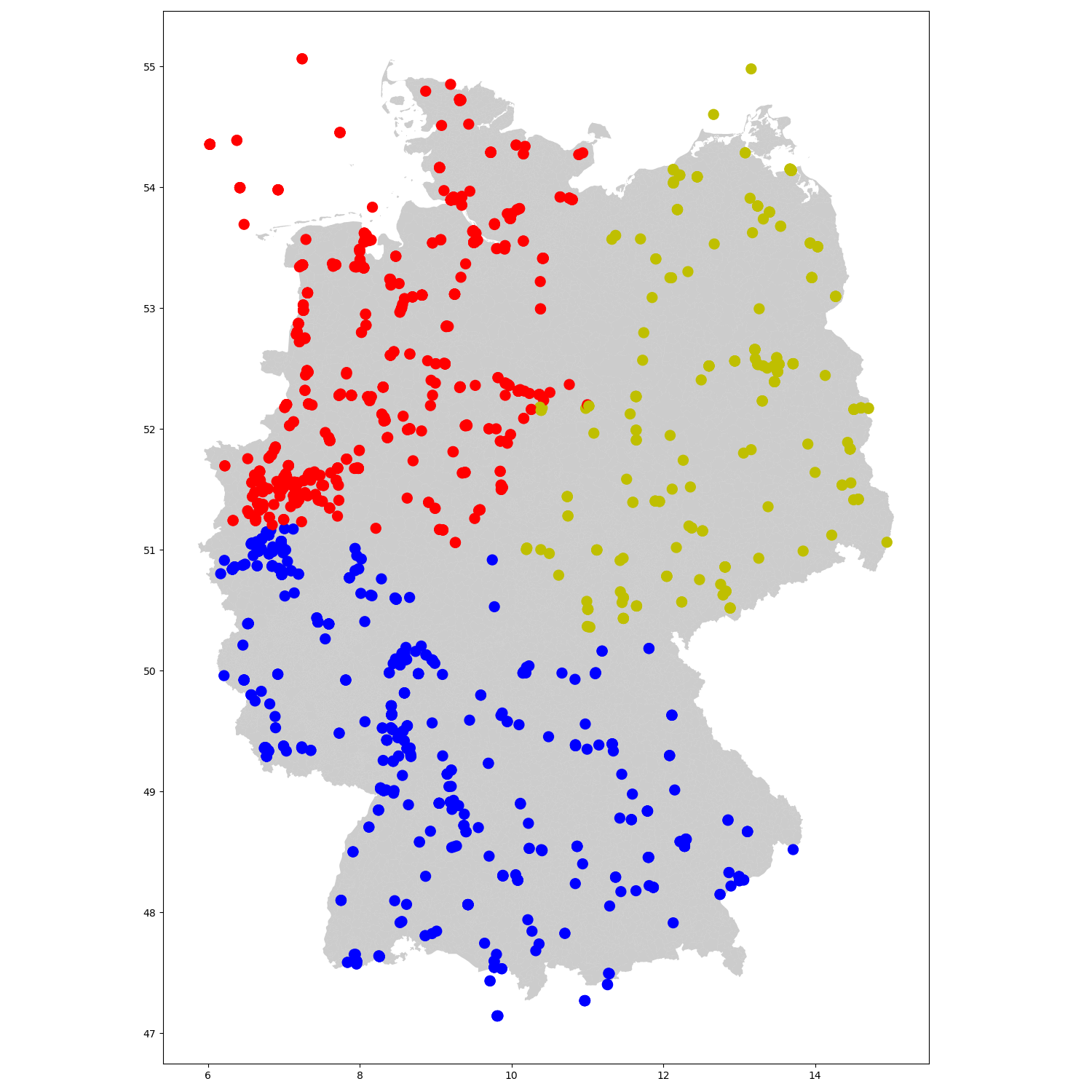}
		\caption{\small 3 Zones}    
		\label{fig:DE3}
	\end{subfigure}
	\begin{subfigure}[b]{0.23\textwidth}   
		\centering 
		\includegraphics[width=\textwidth]{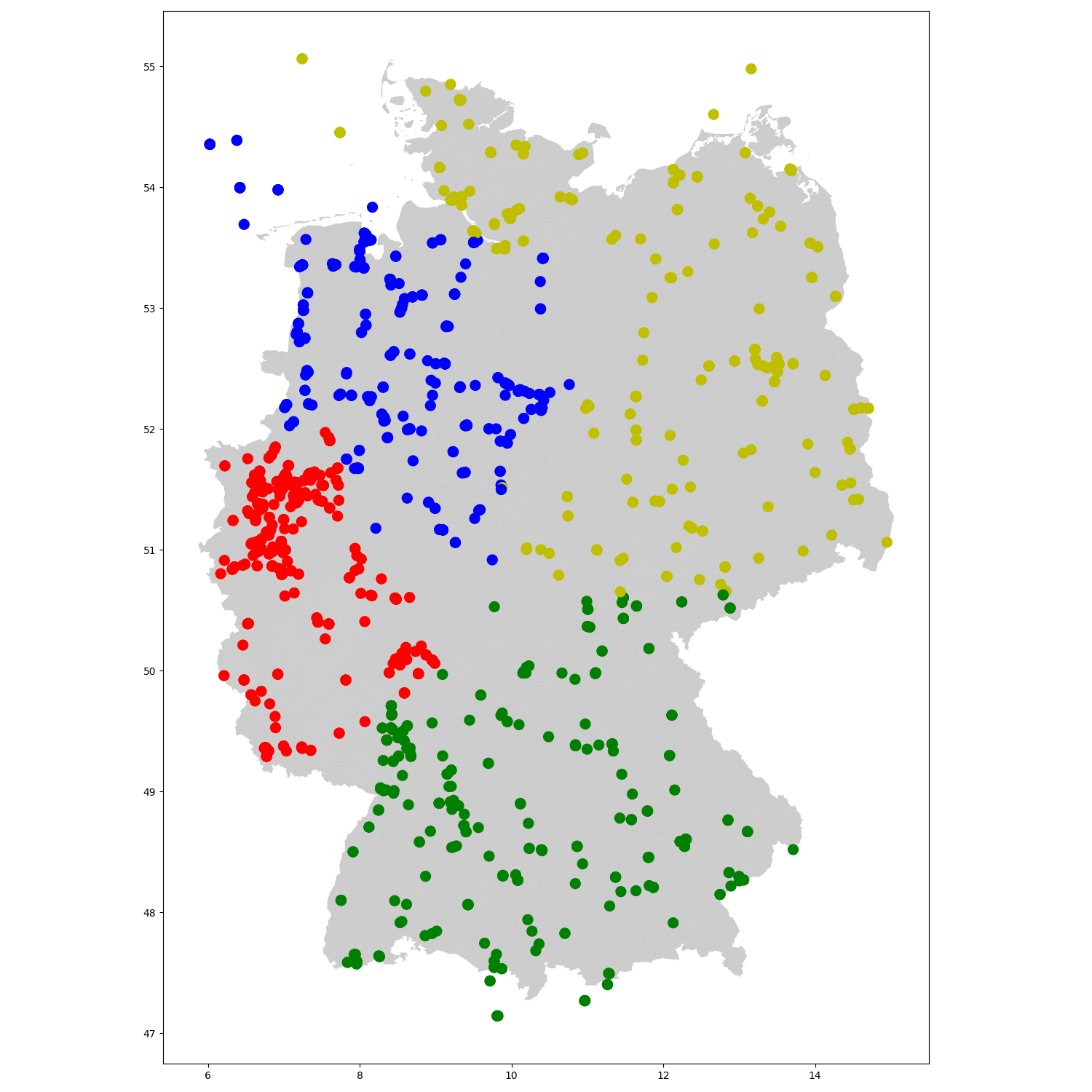}
		\caption{\small 4 Zones}  
		\label{fig:DE4}
	\end{subfigure}
	\caption{Zonal Configurations} 
	\label{fig:zone_plot}
\end{figure*}

Price distributions are generally similar across the different zonal configurations for a specific pricing rule. Cross-zonal flow constraints are rarely tight and thus zonal prices are identical to national (single-zone) prices in many hours. This observation suggests that splits into only a few zones, on average, have little impact on prices and thus provide few locational incentives. Table \ref{tab:zon_prices_drilldown} illustrates that, on average, no major price discrepancies between zones can be expected. The zones are depicted in Figure \ref{fig:zone_plot}.

\begin{table}[!htp]
	\centering
	\begin{tabular}{c|cccccc}
		in EUR/MWh & National & 2 Zones (k) & 2 Zones (s) & 3 Zones & 4 Zones \\
		\hline
		\textcolor{Blue}{Zone 1} & 45.92 (19.76) & 46.88 (19.69) & 48.36 (19.87) & 45.42 (19.68) & 42.43 (19.22) \\
		\textcolor{OrangeRed}{Zone 2} & & 44.02 (19.30) & 44.76 (19.55) & 44.28 (19.37) & 45.58 (19.96) \\
		\textcolor{Goldenrod}{Zone 3} & & & & 45.02 (19.28) & 44.08 (19.05) \\
		\textcolor{OliveGreen}{Zone 4} & & & & & 45.38 (19.86) \\ 
	\end{tabular}
	\caption{IP Zonal Prices Average and Standard Deviation per Zone}
	\label{tab:zon_prices_drilldown}
\end{table}

A more in-depth examination of zonal pricing reveals the impact of cross-zonal flows in equalizing prices across different zones. Specifically, while southern zones face a relative shortage of generation capacity compared to demand, the capacity of cross-zonal transmission lines is typically high enough to transmit electricity from northern zones. Our sensitivity analysis revealed that if cross-zonal flows were limited or prohibited, zonal prices can display differences of more than 30 EUR/MWh between northern and southern regions on some days. The small differences in zonal prices observed in Table \ref{tab:zon_prices_drilldown} are due to the cross-zonal transmission capacity. As described in Section \ref{sec:zonal-market-clearing}, to compute a zonal solution, we model the interconnectors individually but we do not consider critical network elements and loop flows, which might decrease the available capacity between zones in practice.

Under nodal pricing (ca. 40,000 prices per day for 1670 nodes -- compared to, e.g., 24 national prices per day) more price outliers were observed than under zonal pricing. 
For instance, 0.13\% of IP \textcolor{black}{(Integer Programming)} prices had to be capped 100 EUR/MWh under the zonal configurations, compared to 0.25\% under nodal pricing. Without capping, the maximum observed IP and Join price would be as high as 62,000 EUR/MWh for an hour of extreme scarcity. Note that SDAC also imposes a price cap of 5,000 EUR/MWh for such periods. CH prices, in contrast, are less prone to outliers, peaking at 423 EUR/MWh. 
Thus tight flow constraints can increase price levels in the short run. The zonal allocation rules oversimplify the transmission network to an extent that nodal prices can hardly be lower, even without congestion. In contrast, when congestion is present in parts of the grid, nodal prices will be higher, resulting in higher average and median prices. Upon further analysis, it was found that nodal prices have high sensitivity to transmission line susceptances, a phenomenon previously described in \citet{Bichler.2023}.

\begin{figure*}[!htp]
	\centering
	\begin{subfigure}[b]{0.32\textwidth}
		\centering
		\includegraphics[width=\textwidth]{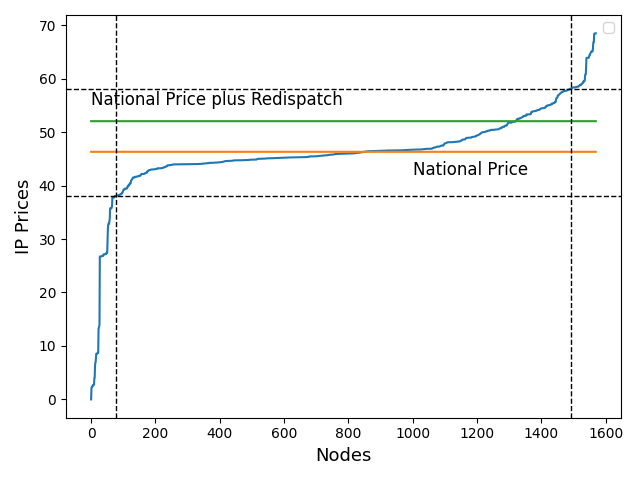}
		\caption{\small IP Prices}   
		\label{fig:sorted_ip}
	\end{subfigure}
	\begin{subfigure}[b]{0.32\textwidth}  
		\centering 
		\includegraphics[width=\textwidth]{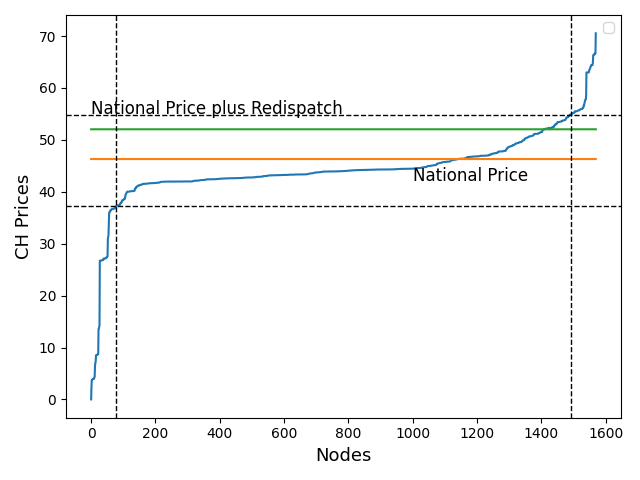}
		\caption{\small CH Prices}
		\label{fig:sorted_ch}
	\end{subfigure}
	\begin{subfigure}[b]{0.32\textwidth}   
		\centering 
		\includegraphics[width=\textwidth]{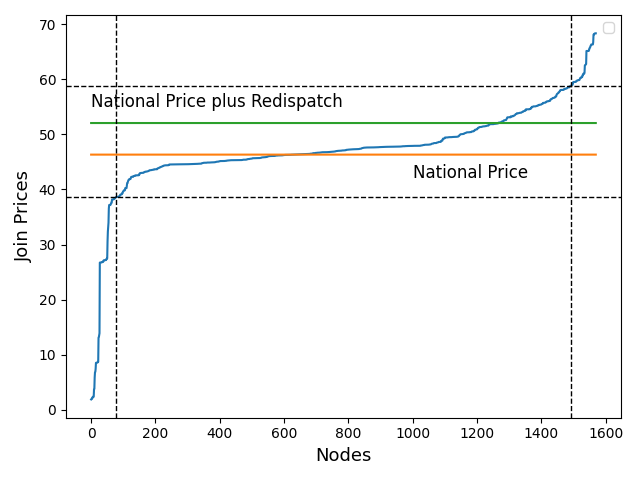}
		\caption{\small Join Prices}    
		\label{fig:sorted_join}
	\end{subfigure}
	\caption{Sorted Nodal Prices [EUR/MWh]} 
	\label{fig:sorted_nodal_prices}
\end{figure*}

Figure \ref{fig:sorted_nodal_prices} illustrates nodal IP, CH, and Join prices, averaged over all hours and sorted in ascending order. The dashed horizontal and vertical lines mark the 5th and 95th percentiles for prices and nodes, respectively. Additionally, the figure shows the average national prices and the national average including a uniform EUR/MWh price adder for redispatch costs. This price adder is calculated as the total redispatch costs (ca. EUR 478 million) divided by the total demand (ca. 83.3 TWh), and equals 5.73 EUR/MWh. Notably, with redispatch taken into account, higher average prices are expected at some of the nodes. For the majority of nodes, however, the change in average prices is minor or even negative under nodal pricing.

\begin{figure*}[!htp]
	\centering
	\begin{subfigure}[b]{0.32\textwidth}
		\centering
		\includegraphics[width=\textwidth]{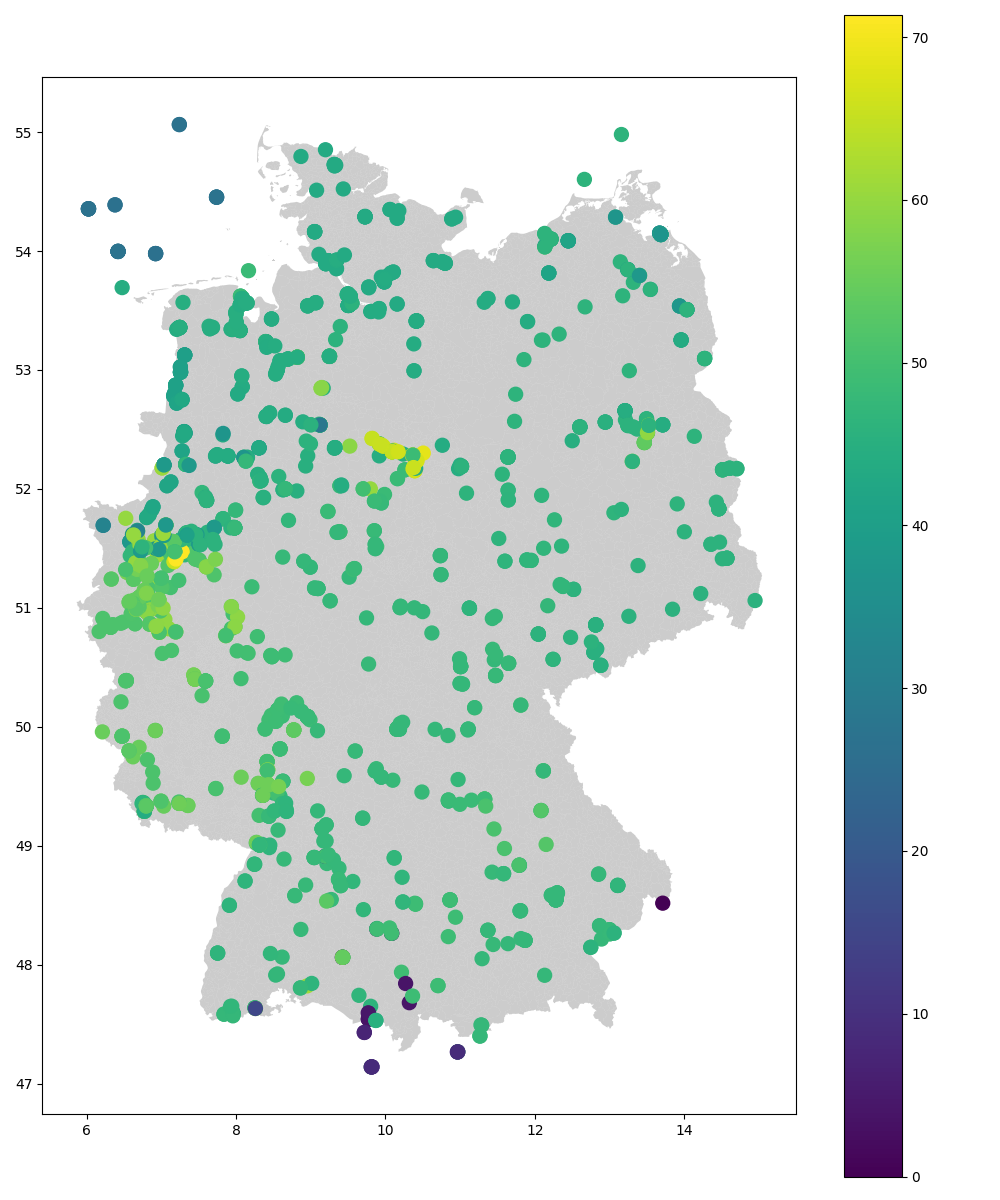}
		\caption{\small IP Prices}   
		\label{fig:map_ip_average}
	\end{subfigure}
	\begin{subfigure}[b]{0.32\textwidth}  
		\centering 
		\includegraphics[width=\textwidth]{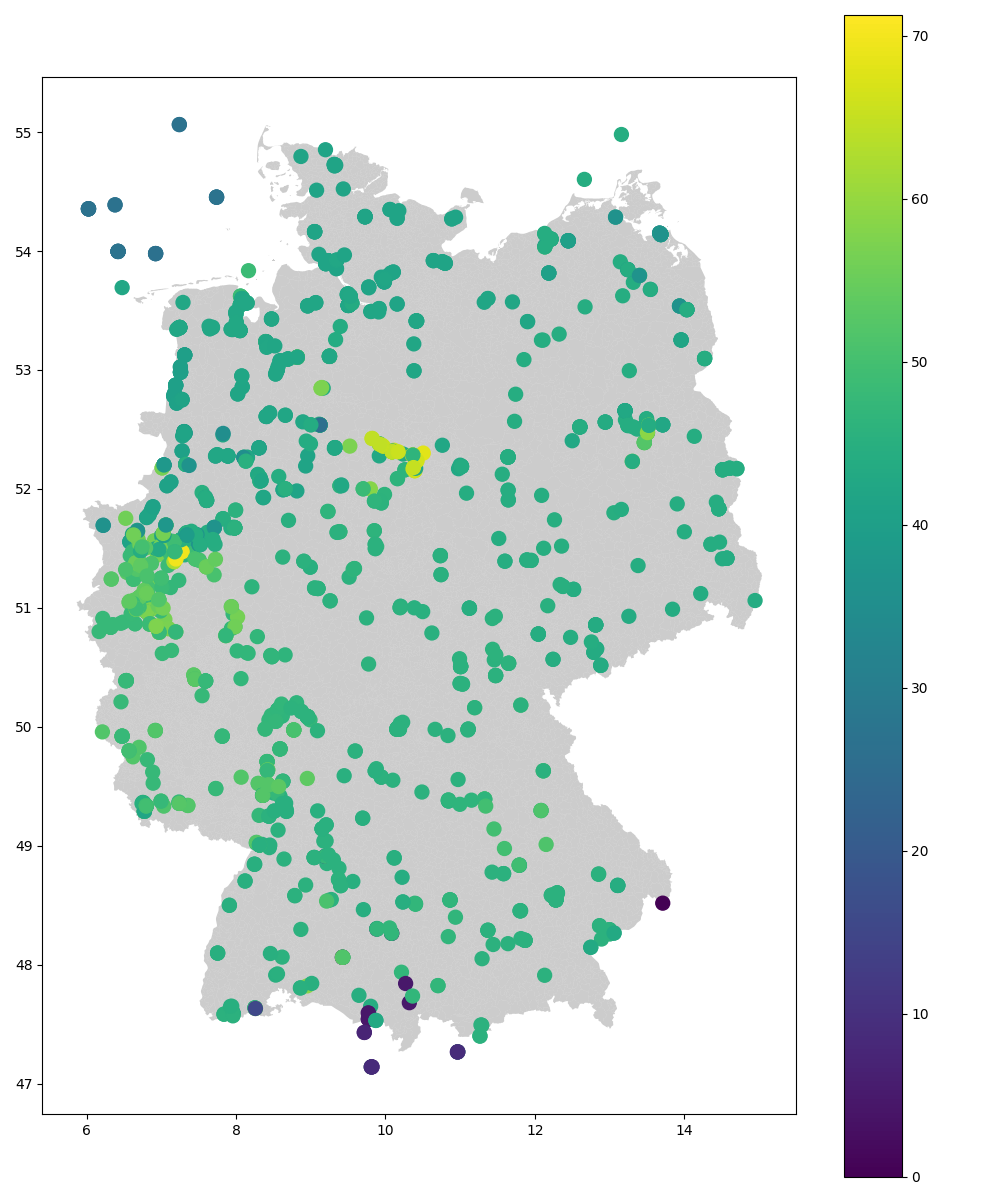}
		\caption{\small CH Prices}
		\label{fig:map_elmp_average}
	\end{subfigure}
	\begin{subfigure}[b]{0.32\textwidth}   
		\centering 
		\includegraphics[width=\textwidth]{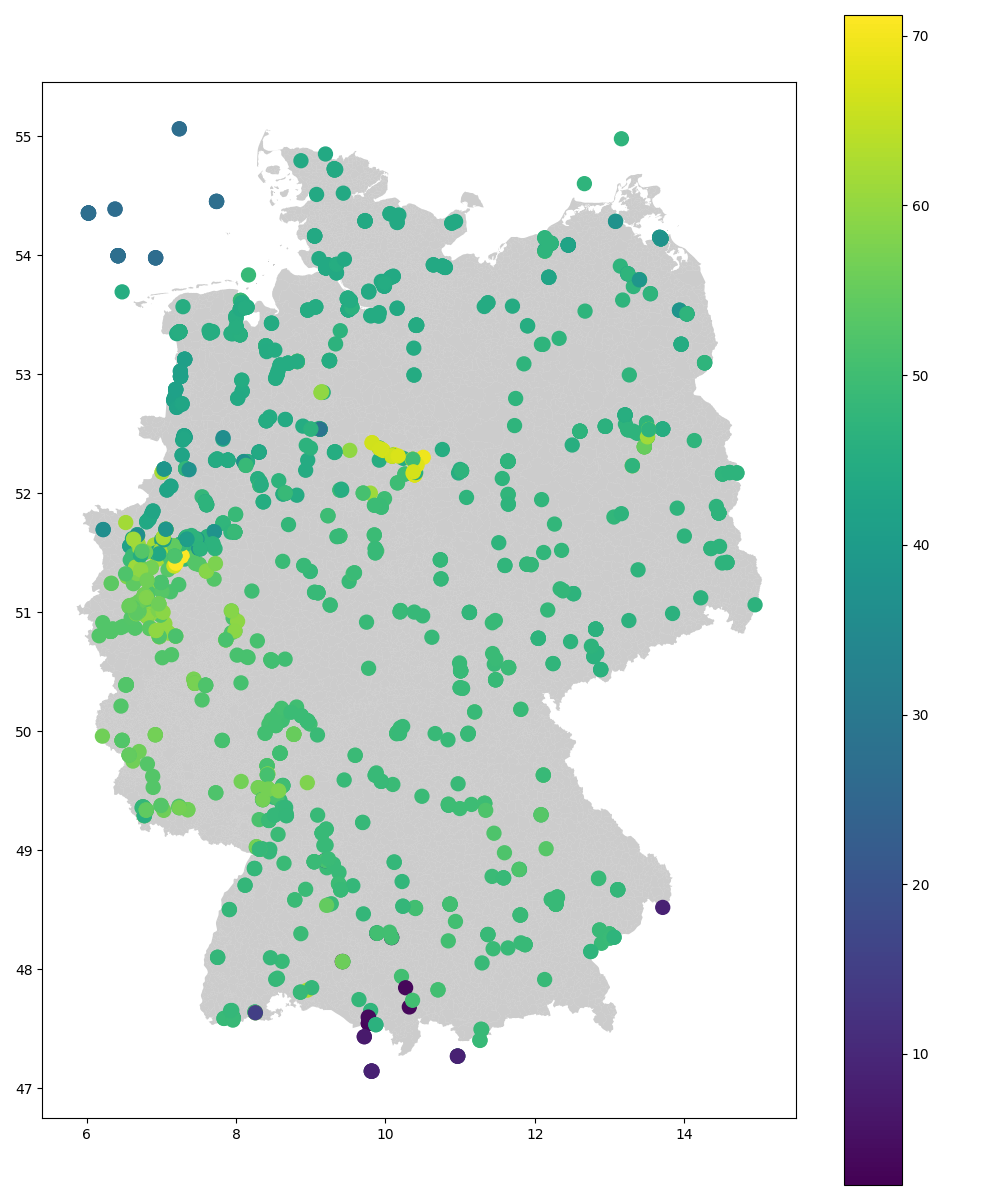}
		\caption{\small Join Prices}    
		\label{fig:map_join_average}
	\end{subfigure}
	\caption{Average Nodal Prices [EUR/MWh]} 
	\label{fig:nodal_prices_plot}
\end{figure*}

Figure \ref{fig:nodal_prices_plot} maps these average prices to their geographical locations. As much of the electricity supply, particularly wind energy, is located in Northern Germany, we observe that average nodal prices tend be higher in Southern Germany. Generally, this price gap is moderate and provides locational incentives for market participants. The choice of pricing rule does not seem to have a major effect on average price levels.  

We observe the highest prices in Northern Central Germany in a region with a high share of expensive biomass energy and the lowest prices in the Alps region with inexpensive hydro energy. The associated nodes are relatively isolated with frequent congestion. This behavior might follow from imprecise estimations of generation costs and line limits described in Section \ref{sec:data}, yet price outliers can also be observed in the ENTSO-E report \citep{ENTSOE.2022}.

\begin{figure*}[!htp]
	\centering
	\begin{subfigure}[b]{0.32\textwidth}
		\centering
		\includegraphics[width=\textwidth]{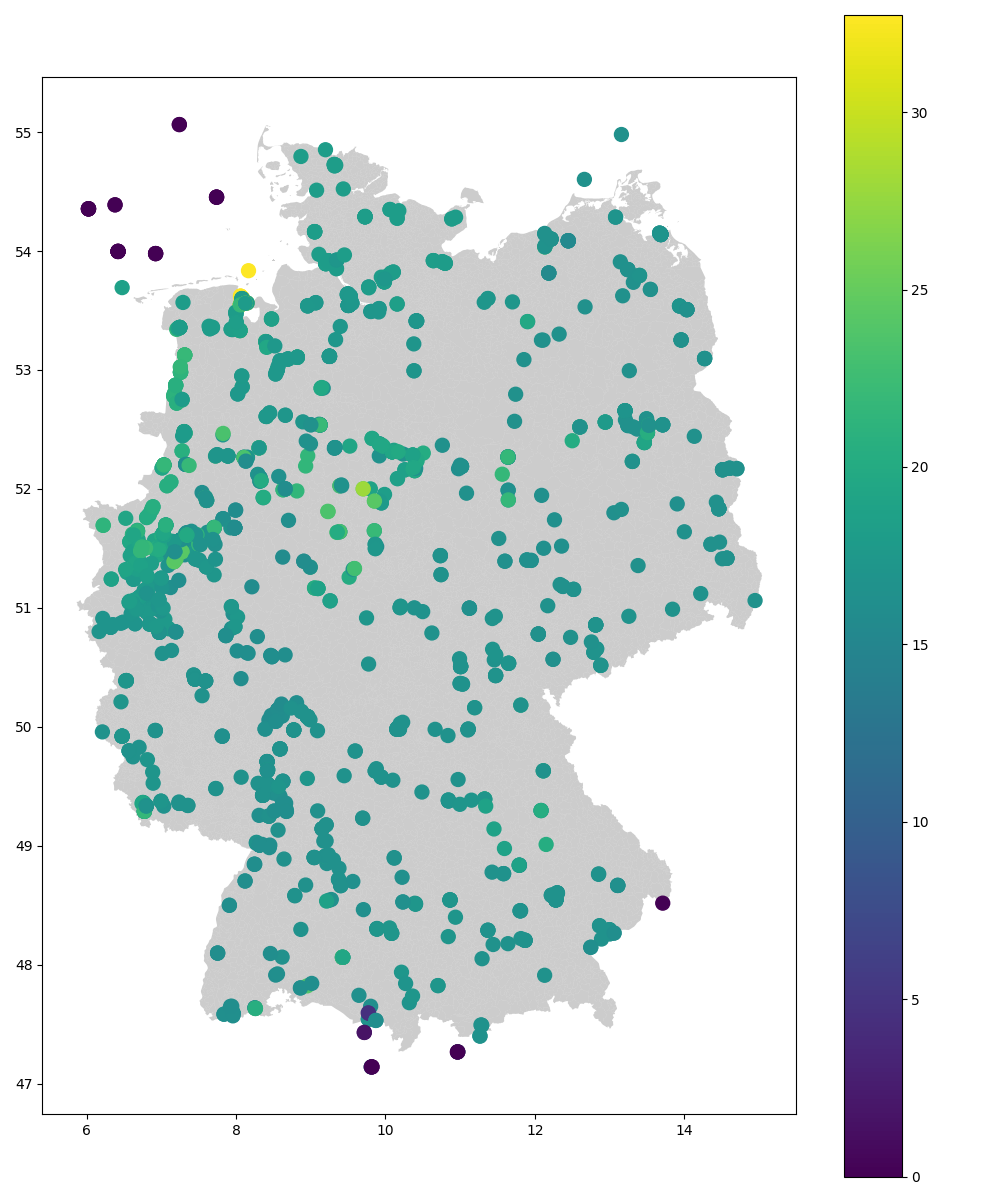}
		\caption{\small IP Prices}   
		\label{fig:ip_std_dev_map}
	\end{subfigure}
	\begin{subfigure}[b]{0.32\textwidth}  
		\centering 
		\includegraphics[width=\textwidth]{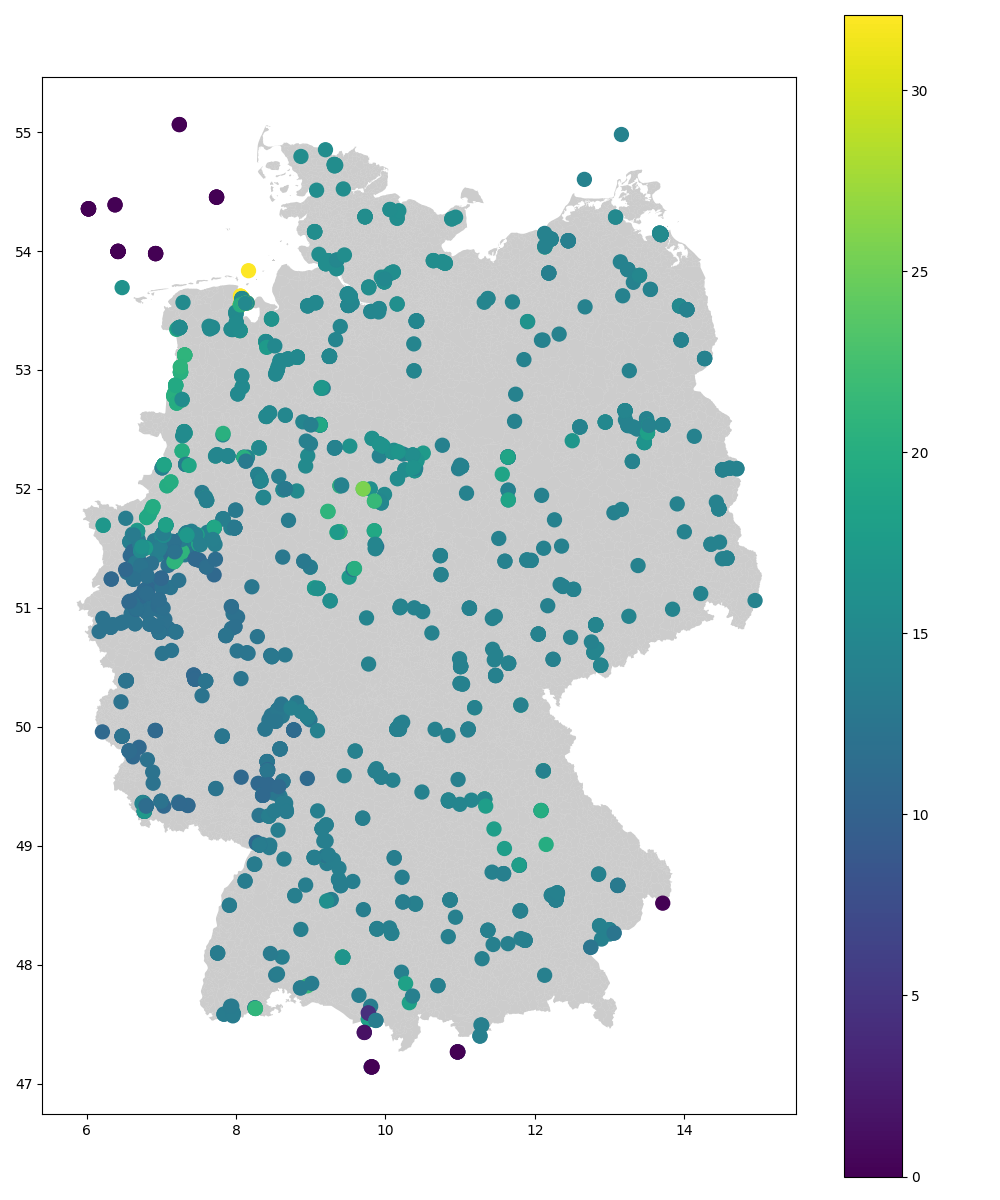}
		\caption{\small CH Prices}
		\label{fig:ch_std_dev_map}
	\end{subfigure}
	\begin{subfigure}[b]{0.32\textwidth}   
		\centering 
		\includegraphics[width=\textwidth]{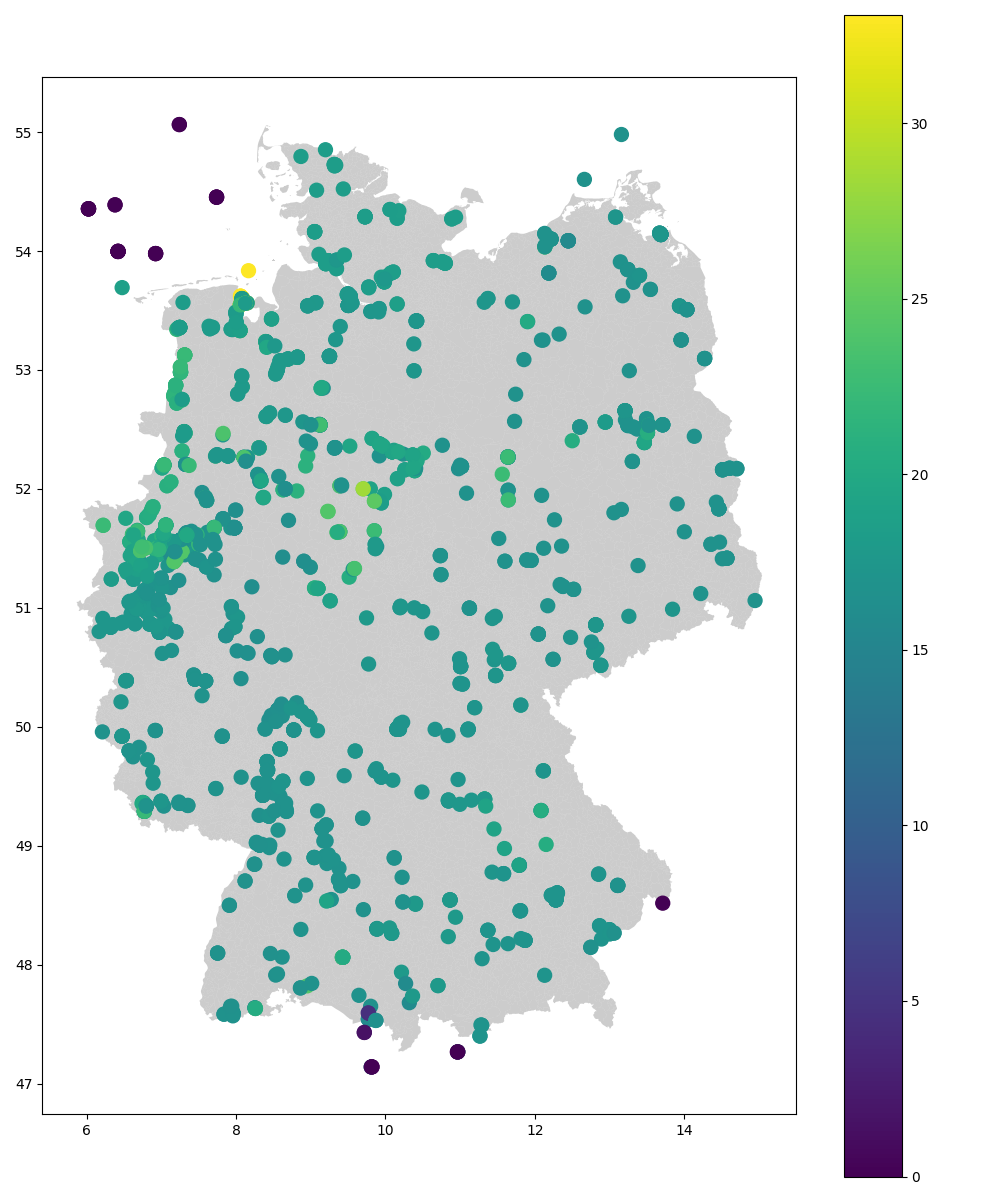}
		\caption{\small Join Prices}    
		\label{fig:join_std_dev_map}
	\end{subfigure}
	\caption{Standard Deviation Nodal Prices [EUR/MWh]} 
	\label{fig:nodal_std_dev_plot}
\end{figure*}

\begin{figure*}[!htp]
	\centering
	\begin{subfigure}[b]{0.32\textwidth}
		\centering
		\includegraphics[width=\textwidth]{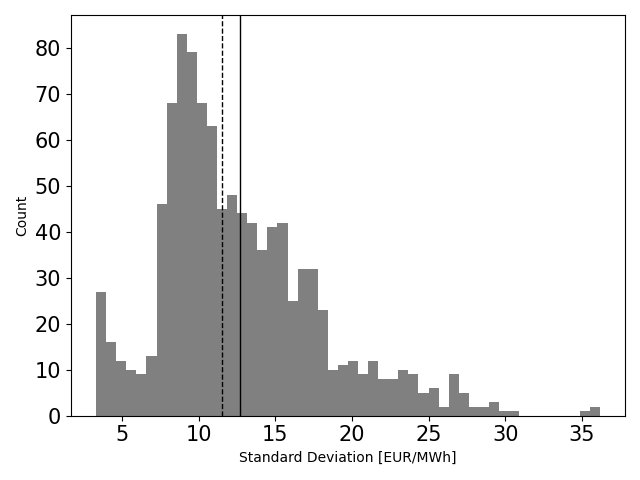}
		\caption{\small IP Prices}   
		\label{fig:ip_std_dev_prices_per_hour}
	\end{subfigure}
	\begin{subfigure}[b]{0.32\textwidth}  
		\centering 
		\includegraphics[width=\textwidth]{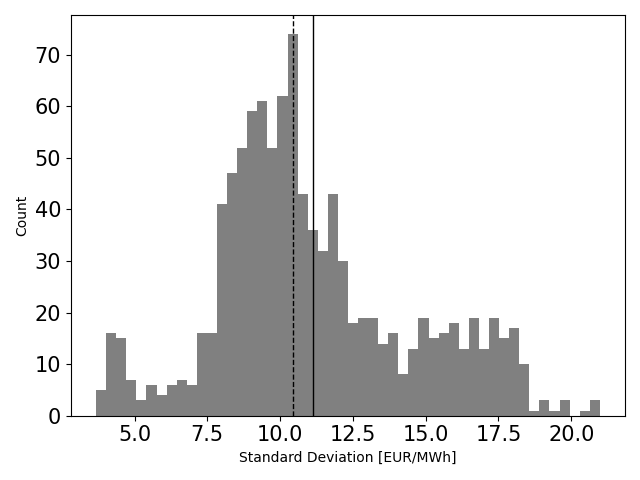}
		\caption{\small CH Prices}
		\label{fig:ch_std_dev_prices_per_hour}
	\end{subfigure}
	\begin{subfigure}[b]{0.32\textwidth}   
		\centering 
		\includegraphics[width=\textwidth]{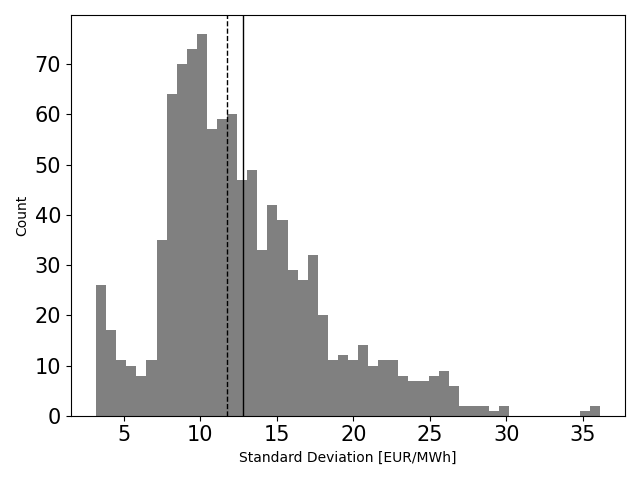}
		\caption{\small Join Prices}    
		\label{fig:join_std_dev_prices_per_hour}
	\end{subfigure}
	\caption{Histograms of Hourly Standard Deviations Across Nodes [EUR/MWh]} 
	\label{fig:hist_std_dev_hour}
\end{figure*}

Nodal prices typically vary across locations for each hour, and thereby set locational incentives for flexible units or storage. As seen in Figure \ref{fig:nodal_std_dev_plot}, standard deviations of nodal prices resemble each other across nodes and pricing rules. Figure \ref{fig:hist_std_dev_hour} illustrates the distribution of hourly standard deviations of prices across nodes, with mean and median standard deviations as solid and dashed lines, respectively. The nodal price variances can be substantial. For example, in the extreme first hour of 2009/11/23, IP prices vary across the country between 0 EUR/MWh and 100 EUR/MWh. This sets incentives for demand response and short-term adjustments in particular in energy-intensive industrial production. Persistent high nodal prices signal transmission bottlenecks and direct investments in grid expansion. 

\begin{figure*}[!htp]
	\centering
	\begin{subfigure}[b]{0.32\textwidth}
		\centering
		\includegraphics[width=\textwidth]{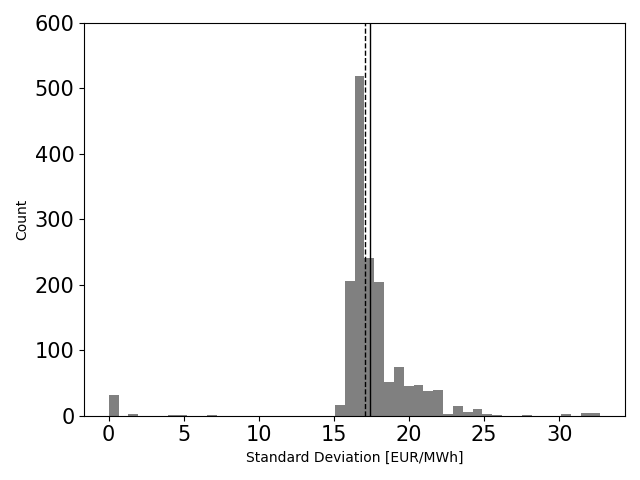}
		\caption{\small IP Prices}   
		\label{fig:ip_std_dev_prices_per_node}
	\end{subfigure}
	\begin{subfigure}[b]{0.32\textwidth}  
		\centering 
		\includegraphics[width=\textwidth]{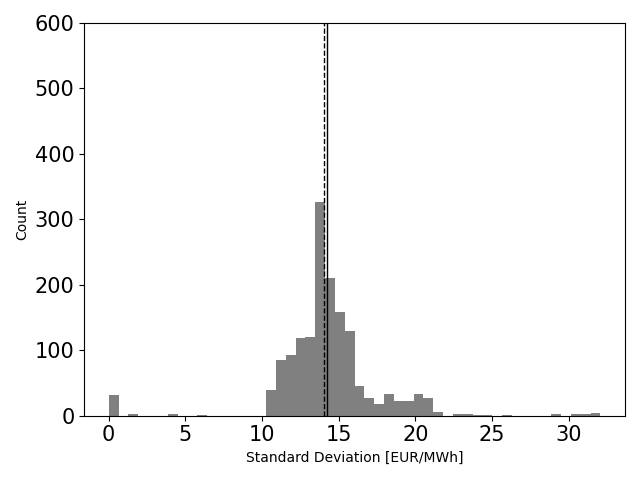}
		\caption{\small CH Prices}
		\label{fig:ch_std_dev_prices_per_node}
	\end{subfigure}
	\begin{subfigure}[b]{0.32\textwidth}   
		\centering 
		\includegraphics[width=\textwidth]{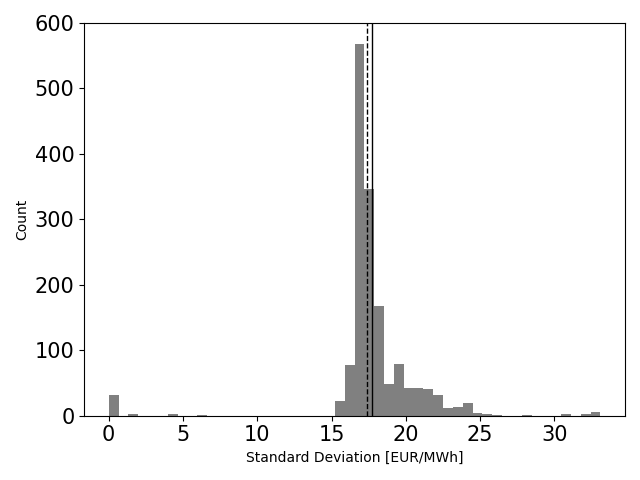}
		\caption{\small Join Prices}    
		\label{fig:join_std_dev_prices_per_node}
	\end{subfigure}
	\caption{Histograms of Nodal Standard Deviations Across Hours [EUR/MWh]} 
	\label{fig:hist_std_dev_node}
\end{figure*}

A primary goal in the design of new bidding zones is the reduction of price variance within a zone \citep{ACER.2022.BZRConfig}, which signals congestion. 
Tables \ref{tab:volas} and \ref{tab:zon_prices_drilldown} indicate that the standard deviation of zonal prices does not decrease with the number of zones. However, varying prices can have two main reasons: congestion and temporal changes in supply and demand. Figure \ref{fig:hist_std_dev_hour} considers hourly standard deviations of prices across all nodes, meaning it illustrates the price variance based on grid congestions in each hour, which we refer to as \textit{congestion-based standard deviation}. In contrast, Figure \ref{fig:hist_std_dev_node} includes histograms of the standard deviation of nodal prices across all hours, meaning it illustrates the price variance based on temporal effects of supply and demand at each node, which we refer to as \textit{time-based standard deviation}. Together, Figures \ref{fig:hist_std_dev_hour} and \ref{fig:hist_std_dev_node} represent a composition of the standard deviations reported for nodal prices in Table \ref{tab:volas}. Note that prices at a single node can vary substantially over time due to changes in supply and demand. This time-based standard deviation exceeds the congestion-based standard deviation. For example, IP prices exhibit a mean congestion-based standard deviation of EUR/MWh 12.68, compared to a mean time-based standard deviation of EUR/MWh 17.39. These numbers provide evidence that temporal fluctuations of supply and demand have greater impact on price variance compared to congestion.

Tables \ref{tab:cong_std_dev} and \ref{tab:time_std_dev} provide a more detailed variance decomposition for IP prices under all zonal configurations. In particular, Table \ref{tab:cong_std_dev} indicates the congestion-based standard deviation of nodal IP prices, averaged over all nodes assigned to a zone. Equivalently, Table \ref{tab:time_std_dev} summarizes the time-based standard deviations grouped by zone. Specifically, the time-based standard deviation always exceeds the congestion-based standard deviation, implying that temporal effects drive price variance more than congestion effects. At the same time, the time-based standard deviation varies only little across allocation rules, suggesting that nation-wide fluctuations in supply and demand, on average, affect all price zones equally. In contrast, the congestion-based standard deviation has a slightly decreasing trend with more zones, indicating that zonal splits can reduce price variance at least for some of the zones (e.g., Zone 3). This effect, however, is not consistent across all zones, implying that the cross-zonal lines fail to encompass all congested lines. Consequently, the computed dispatch does not account for the remaining intra-zonal congestion, necessitating costly redispatch under all zonal models, as shown in Table \ref{tab:costs}. \cite{Dobos.2024} demonstrate that the proposed zone configurations lack stability and fail to effectively segregate nodes along congested lines 
As a result, the congestion-based standard deviation exhibits minimal reduction as the number of zones increases, and redispatch costs remain high.

\textcolor{black}{Before moving to the next section, let us emphasize that day-ahead prices alone cannot fully determine the desirability of different market designs or bidding zone configurations. These prices do not reflect: (1) the magnitude of redispatch costs, (2) network tariffs (\textit{Netzentgelte}), which TSOs use to cover redispatch costs and maintain network infrastructure, (3) congestion revenues received by TSOs, and (4) payments required to address non-convexities in generation profiles (see Section \ref{sec:pricing_on_non_convex_markets}).}

\begin{table}[!htp]
	\centering
	\begin{tabular}{c|ccccc}
		in EUR/MWh & National & 2 Zones (k) & 2 Zones (s) & 3 Zones & 4 Zones \\
		\hline
		\textcolor{Blue}{Zone 1} & 12.19 & 12.77 & 12.71 & 11.99 & 12.25 \\
		\textcolor{OrangeRed}{Zone 2} & & 10.25 & 10.26 & 12.32 & 11.16 \\
		\textcolor{Goldenrod}{Zone 3} & & & & 6.27 & 6.07 \\
		\textcolor{OliveGreen}{Zone 4} & & & & & 13.21
	\end{tabular}
	\caption{Average Congestion-Based Standard Deviation of Nodal IP Prices}
	\label{tab:cong_std_dev}
\end{table}

\begin{table}[!htp]
	\centering
	\begin{tabular}{c|ccccc}
		in EUR/MWh & National & 2 Zones (k) & 2 Zones (s) & 3 Zones & 4 Zones \\
		\hline
		\textcolor{Blue}{Zone 1} & 17.39 & 17.09 & 17.39 & 16.58 & 18.28 \\
		\textcolor{OrangeRed}{Zone 2} & & 17.65 & 17.40 & 18.21 & 17.56 \\
		\textcolor{Goldenrod}{Zone 3} & & & & 17.00 & 17.20 \\
		\textcolor{OliveGreen}{Zone 4} & & & & & 16.12
	\end{tabular}
	\caption{Average Time-Based Standard Deviation of Nodal IP Prices}
	\label{tab:time_std_dev}
\end{table}

\subsection{GLOCs, LLOCs, MWPs}

The average daily GLOCs, LLOCs, and MWPs are summarized in Tables \ref{tab:gloc}--\ref{tab:mwp}, respectively. 
\textcolor{black}{Recall from Section \ref{sec:pricing_on_non_convex_markets} that GLOCs represent the opportunity costs incurred if agents are allowed to deviate from the current allocation to achieve their individual maximum profit, given the market-clearing prices. LLOCs are a subclass of GLOCs quantifying the opportunity costs obtained when agents are not allowed to change their commitment status. MWPs are the minimum compensation provided to market agents, ensuring they do not incurr losses (i.e., negative utility) given the market-clearing prices. GLOCs and LLOCs are not paid in practice. MWPs, however, are paid in US electricity markets and are generally not covered directly by revenues obtained from the buyers through energy purchases in the day-ahead market.}

\begin{table}[!htp]
	\centering
	\begin{tabular}{c|cccccc}
		in EUR & National & 2 Zones (k) & 2 Zones (s) & 3 Zones & 4 Zones &  Nodal \\
		\hline
		IP & 2,975,070.52 & 2,666,810.11 & 3,332,450.37 & 2,563,874.88 & 3,130,778.9 & 5,469,460.3\\
		CH & 124,235.5 & 162,896.45 & 520,529.19 & 141,048.26 & 717,983.02 & 382,104.15 \\
		Join & 981,496.35 & 1,880,938.84 & 1,001,437.75 & 1,801,132.46 & 2,616,044.17 & 4,792,706.62 \\
		Euphemia & 1,002,357.56 & & & & &
	\end{tabular}
	\caption{Average Daily GLOCs}
	\label{tab:gloc}
\end{table}

\begin{table}[!htp]
	\centering
	\begin{tabular}{c|cccccc}
		in EUR & National & 2 Zones (k) & 2 Zones (s) & 3 Zones & 4 Zones &  Nodal \\
		\hline
		IP & 0.00 & 0.00 & 0.00 & 0.00 & 0.00 & 0.00 \\
		CH & 11,388.11 & 31,280.46 & 109,165.0 & 22,230.84 & 174,685.4 & 184,626.79 \\
		Join & 883.43 & 3,539.37 & 1,432.85 & 1,473.60 & 15,489.43 & 36,927.06 \\
		Euphemia & 0.00 & & & & &
	\end{tabular}
	\caption{Average Daily LLOCs}
	\label{tab:lloc}
\end{table}

\begin{table}[!htp]
	\centering
	\begin{tabular}{c|cccccc}
		in EUR & National & 2 Zones (k) & 2 Zones (s) & 3 Zones & 4 Zones &  Nodal \\
		\hline
		IP & 32,400.08 & 45,448.71 & 30,881.3 & 21,122.54 & 42,156.28 & 202,106.42 \\
		CH & 8,691.04 & 38,029.2 & 78,808.99 & 19,794.25 & 161,948.7 & 114,669.73 \\
		Join & 1,153.58 & 7,375.58 & 1,780.6 & 797.12 & 1,364.48 & 22,105.27 \\
		Euphemia & 0.00 & & & & &
	\end{tabular}
	\caption{Average Daily MWPs}
	\label{tab:mwp}
\end{table}

By definition, CH prices minimize GLOCs, and on average, they are substantially lower compared to both IP and Join prices, irrespective of the chosen allocation rule. Similar orders of magnitude for GLOCs are observed across all zonal allocation rules, including Euphemia.  In contrast, nodal pricing results in the highest GLOCs, aligning with the higher average prices presented in Table \ref{tab:prices}. Higher prices imply greater forgone profits for generators and increase GLOCs. Generators in nodal markets face stronger incentives to deviate from the optimal outcome, necessitating penalties -- a common practice in many U.S. markets \citep{Bichler.2021}.

As a natural subset of GLOCs, LLOCs follow a similar pattern, with the highest values occurring under nodal pricing, once again correlated with higher average prices. IP pricing ensures zero LLOCs, as discussed in Section \ref{subsec:IP}, ensuring reliable congestion signals, \textcolor{black}{meaning that price differences between adjacent nodes arise only if the line connecting them is congested}. In this case, the price difference precisely reflects the marginal value of transmission capacity. In contrast, CH pricing consistently results in the highest LLOCs, supporting observations that congestion signals may be flawed \citep{Schiro.2016}. This is especially relevant under a nodal allocation rule considering all transmission lines. The Join pricing rule yields substantially fewer LLOCs than CH pricing, indicating a superior quality of congestion signals.

In terms of average MWPs to compensate generator losses, only the Euphemia algorithm generates zero MWPs, albeit at the expense of welfare losses. An interesting observation is that the welfare losses of Euphemia -- as discussed in Section \ref{subsec:results_costs} -- exceed, on average, the MWPs required under a national welfare-maximizing allocation \textcolor{black}{(i.e., a zonal model as described in Section \ref{sec:zonal-market-clearing} with a single zone)}, irrespective of the pricing rule. This implies that the additional generated welfare could potentially cover the MWPs under a national allocation, resulting in a net welfare gain. Under nodal pricing, MWPs tend to increase compared to zonal pricing, particularly for IP pricing. This can be attributed to the fact that under a national allocation rule, there exists a single hourly IP price set at the variable cost of the marginal generator. With positive fixed costs and constant variable costs, such a generator will inevitably incur a loss and necessitate MWPs under IP pricing. In contrast, nodal pricing involves several hourly IP prices and price-setting generators across the network, leading to higher overall MWPs. Conversely, the Join pricing rule generates lower MWPs than IP and CH pricing, constituting a negligible share of the total costs outlined in Table \ref{tab:costs}.

In summary, GLOCs, LLOCs, and MWPs follow similar patterns across different zonal configurations, but increase with nodal pricing. Concerning pricing rules, IP, CH, and Euphemia each minimize a specific class of lost opportunity costs. However, the Join pricing rule stands out by striking a remarkable balance between low MWPs and LLOCs. At the expense of slightly increased GLOCs, Join prices facilitate marginal side-payments and maintain a high quality of congestion signals.

\subsection{Discussion}

Our numerical experiments with the BZR data set provide insights into different market clearing models, pricing rules, and their short-term economic implications on day-ahead markets. Nodal pricing tends to result in slightly higher average prices compared to zonal pricing, irrespective of the chosen zonal configuration. However, in zonal pricing, transmission constraints can be violated and necessitate costly redispatch measures. As a result, a zonal pricing selects an inefficient dispatch, and even if redispatch is conducted at minimal costs, the total system costs will be significantly higher than under nodal pricing. Given the steep increase of redispatch costs in recent years, this effect is likely to magnify in the future.

Interestingly, the choice of zonal configuration did not substantially impact prices. All zonal configurations suggested by ACER in the BZR simplify the network to an extent that opting for two, three, or four zones yields similar results regarding the average prices and the price standard deviation. 
Analyses of individual days show that congestion arises, but not necessarily at the cross-zonal lines but within zones. 
While nodal prices may vary between nodes, these differences incentivize short-term demand response on nodes where this is most helpful for overall system stability. If such price differences persist over more extended time periods, they also set investment incentives for generators and grid expansion. Note that our analysis does not take such incentives into account and assumes static demand. \textcolor{black}{Nevertheless, long-run effects related to inefficient placement and technology choice for production, consumption and grid expansion are important factors to consider when evaluating a market clearing model.} 

Given the absence of Walrasian equilibria, any pricing rule -- regardless of zonal or nodal pricing -- must balance economic properties such as welfare gains and the participants' lost opportunity costs. \textcolor{black}{The currently used Euphemia algorithm leads to welfare losses and is computationally expensive. It avoids paradoxically accepted bids but may result in paradoxically rejected bids. Our experiments show that the welfare losses of Euphemia exceed the alternative costs of make-whole payments. Moreover, non-uniform pricing rules, such as CH pricing, achieve low make-whole payments for paradoxically accepted bids while maintaining the optimal allocation and scaling in polynomial time.} The findings provide support for non-uniform pricing as it is currently being discussed in the context of the Capacity Allocation and Congestion Management (CACM) Regulations \citep{AllNEMOCommittee.2023.Nonuniform}. 
CH pricing minimizes GLOCs, but our results confirm previous findings that high MWPs are necessary. IP pricing, the prevailing rule in U.S. markets, ensures effective congestion signals but requires very high MWPs. The Join pricing rule strikes a balance between LLOCs and MWPs, ensuring low side-payments and good congestion signals. 

\color{black}
Finally, we note that a a potential split of the German BZ has raised concerns about long-term regional impacts. Literature shows that energy-intensive industries in the south would likely face higher electricity prices, threatening their competitiveness \citep{Growitsch2014, Traber2015}. RES providers in the north may benefit from reduced curtailment but could experience lower and more volatile prices, weakening investment incentives \citep{Kunz2015, Hirth2019}. A zone split may reduce liquidity in the German futures market, complicating hedging and increasing risk premiums.
While such distributional aspects are not the focus of our study, they need to be considered when evaluating the effects of a zonal split.
\color{black}

\section{Conclusion} \label{sec:conclusion}
The discussion surrounding zonal and nodal pricing has a long history. The European electricity market has employed zonal pricing for many years, often with large, nationwide price zones. However, the ongoing energy transition requires a reevaluation of price zones. Based on recent data released by ENTSO-E in the context of the EU Bidding Zone Review process, we compare \textcolor{black}{the short-run effects of} various zonal and nodal pricing rules for the German power market. Our findings indicate that, in terms of system costs and price signals, a nodal allocation rule with non-uniform pricing leads to the lowest total costs. Costly redispatch is mitigated, and the average prices increase only slightly. Additionally, pricing rules such as Join pricing require very low side-payments, while efficiently signaling transmission bottlenecks. In contrast, the proposed zonal configurations only differ marginally regarding system costs and the differences in average prices across zones are low. Also, the effect of zonal splits on price standard deviations is low, and we do not find evidence for significant reductions in redispatch costs resulting from such splits. The study does not say that there cannot be congestion on cross-zonal lines leading to significant price differences between zones at certain times. However, in the BZR dataset that was intended to inform ACER based on realistic demand and supply scenarios for the target year 2025, we do not find significant price differences. Fears of substantial price differences after a zonal split might be overrated based on this comprehensive and up-to-date dataset. Importantly, we could provide a fair comparison with nodal pricing rules and show that the welfare gains are around 5-6\%, depending on the zonal configuration.


\vspace{-1cm}
\begin{singlespace}
\bibliographystyle{apalike-ejor}
\bibliography{bibliography}

\begin{thebibliography}{}

\bibitem[ACER, 2022a]{acer-clustering-algo}
ACER (2022a).
\newblock {\em {ACER's Decision on the alternative bidding zone configurations to be considered in the bidding zone review process.} {Annex IV.} {Description of the clustering algorithms}}.
\newblock \url{https://www.acer.europa.eu/Individual%20Decisions_annex/ACER%20Decision%2011-2022%20on%20alternative%20BZ%20configurations%20-%20Annex%20IV.pdf}.
\newblock Accessed: 2024-02-09.

\bibitem[ACER, 2022b]{ACER.2022.Decision}
ACER (2022b).
\newblock {\em Decision no 11/2022 of the european union agency for the cooperation of energy regulators}.
\newblock \url{https://www.acer.europa.eu/sites/default/files/documents/Individual%20Decisions/ACER%20Decision%2011-2022%20on%20alternative%20BZ%20configurations.pdf}

\bibitem[ACER, 2022c]{ACER.2022.BZRConfig}
ACER (2022c).
\newblock {\em List of alternative bidding zone configurations to be considered for the bidding zone review}.

\bibitem[ACER, 2025]{ENTSOE.2025.Annex2}
ACER (2025).
\newblock {\em {Annex II CE BZRR Input Data}}.
\newblock \url{https://eepublicdownloads.blob.core.windows.net/public-cdn-container/clean-documents/Network%20codes%20documents/NC%20CACM/BZR/2025/Annex_2_CE_BZRR_input_data.pdf}.
\newblock Accessed: 2025-05-07.

\bibitem[Ahunbay et~al., 2024]{Ahunbay.2024OR}
Ahunbay, M.~c., Bichler, M., \& Kn\"{o}rr, J. (2024).
\newblock Pricing optimal outcomes in coupled and non-convex markets: Theory and applications to electricity markets.
\newblock {\em Operations Research}.
\newblock \url{https://doi.org/10.1287/opre.2023.0401}

\bibitem[Ahunbay et~al., 2022]{Ahunbay.2022}
Ahunbay, M.~{\c{S}}., Bichler, M., \& Kn{\"o}rr, J. (2022).
\newblock {\em Pricing optimal outcomes in coupled and non-convex markets: Theory and applications to electricity markets}.
\newblock \url{http://arxiv.org/pdf/2209.07386v1}

\bibitem[{All NEMO Committee}, 2022]{AllNEMOCommittee.2022}
{All NEMO Committee} (2022).
\newblock {\em {CACM} annual report 2021}.
\newblock \url{https://www.nemo-committee.eu/assets/files/nemo_CACM_Annual_Report_2021_220630-4e7321983974b812f28730a301c9f7d9.pdf}

\bibitem[{All NEMO Committee}, 2023]{AllNEMOCommittee.2023.Nonuniform}
{All NEMO Committee} (2023).
\newblock {\em Non-uniform pricing: Explanatory note}.
\newblock \url{https://www.nemo-committee.eu/assets/files/sdac-non-uniform-pricing-explanatory-note.pdf}

\bibitem[Ambrosius et~al., 2022]{ambrosius2022risk}
Ambrosius, M., Egerer, J., Grimm, V., \& van~der Weijde, A.~H. (2022).
\newblock Risk aversion in multilevel electricity market models with different congestion pricing regimes.
\newblock {\em Energy Economics}, 105, 105701.

\bibitem[Ambrosius et~al., 2020]{ambrosius2020endogenous}
Ambrosius, M., Grimm, V., Kleinert, T., Liers, F., Schmidt, M., \& Z{\"o}ttl, G. (2020).
\newblock Endogenous price zones and investment incentives in electricity markets: An application of multilevel optimization with graph partitioning.
\newblock {\em Energy Economics}, 92, 104879.

\bibitem[Aravena et~al., 2021]{aravena2021transmission}
Aravena, I., Leté, Q., Papavasiliou, A., \& Smeers, Y. (2021).
\newblock Transmission capacity allocation in zonal electricity markets.
\newblock {\em Operations Research}, 69, 1240--1255.
\newblock \url{https://doi.org/10.1287/opre.2020.2082}

\bibitem[Aravena \& Papavasiliou, 2016]{aravena2016renewable}
Aravena, I. \& Papavasiliou, A. (2016).
\newblock Renewable energy integration in zonal markets.
\newblock {\em IEEE Transactions on Power Systems}, 32, 1--1.
\newblock \url{https://doi.org/10.1109/TPWRS.2016.2585222}

\bibitem[Arrow \& Debreu, 1954]{arrow1954existence}
Arrow, K.~J. \& Debreu, G. (1954).
\newblock Existence of an equilibrium for a competitive economy.
\newblock {\em Econometrica: Journal of the Econometric Society}, 265--290.

\bibitem[Aurora, 2023]{AuroraStudy2023}
Aurora (2023).
\newblock {\em {Power Market Impact of Splitting the German Bidding Zone}}.
\newblock \url{https://auroraer.com/insight/power-market-impact-of-splitting-the-german-bidding-zone/}

\bibitem[Baldwin \& Klemperer, 2019]{baldwin2019understanding}
Baldwin, E. \& Klemperer, P. (2019).
\newblock Understanding preferences: demand types, and the existence of equilibrium with indivisibilities.
\newblock {\em Econometrica}, 87(3), 867--932.

\bibitem[Bertsch et~al., 2016]{Bertsch.2016}
Bertsch, J., Hagspiel, S., \& Just, L. (2016).
\newblock Congestion management in power systems.
\newblock {\em Journal of Regulatory Economics}, 50(3), 290--327.

\bibitem[Bichler et~al., 2022]{Bichler.2021}
Bichler, M., Knoerr, J., \& Maldonado, F. (2022).
\newblock Pricing in non-convex markets: How to price electricity in the presence of demand response.
\newblock {\em Information Systems Research}, 1(to appear).

\bibitem[Bichler \& Kn{\"o}rr, 2023]{Bichler.2023}
Bichler, M. \& Kn{\"o}rr, J. (2023).
\newblock Getting prices right on electricity spot markets: On the economic impact of advanced power flow models.
\newblock {\em Energy Economics}, 126, 106968.

\bibitem[Bikhchandani \& Mamer, 1997]{bikhchandani1997competitive}
Bikhchandani, S. \& Mamer, J.~W. (1997).
\newblock Competitive equilibrium in an exchange economy with indivisibilities.
\newblock {\em Journal of Economic Theory}, 74(2), 385--413.

\bibitem[Bikhchandani \& Ostroy, 2002]{bikhchandani2002package}
Bikhchandani, S. \& Ostroy, J.~M. (2002).
\newblock The package assignment model.
\newblock {\em Journal of Economic Theory}, 107(2), 377--406.

\bibitem[Breuer et~al., 2013]{breuer2013}
Breuer, C., Seeger, N., \& Moser, A. (2013).
\newblock Determination of alternative bidding areas based on a full nodal pricing approach.
\newblock 1--5.
\newblock \url{https://doi.org/10.1109/PESMG.2013.6672466}

\bibitem[Bundesnetzagentur, 2022]{Bundesnetzagentur.2022}
Bundesnetzagentur (2022).
\newblock {\em Zahlen zu netzengpassmanagementma{\ss}nahmen -- gesamtes jahr 2022}.
\newblock \url{(https://www.bundesnetzagentur.de/SharedDocs/Downloads/DE/Sachgebiete/Energie/Unternehmen_Institutionen/Versorgungssicherheit/Engpassmanagement/Ganzjahreszahlen2022.pdf}

\bibitem[Bundesnetzagentur, 2024]{Bundesnetzagentur.2024}
Bundesnetzagentur (2024).
\newblock {\em Bundesnetzagentur ver{\"o}ffentlicht daten zum strommarkt 2023}.
\newblock \url{https://www.bundesnetzagentur.de/SharedDocs/Pressemitteilungen/DE/2024/20240103_SMARD.html}

\bibitem[Burstedde, 2012]{burstedde2012}
Burstedde, B. (2012).
\newblock From nodal to zonal pricing: A bottom-up approach to the second-best.
\newblock {\em 2012 9th International Conference on the European Energy Market}, 1--8.
\newblock \url{https://doi.org/10.1109/EEM.2012.6254665}

\bibitem[{California ISO}, 2024]{caiso2024operating}
{California ISO} (2024).
\newblock {\em {Day-Ahead Market. Operating Procedure}}.
\newblock \url{https://www.caiso.com/documents/1210.pdf}

\bibitem[Christie et~al., 2000]{christie2000transmission}
Christie, R., Wollenberg, B., \& Wangensteen, I. (2000).
\newblock Transmission management in the deregulated environment.
\newblock {\em Proceedings of the IEEE}, 88(2), 170--195.
\newblock \url{https://doi.org/10.1109/5.823997}

\bibitem[Diers et~al., 2023]{EWITHEMAStudy2023}
Diers, H., Emelianova, P., Kienscherf, P.~A., Walde, M., \& Zinke, J. (2023).
\newblock {\em {Price impact of a German bidding zone split}}.
\newblock \url{https://www.ewi.uni-koeln.de/en/publications/price-impact-of-a-german-bidding-zone-split/}

\bibitem[Dobos et~al., 2025]{Dobos.2024}
Dobos, T., Bichler, M., \& Knörr, J. (2025).
\newblock Challenges in finding stable price zones in european electricity markets: Aiming to square the circle?
\newblock {\em Applied Energy}, 382, 125315.
\newblock \url{https://doi.org/https://doi.org/10.1016/j.apenergy.2025.125315}

\bibitem[Eicke \& Schittekatte, 2022]{eicke2022fighting}
Eicke, A. \& Schittekatte, T. (2022).
\newblock Fighting the wrong battle? a critical assessment of arguments against nodal electricity prices in the european debate.
\newblock {\em Energy Policy}, 170, 113220.

\bibitem[ENTSO-E, 2022]{ENTSOE.2022}
ENTSO-E (2022).
\newblock {\em Report on the locational marginal pricing study of the bidding zone review process}.
\newblock \url{https://eepublicdownloads.blob.core.windows.net/public-cdn-container/clean-documents/Publications/Market%20Committee%20publications/ENTSO-E%20LMP%20Report_publication.pdf}

\bibitem[ENTSO-E, 2023]{ENTSOE.2023.BZRData}
ENTSO-E (2023).
\newblock {\em Bidding zone review}.
\newblock \url{https://www.entsoe.eu/network_codes/bzr/}

\bibitem[ENTSO-E, 2024]{ENTSOE.2024}
ENTSO-E (2024).
\newblock {\em Bidding zone review consultative group (bzr cg)}.
\newblock \url{https://eepublicdownloads.blob.core.windows.net/public-cdn-container/clean-documents/Network%20codes%20documents/NC%20EB/2024/241105_BZR_CG_ENTSO-E_slides.pdf}

\bibitem[ENTSO-E, 2025]{ENTSOE.2025MainReport}
ENTSO-E (2025).
\newblock {\em Main report, bidding zone review of the 2025 target year}.
\newblock \url{https://eepublicdownloads.blob.core.windows.net/public-cdn-container/clean-documents/Network%20codes%20documents/NC%20CACM/BZR/2025/Bidding_Zone_Review_of_the_2025_Target_Year.pdf}

\bibitem[{European Commission}, 2024]{EuropeanCommission2024Climate}
{European Commission} (2024).
\newblock {\em {2040 climate target}}.
\newblock \url{https://climate.ec.europa.eu/eu-action/climate-strategies-targets/2040-climate-target_en}

\bibitem[{Federal Ministry of Education and Research}, 2024]{energiewende}
{Federal Ministry of Education and Research} (2024).
\newblock {\em German energy transition}.
\newblock \url{https://www.bmbf.de/bmbf/en/research/energy-and-economy/german-energy-transition/german-energy-transition_node.html}

\bibitem[Felling \& Weber, 2018]{weber2018consistent}
Felling, T. \& Weber, C. (2018).
\newblock Consistent and robust delimitation of price zones under uncertainty with an application to central western europe.
\newblock {\em Energy Economics}, 75, 583--601.
\newblock \url{https://doi.org/https://doi.org/10.1016/j.eneco.2018.09.012}

\bibitem[Gerhardt et~al., 2025]{AgoraFraunhofer2025}
Gerhardt, N., Knorr, K., Harms, Y., und Diana~Bottger, J.~K., Godron, P., Lenck, T., Hartz, K., Huneke, F., \& Schaber, K. (2025).
\newblock {\em {Lokale Strompreise}}.
\newblock \url{https://www.agora-energiewende.de/publikationen/lokale-strompreise}

\bibitem[Grainger \& Stevenson, 1994]{Grainger.1994}
Grainger, J.~J. \& Stevenson, W.~D. (1994).
\newblock {\em Power System Analysis}.
\newblock Electrical engineering series. McGraw-Hill.

\bibitem[Green, 2007]{Green.2007}
Green, R. (2007).
\newblock Nodal pricing of electricity: how much does it cost to get it wrong?
\newblock {\em Journal of Regulatory Economics}, 31(2), 125--149.
\newblock \url{https://doi.org/10.1007/s11149-006-9019-3}

\bibitem[Gribik et~al., 2007]{gribik2007market}
Gribik, P.~R., Hogan, W.~W., Pope, S.~L., et~al. (2007).
\newblock Market-clearing electricity prices and energy uplift.
\newblock {\em Cambridge, MA}.

\bibitem[Grimm et~al., 2019]{grimm2019optimal}
Grimm, V., Kleinert, T., Liers, F., Schmidt, M., \& Z{\"o}ttl, G. (2019).
\newblock Optimal price zones of electricity markets: a mixed-integer multilevel model and global solution approaches.
\newblock {\em Optimization methods and software}, 34(2), 406--436.

\bibitem[Grimm et~al., 2016]{grimm2016transmission}
Grimm, V., Martin, A., Schmidt, M., Weibelzahl, M., \& Z{\"o}ttl, G. (2016).
\newblock Transmission and generation investment in electricity markets: The effects of market splitting and network fee regimes.
\newblock {\em European Journal of Operational Research}, 254(2), 493--509.

\bibitem[Grimm et~al., 2021]{grimm2021impact}
Grimm, V., R{\"u}ckel, B., S{\"o}lch, C., \& Z{\"o}ttl, G. (2021).
\newblock The impact of market design on transmission and generation investment in electricity markets.
\newblock {\em Energy Economics}, 93, 104934.

\bibitem[Grimm et~al., 2022]{grimm2022emissions}
Grimm, V., S{\"o}lch, C., \& Z{\"o}ttl, G. (2022).
\newblock Emissions reduction in a second-best world: On the long-term effects of overlapping regulations.
\newblock {\em Energy Economics}, 109, 105829.

\bibitem[Growitsch et~al., 2014]{Growitsch2014}
Growitsch, C., Nick, S., \& Pokorny, S. (2014).
\newblock Efficient zonal configurations for european electricity markets—evidence from an empirical evaluation.
\newblock {\em Energy Economics}, 44, 387--395.

\bibitem[Hao \& Xu, 2008]{Hao.2008}
Hao, J. \& Xu, W. (2008).
\newblock Extended transmission line loadability curve by including voltage stability constrains.
\newblock {\em 2008 IEEE Canada Electric Power Conference}, 1--5.

\bibitem[Hinojosa, 2020]{hinojosa2020comparing}
Hinojosa, V.~H. (2020).
\newblock Comparing corrective and preventive security-constrained dcopf problems using linear shift-factors.
\newblock {\em Energies}, 13(3).
\newblock \url{https://doi.org/10.3390/en13030516}

\bibitem[Hirth \& Schlecht, 2019]{Hirth2019}
Hirth, L. \& Schlecht, I. (2019).
\newblock Market design for variable renewables: Taking storage and flexibility into account.
\newblock {\em Energy Economics}, 80, 524--538.

\bibitem[Hirth \& Schlecht, 2020]{LionHirth.2020}
Hirth, L. \& Schlecht, I. (2020).
\newblock {\em Market-based redispatch in zonal electricity markets: The preconditions for and consequence of inc-dec gaming}.
\newblock \url{http://hdl.handle.net/10419/222925}

\bibitem[Hogan \& Ring, 2003]{hogan2003minimum}
Hogan, W.~W. \& Ring, B.~J. (2003).
\newblock On minimum-uplift pricing for electricity markets.
\newblock {\em Electricity Policy Group}.

\bibitem[Holmberg \& Tangerås, 2023]{Holmberg2023survey}
Holmberg, P. \& Tangerås, T. (2023).
\newblock A survey of capacity mechanisms: Lessons for the swedish electricity market.
\newblock {\em The Energy Journal}, 44(6), 275--304.
\newblock \url{https://doi.org/10.5547/01956574.44.6.phol}

\bibitem[H{\"o}rsch et~al., 2018]{horsch2018pypsa}
H{\"o}rsch, J., Hofmann, F., Schlachtberger, D., \& Brown, T. (2018).
\newblock Pypsa-eur: An open optimisation model of the european transmission system.
\newblock {\em Energy strategy reviews}, 22, 207--215.

\bibitem[Hua \& Baldick, 2017]{Hua.2017}
Hua, B. \& Baldick, R. (2017).
\newblock A convex primal formulation for convex hull pricing.
\newblock {\em IEEE Transactions on Power Systems}, 32(5), 3814--3823.

\bibitem[IRENA, 2022]{IRENA.2022}
IRENA (2022).
\newblock {\em Renewable power generation costs in 2021}.

\bibitem[Kelso \& Crawford, 1982]{Kelso82}
Kelso, A.~S. \& Crawford, V.~P. (1982).
\newblock Job matching, coalition formation , and gross substitute.
\newblock {\em Econometrica}, 50, 1483--1504.

\bibitem[Kost et~al., 2021]{Kost.2021}
Kost, C., Shammugam, S., Fluri, V., Peper, D., {Davoodi Memar}, A., \& Schlegl, T. (2021).
\newblock {\em Levelized cost of electricity - renewable energy technologies}.

\bibitem[Kumar et~al., 2004]{kumar2004}
Kumar, A., Srivastava, S., \& Singh, S. (2004).
\newblock A zonal congestion management approach using real and reactive power rescheduling.
\newblock {\em IEEE Transactions on Power Systems}, 19(1), 554--562.
\newblock \url{https://doi.org/10.1109/TPWRS.2003.821448}

\bibitem[Kunz \& Zerrahn, 2015a]{Kunz2015benefits}
Kunz, F. \& Zerrahn, A. (2015a).
\newblock Benefits of coordinating congestion management in electricity transmission networks: Theory and application to germany.
\newblock {\em Utilities Policy}, 37, 34--45.
\newblock \url{https://doi.org/https://doi.org/10.1016/j.jup.2015.09.009}

\bibitem[Kunz \& Zerrahn, 2015b]{Kunz2015}
Kunz, F. \& Zerrahn, A. (2015b).
\newblock Grid vs. market costs of electricity transmission expansion—a welfare analysis.
\newblock {\em Energy Journal}, 36(4), 49--70.

\bibitem[Kłos et~al., 2014]{klos2014}
Kłos, M., Wawrzyniak, K., Jakubek, M., \& Oryńczak, G. (2014).
\newblock The scheme of a novel methodology for zonal division based on power transfer distribution factors.
\newblock {\em IECON 2014 - 40th Annual Conference of the IEEE Industrial Electronics Society}, 3598--3604.
\newblock \url{https://doi.org/10.1109/IECON.2014.7049033}

\bibitem[Leuthold et~al., 2008]{Leuthold2008efficient}
Leuthold, F., Weigt, H., \& {von Hirschhausen}, C. (2008).
\newblock Efficient pricing for european electricity networks – the theory of nodal pricing applied to feeding-in wind in germany.
\newblock {\em Utilities Policy}, 16(4), 284--291.
\newblock \url{https://doi.org/https://doi.org/10.1016/j.jup.2007.12.003}.
\newblock European Regulatory Perspectives

\bibitem[Liberopoulos \& Andrianesis, 2016]{liberopoulos2016critical}
Liberopoulos, G. \& Andrianesis, P. (2016).
\newblock Critical review of pricing schemes in markets with non-convex costs.
\newblock {\em Operations Research}, 64(1), 17--31.

\bibitem[Marien et~al., 2013]{marien2013importance}
Marien, A., Luickx, P., Tirez, A., \& Woitrin, D. (2013).
\newblock Importance of design parameters on flowbased market coupling implementation.
\newblock {\em 2013 10th International Conference on the European Energy Market (EEM)}, 1--8.
\newblock \url{https://doi.org/10.1109/EEM.2013.6607298}

\bibitem[Meeus et~al., 2009]{Meeus.2009}
Meeus, L., Verhaegen, K., \& Belmans, R. (2009).
\newblock Block order restrictions in combinatorial electric energy auctions.
\newblock {\em European Journal of Operational Research}, 196(3), 1202--1206.

\bibitem[MISO, 2023]{MISO.2019}
MISO (2023).
\newblock {\em Schedule 29{A}: {ELMP} for energy and operating reserve market: Ex-post pricing formulations}.
\newblock \url{https://docs.misoenergy.org/legalcontent/Schedule_29-A_-_ELMP_for_Energy_and_Operating_Reserve_Market.pdf}

\bibitem[Molzahn \& Hiskens, 2019]{Molzahn.2019}
Molzahn, D.~K. \& Hiskens, I.~A. (2019).
\newblock A survey of relaxations and approximations of the power flow equations.
\newblock {\em Foundations and Trends{\circledR} in Electric Energy Systems}, 4(1-2), 1--221.

\bibitem[Moreira et~al., 2006]{Moreira.2006}
Moreira, F.~S., Ohishi, T., \& {Da Silva Filho}, J.~I. (2006).
\newblock Influence of the thermal limits of transmission lines in the economic dispatch.
\newblock {\em 2006 IEEE Power Engineering Society General Meeting}, 6 pp.

\bibitem[{NEMO Committee}, 2019]{NEMOCommittee.2019}
{NEMO Committee} (2019).
\newblock {\em {EUPHEMIA}~public description: Single price coupling algorithm}.
\newblock \url{https://www.epexspot.com/document/40503/Euphemia%20Public%20Description}

\bibitem[{NEMO Committee}, 2024]{nemo2020euphemia}
{NEMO Committee} (2024).
\newblock {\em {EUPHEMIA} public description}.
\newblock \url{https://www.nemo-committee.eu/assets/files/euphemia-public-description.pdf}

\bibitem[Neuhoff et~al., 2013]{Neuhoff.2013}
Neuhoff, K., Barquin, J., Bialek, J.~W., Boyd, R., Dent, C.~J., Echavarren, F., Grau, T., von Hirschhausen, C., Hobbs, B.~F., Kunz, F., Nabe, C., Papaefthymiou, G., Weber, C., \& Weigt, H. (2013).
\newblock Renewable electric energy integration: Quantifying the value of design of markets for international transmission capacity.
\newblock {\em Energy Economics}, 40, 760--772.

\bibitem[Newbery, 2023]{newbery2023high}
Newbery, D.~M. (2023).
\newblock High renewable electricity penetration: Marginal curtailment and market failure under “subsidy-free” entry.
\newblock {\em Energy Economics}, 126, 107011.
\newblock \url{https://doi.org/https://doi.org/10.1016/j.eneco.2023.107011}

\bibitem[Newbery \& Biggar, 2024]{newbery2024marginal}
Newbery, D.~M. \& Biggar, D.~R. (2024).
\newblock Marginal curtailment of wind and solar pv: Transmission constraints, pricing and access regimes for efficient investment.
\newblock {\em Energy Policy}, 191, 114206.
\newblock \url{https://doi.org/https://doi.org/10.1016/j.enpol.2024.114206}

\bibitem[O'Neill et~al., 2005]{oneill2005efficient}
O'Neill, R.~P., Sotkiewicz, P.~M., Hobbs, B.~F., Rothkopf, M.~H., \& Stewart, W.~R. (2005).
\newblock Efficient market-clearing prices in markets with nonconvexities.
\newblock {\em European Journal of Operational Research}, 1(164), 269--285.

\bibitem[PJM, 2018]{PJM.2018}
PJM (2018).
\newblock {\em {FERC} {D}ocket {EL} 18-34-000: Fast start resources}.
\newblock \url{https://www.pjm.com/-/media/committees-groups/task-forces/epfstf/20180118/20180118-fast-start-pricing.ashx}

\bibitem[Pollitt, 2023]{Pollitt2023}
Pollitt, M.~G. (2023).
\newblock {\em {Locational Marginal Prices (LMPs) for Electricity in Europe? The Untold Story}}.
\newblock \url{https://www.jbs.cam.ac.uk/wp-content/uploads/2023/12/eprg-wp2318.pdf}

\bibitem[Schiro et~al., 2016]{Schiro.2016}
Schiro, D.~A., Zheng, T., Zhao, F., \& Litvinov, E. (2016).
\newblock Convex hull pricing in electricity markets: Formulation, analysis, and implementation challenges.
\newblock {\em IEEE Transactions on Power Systems}, 31(5), 4068--4075.

\bibitem[Schmitt, 2023]{Schmitt.2023}
Schmitt, N. (2023).
\newblock {\em Towards a central dispatch model for european electricity markets}.

\bibitem[Schr{\"o}der et~al., 2013]{Schroder.2013}
Schr{\"o}der, A., Kunz, F., Meiss, R., Mendelevitch, R., \& von Hirschhausen, C. (2013).
\newblock {\em Current and prospective costs of electricity generation until 2050}.

\bibitem[Simshauser, 2025]{Simshauser2025competition}
Simshauser, P. (2025).
\newblock Competition vs. coordination: Optimising wind, solar and batteries in renewable energy zones.
\newblock {\em Energy Economics}, 143, 108279.
\newblock \url{https://doi.org/https://doi.org/10.1016/j.eneco.2025.108279}

\bibitem[{St. Clair}, 1953]{St.Clair.1953}
{St. Clair}, H.~P. (1953).
\newblock Practical concepts in capability and performance of transmission lines [includes discussion].
\newblock {\em Transactions of the American Institute of Electrical Engineers. Part III: Power Apparatus and Systems}, 72(6), 1152--1157.

\bibitem[Stoft, 1997]{stoft1997}
Stoft, S. (1997).
\newblock Transmission pricing zones: simple or complex?
\newblock {\em The Electricity Journal}, 10(1), 24--31.
\newblock \url{https://doi.org/https://doi.org/10.1016/S1040-6190(97)80294-1}

\bibitem[Stott et~al., 2009]{Stott.2009}
Stott, B., Jardim, J., \& Alsac, O. (2009).
\newblock {DC} power flow revisited.
\newblock {\em IEEE Transactions on Power Systems}, 24(3), 1290--1300.

\bibitem[Taheri \& Molzahn, 2024]{taheri2024acfeasibility}
Taheri, B. \& Molzahn, D.~K. (2024).
\newblock Ac power flow feasibility restoration via a state estimation-based post-processing algorithm.
\newblock {\em Electric Power Systems Research}, 235, 110642.
\newblock \url{https://doi.org/https://doi.org/10.1016/j.epsr.2024.110642}

\bibitem[{The Federal Government}, 2024]{endingcoal}
{The Federal Government} (2024).
\newblock {\em Ending coal-generated power}.
\newblock \url{https://www.bundesregierung.de/breg-en/service/archive/kohleausstiegsgesetz-1717014}

\bibitem[Thomassen et~al., 2024]{thomassen2024redispatch}
Thomassen, G., A, F., R, C., D, P.~C., \& S, V. (2024).
\newblock Redispatch and congestion management.
\newblock Scientific analysis or review,Policy assessment,Anticipation and foresight KJ-NA-31-924-EN-N (online),KJ-NB-31-924-EN-N, Luxembourg (Luxembourg).
\newblock \url{https://doi.org/10.2760/853898 (online),10.2760/871}

\bibitem[Tiedemann et~al., 2024]{tiedemann2024gebotszonenteilung}
Tiedemann, S., Stiewe, C., Kratzke, C., Hirth, L., Jentsch, M., Damm, N., Gerhardt, N., \& Pape, C. (2024).
\newblock Gebotszonenteilung: Auswirkungen auf den marktwert der erneuerbaren energien im jahr.

\bibitem[Townsend et~al., 2014]{townsend2014mwps}
Townsend, A., otterud, A., evin, T., Milligan, M., \& Bloom, A. (2014).
\newblock {\em Evolution of wholesale electricity market design with increasing levels of renewable generation}.
\newblock \url{https://doi.org/10.2172/1159375}

\bibitem[Traber \& Kemfert, 2015]{Traber2015}
Traber, T. \& Kemfert, C. (2015).
\newblock Gone with the wind?—electricity market prices and incentives to invest in thermal power plants under increasing wind energy supply.
\newblock {\em Energy Economics}, 40, 594--602.

\bibitem[Trepper et~al., 2015a]{Trepper.2015}
Trepper, K., Bucksteeg, M., \& Weber, C. (2015a).
\newblock Market splitting in germany -- new evidence from a three-stage numerical model of europe.
\newblock {\em Energy Policy}, 87, 199--215.

\bibitem[Trepper et~al., 2015b]{trepper2015}
Trepper, K., Bucksteeg, M., \& Weber, C. (2015b).
\newblock Market splitting in germany – new evidence from a three-stage numerical model of europe.
\newblock {\em Energy Policy}, 87, 199--215.
\newblock \url{https://doi.org/https://doi.org/10.1016/j.enpol.2015.08.016}

\bibitem[Voswinkel et~al., 2019]{voswinkel2019}
Voswinkel, S., Felten, B., Felling, T., \& Weber, C. (2019).
\newblock Flow-based market coupling: What drives welfare in europe's electricity market design?
\newblock HEMF Working Paper 08/2019, Essen.
\newblock \url{https://hdl.handle.net/10419/201591}

\bibitem[Weinhold \& Mieth, 2024]{weinhold2023uncertainty}
Weinhold, R. \& Mieth, R. (2024).
\newblock Uncertainty-aware capacity allocation in flow-based market coupling.
\newblock {\em IEEE Transactions on Power Systems}, 39(1), 147--159.
\newblock \url{https://doi.org/10.1109/TPWRS.2023.3265320}

\bibitem[Wyrwoll et~al., 2018]{wyrwoll2018}
Wyrwoll, L., Kollenda, K., Müller, C., \& Schnettler, A. (2018).
\newblock Impact of flow-based market coupling parameters on european electricity markets.
\newblock {\em 2018 53rd International Universities Power Engineering Conference (UPEC)}, 1--6.
\newblock \url{https://doi.org/10.1109/UPEC.2018.8541904}

\bibitem[Yang et~al., 2019]{Yang.2019}
Yang, Z., Zheng, T., Yu, J., \& Xie, K. (2019).
\newblock A unified approach to pricing under nonconvexity.
\newblock {\em IEEE Transactions on Power Systems}, 34(5), 3417--3427.

\end{thebibliography}
\end{singlespace}
\clearpage

\appendix

\section{Market Clearing Problem - Notation} \label{app:dcopf}

\begin{table}[H]\centering
	\color{black}
	\def\sym#1{\ifmmode^{#1}\else\(^{#1}\)\fi}
	\begin{tabular}{l*{2}{l}}
	\toprule
	Parameters      &     &       \\
	\midrule
	$B, S, N, T$            &   & Sets of buyers, sellers, network nodes, periods \\
	$n^*$            &   & Reference node \\
	$z^*$            &   & Reference zone \\
	$n_i$  & $B \cup S \rightarrow N $ &  Node in which agent $i \in B \cup S$ is located\\
	$g_s$            &  [EUR/MWh] & Marginal cost of seller $s \in S$ \\
	$h_s$ & [EUR/MWh] & No-load cost of seller $s \in S$ \\
	$\underline{R}_s$ & & Minimum uptime of seller $s \in S$ (periods) \\
	$\underline{P}_{st}, \overline{P_{st}}$ & [MWh] & Minimum and maximum output of seller $s \in S$ in period $t\in T$ \\
	$P_{bt}$ & [MWh] & Demand of buyer $b \in B$ in period $t\in T$ \\
	$\underline{F}_{nm}, \overline{F}_{nm}$ & [MWh] & Minimum and maximum flow on the line connecting $n, m \in N$\\ 
	$B_{nm}$ & [pu] & Susceptance of the line connecting $n, m \in N$ \\
	\midrule
	Variables      &     &       \\
	\midrule
	$x_{b,nt}$ & [MW] & Total consumption of buyer $b \in B$ in node $n \in N$, period $t \in T$ \\ 
	$y_{s, nt} \geq 0$ & [MW] & Total production of seller $s \in S$ in node $n \in N$, period $t\in T$ \\
	$u_{st} \in \{0, 1\}$ & & Commitment of seller $s \in S$ in $t\in T$\\
	$\phi_{st} \geq 0$ & & Startup indicator of seller $s \in S$ in period $t\in T$\\
	$\theta_{nt}$ & & Voltage angle for node $n \in N$ and period $t \in T$ \\
	\midrule
	\end{tabular}
	\caption{Notation}
\end{table}

\section{Bidding Zone Review Data} \label{app:data}

In September 2022, ENTSO-E, on behalf of all European TSOs, released various data sets related to their locational pricing study \citep{ENTSOE.2022, ENTSOE.2023.BZRData}. The input data structure along two primary datasets: the \textit{grid model} as a basis for the transmission network and the \textit{input files} to model generation and demand. Data on reserves, storage, and imports/exports were not considered in our study.
Along with these input data, results files such as locational prices and cleared generation, demand, and storage were released. The publication refers to a total of 24 weeks over three representative climate years (1989, 1995, and 2009). Only the climate information from 1989, 1995, and 2009 is used, while generation and demand scenarios were generated for the target year 2025 \citep{ENTSOE.2022}.

Unfortunately, some parts of the data were aggregated before their publication (e.g., local demand to country-wide demand), and some data were not revealed (e.g., a mapping of generators in the \textit{input files} to their location in the \textit{grid model}). Without access to these proprietary data, the outcome of the ENTSO-E study cannot be exactly reproduced. However, we could assign most of generators to locations via OpenStreetMap.\footnote{\url{https://www.openstreetmap.de/}} 

Let us discuss the data in more depth. The topology of the grid, defined by its nodes and connections, dictates the paths electricity takes from generation to consumption, with substations facilitating the necessary voltage transformations along the way. Substations act as critical nodes in the network that manage voltage levels for efficient power flow. A substation might handle multiple voltage levels. For example, a transmission substation might receive power at a very high voltage (e.g., 220kV or higher) and step it down to a lower high voltage (e.g., 110kV) for further transmission or to a medium voltage (e.g., 10-35kV) for distribution to urban or industrial areas. 

The ENTSO-E dataset provides a model of Germany's power grid, detailing 834 substations. Each of these substations encompasses one or more voltage levels, summing up to 1697 voltage levels at substations in the dataset. These voltage levels cover the spectrum from transmission ($\geq$~220kV) to distribution ($\leq$~110kV) and are composed of several topological nodes. Altogether, there are 2898 topological nodes listed in the data sets. Additionally, the dataset includes details for 100 substations located in the neighboring countries, which have been integrated into our analysis. For our study, we manually determined the geographical locations of all substations and computed prices for all voltage levels on these substations. After some pre-processing, which involved removing disconnected nodes, breakers, and disconnectors, our German grid model consists of 1670 nodes and 2407 lines \citep{Schmitt.2023}. Technically, supply and demand on different voltage levels of substations do not necessarily need to be the same. However, it turned out that prices computed for voltage levels on a substation were identical or very close in the ENTSO-E dataset, which is also the case for the nodal prices provided by ENTSO-E. Overall, the average prices that we computed were not significantly different from those in the BZR dataset for substations. There were differences on individual nodes, which is probably due to the fact that the prices computed by ENTSO-E are based on the duals of a linear programming relaxation, while we use the mixed integer linear program and pricing rules as they are defined above. 

For the grid constraints of our DCOPF in (\ref{eq:dcopf}), we derive the per-unit susceptance $B_{nm}$ for each line from the given actual susceptance and base admittance \citep{Grainger.1994}. We also induce transmission limits on each power line, which we infer as the minimum of each line's angular stability limit (inferred from its surge impedance loading and the St. Clair curve \citep{St.Clair.1953}), voltage drop limit (inferred from voltage and reactance \citep{Hao.2008}), and thermal limit (inferred from voltage and current on a three-phase system \citep{Moreira.2006}). We assume these transmission limits to be constant over time. 
Missing data, e.g., on the length and susceptance of lines, were replaced by the data in the JAO Static Grid Model where possible, and else inferred from the mean of all other lines. 
As discussed earlier, we reserved a buffer of 20\% on the transmission limits to account for secure operations of the electricity grid.

For the demand side, the \textit{input files} provide hourly aggregated load profiles for the entire country of Germany.
Neither the nodal distribution of demand nor demand valuations are provided. Demand-side response was considered in the ENTSO-E study, but no related data was released. We distributed demand proportional to each consumer's base load that is available as part of the \textit{grid model}. We thereby assume that demand fluctuations over time occur uniformly across all consumers \citep{Schmitt.2023}. We further assume that demand is price-inelastic, as was done in the study by ENTSO-E.

The \textit{input files} pertaining to the supply side are more granular and separated by generator type, i.e., Hard Coal, Lignite, Gas, Light Oil, Solar, Hydro Run-Of-River, Hydro Reservoir, Hydro PumpOL, Hydro PumpCL, Onshore Wind, Offshore Wind, Other Non-RES, and Other RES. They contain operational characteristics of thermal units, derived from the Pan-European Market Modeling Database (PEMMDB), and hourly aggregated output of renewable energy sources, derived from Pan European Climate Database (PECD). The operational characteristics include -- among others -- minimum / maximum power and minimum runtime requirements that we use to parameterize our generator cost functions. The data also includes must-run obligations, start-up / shut-down times, and ramp rates, which may be used to further extend the model. Similarly, data on storage or demand response could leverage more detailed models.

As the maximum power of each seller is given as a fixed quantity (denoting its nominal capacity), we need to manually incorporate the variability of renewable energy sources. We thus distribute the aggregate dispatched energy in each hour equally to all plants of the same type, proportional to their nominal capacities, and set the maximum power to this value. The underlying assumptions are that renewables are always dispatched at maximum power and that meteorological conditions are equal across Germany. Because hydro power plants are dispatchable, they are excluded from this logic. \textcolor{black}{Out of 4537 generation units that we consider, 1709 ($\approx$37\%) are subject to unit-commitment rigidities. These units have minimum uptime and minimum operating power requirements greater than 0.}
The data also do not include information on variable or fixed costs for renewables, which we derived from literature \citep{Kost.2021,IRENA.2022,ENTSOE.2022}. For thermal units, variable cost information are provided, while fixed cost information were again obtained from literature \citep{Kost.2021, Schroder.2013,Bundesnetzagentur.2022}. 
\textcolor{black}{We refer the reader to \ref{app:generation-data} for more details.}

An additional challenge is the mapping of generators to corresponding network nodes. The grid data specify the broad category of generating units at each node (e.g., thermal, hydro, external injections) and their nominal capacities. While renewable resources could be mapped to the seller data by matching nominal capacities, the grid data lacks additional information regarding thermal plant types. We assigned thermal generators in the grid to their operational characteristics (such as minimum uptime constraints) \citep{Schmitt.2023}. Firstly, we match identifiers with the \textit{Kraftwerksliste} (power plant list) provided by the Bundesnetzagentur to obtain each generator's broad thermal generation type (e.g., lignite, gas, etc.). As nominal capacities do not match between grid and seller data, and seller data have a higher granularity (e.g., Gas CCGT, Gas OCGT, etc.), we solve an integer program to assign each entity in the grid model to exactly one corresponding seller of the same plant type, with the objective to minimize the aggregated differences in capacities and number of units per plant type. We provide this integer program and the outcome of the generation matching in \ref{app:mapping}. 
While this assignment might not be exact, we argue that smaller mistakes in the assignment of generators to nodes have little impact on the more general questions raised in our paper that address the benefits of different zonal and nodal configurations. 
The data are accessible through the ENTSO-E website.

\textcolor{black}{Importantly, the data we use in our study reflects the target year 2025, while the fraction of the generator types is expected to change as Germany continues its energy transition (Energiewende) toward a greener power supply. According to the \citep{energiewende}, renewable energy sources are projected to account for up to 80\% of gross electricity consumption by 2050.  In addition, coal-fired power plants are set to be completely phased out by 2038 \citep{endingcoal}, while the share of combined-cycle gas turbine (CCGT) plants is expected to increase temporarily to compensate for the reduction in coal and to support the intermittent nature of renewable energy sources.}

\subsection{Generation Data} \label{app:generation-data}

In what follows we outline how we derive the costs and capacities for the renewable (RES) and non-renewable (Non-RES) energy sources \citep{Schmitt.2023}.
Since the procedures used to establish these costs differ between RES and Non-RES, we discuss them independently.

\subsubsection{Renewables}
To derive the costs for RES, we consider the operating costs (OPEX), fuel and CO$_2$ costs from Table \ref{tab:costs-res}, which are based on \citep{Kost.2021}:

\begin{table}[h!]
	\centering
	\color{black}
	\begin{tabular}{l|c|c|c|c}
	RES & OPEX var 	& Fuel & CO$_2$ & OPEX fix \\ 
     				& [EUR/kWh] & [EUR/kWh] & [EUR/kWh] & [EUR/kW/year] \\ \hline
	Offshore Wind   & 0.008     & 0      & 0          & 70     \\ 
	Onshore Wind    & 0.008     & 0      & 0          & 20     \\ 
	Biomass         & 0.004     & 0.04286 & 0.0026812  & 160    \\
	Photovoltaic    & 0         & 0      & 0          & 13.3   \\ 
	Hydro          	& 0         & 0      & 0          & 36.64  \\ 
	\end{tabular}
	\caption{Costs for RES}
	\label{tab:costs-res}
\end{table}

We assume that RES require high initial investment costs but relatively low fixed costs. This assumption is attributed to the fact that particularly wind and solar power plants do not require ongoing expenses for fuel purchase and transportation, and have minimal maintenance costs due to the absence of moving parts. Therefore, the fixed cost $h_s$ is set to 0 for any renewable energy source $s \in S$.
However, the fixed OPEX cannot be disregarded and are instead incorporated into the variable costs $g_{st}$.

We derive the full load hours (FLH) from Table \ref{tab:res-flh} considering  the generated power per production type for the year 2009 and the sum of capacities that are included in the BZR dataset.

\begin{table}[!htp]
	\centering
	\color{black}
	\begin{tabular}{l|c}
		RES & FLH [h]  \\
		\hline
		Offshore Wind & 3,657.517 \\
		Onshore Wind & 1,864.482 \\
		Biomass & 5,756.5427 \\ 
		Photovoltaic & 986.3497 \\
		Hydro & 4,262.101782     
	\end{tabular}
	\caption{FLH for RES}
	\label{tab:res-flh}
\end{table}

Using the FLH of each production type, we determine the variable cost $c_{st}$ of a renewable unit $s \in S$ in a period $t \in T$ as follows:
\begin{align*}
	c_{st} = \text{OPEX}_{\text{var}_{f(s)}} + \text{Fuel}_{f(s)} + {\text{CO}_{2}}_{f(s)} + \frac{\text{OPEX}_{\text{fix}_{f(s)}}}{\text{FLH}_{f(s)}} 
\end{align*}

Note that we make the simplifying assumption that variable costs are constant for all time periods. Furthermore, while allocating fixed to variable costs, we do not consider the specific fixed cost of each plant, which would lead to different variable costs for each plant. Instead, as we use the capacity and generated power per plant type, we derive fixed costs for an average plant per plant type. Table \ref{tab:costs-res-var} shows the variable costs for the RES considered in this study. Note that in a final step, these values are anonymized for each power plant by drawing the final costs from a normal distribution, where the calculated variable costs serve as the mean and 1\% of these variable costs serve as the standard deviation.

\begin{table}[!htp]
	\centering
	\color{black}
	\begin{tabular}{l|c}
		RES & Variable Costs [EUR/MWh]  \\
		\hline
		Offshore Wind & 27.24 \\
		Onshore Wind & 18.73 \\ 
		Biomass & 77.34 \\ 
		Photovoltaic & 13.84 \\
		Hydro & 8.60       
	\end{tabular}
	\caption{Variable Costs for RES}
	\label{tab:costs-res-var}
\end{table}

The capacity of each RES unit is included in the BZR dataset as a static attribute. This static attribute does not adequately reflect the dynamic nature of RES outputs, which are influenced by weather and time of day. To calculate maximum output, we first determine the total capacity for each RES type and derive a score that represents the ratio of aggregated hourly output to total capacity. The maximum output per period for each RES unit is obtained by multiplying their capacity by this score.

\subsubsection{Non-Renewables}
The BZR dataset includes the variable costs for all thermal plant types. However, the fixed costs are not available, so we use the costs from \citep{Kost.2021}, which are shown in Table \ref{tab:non-res-costs}:

\begin{table}[!htp]
	\centering
	\color{black}
	\begin{tabular}{l|c}
		Technology & Variable Costs [EUR/MWh]  \\
		\hline
		Hard Coal & 22 \\
		Lignite & 32 \\ 
		CCGT & 20
	\end{tabular}
	\caption{Fixed Costs for Non-RES}
	\label{tab:non-res-costs}
\end{table}

Regarding the units that generate electricity from light oil, we refer to an older publication by \cite{Schroder.2013}, as more recent publications on the cost of electricity do not include power plants that generate electricity from oil. Concerning oil, \cite{Schroder.2013} state fixed costs of 6 EUR/kW/year for the year 2010.

Based on these costs for the different technologies, we estimate yearly fixed costs for each generation unit. However, as fixed costs in our market clearing models are incurred during operating hours, these costs must be distributed over those hours. We approximate these costs using the FLH, which relate a power plant's generated power to its capacity and represent the number of hours a unit operates at full capacity over a year. \cite{Kost.2021} specify the FLH for the year 2020 shown in Table \ref{tab:non-res-flh}.

\begin{table}[!htp]
	\centering
	\color{black}
	\begin{tabular}{l|c|c}
		& \multicolumn{2}{c}{FLH [h]} \\
		Technology & Lower Bound & Upper Bound  \\
		\hline
		Hard Coal & 2,600 & 6,200 \\
		Lignite & 5,300 & 7,300 \\
		Gas Turbine & 500 & 3,000 \\
		CCGT & 3,000 & 8,000
	\end{tabular}
	\caption{FLH for Non-RES}
	\label{tab:non-res-flh}
\end{table}

Note that we assume that any power plant that generates electricity from gas and that is not categorized as a CCGT power plant is associated with the gas turbine technology. Furthermore, since neither \cite{Kost.2021} nor \cite{Schroder.2013} provide information on FLH for oil power plants, we take the estimates from a recent monitoring report of the Federal Network Agency.\footnote{\url{https://data.bundesnetzagentur.de/Bundesnetzagentur/SharedDocs/Mediathek/Monitoringberichte/monitoringbericht_energie2021.pdf}} In particular, we consider the installed generation capacity of 4.8 GW in 2020 and the net power generation of 4.3 TWh in 2020 implying a value of 895.83 FLH for an average power plant that produces energy from light oil. Therefore, we consider the values depicted in Table \ref{tab:fixed-opex-non-res} for each of the plant types.

\begin{table}[!htp]
	\centering
	\color{black}
	\begin{tabular}{l|c|c|c}
		& OPEX fix & FLH Lower Bound & FLH Upper Bound		\\
		Fuel/Plant Type & [EUR/kW/year] & [h] & [h] \\
		\hline
		Hard Coal/old 1 & 22 & 2600 & 6200 \\
		Hard Coal/old 2 & 22 & 2600 & 6200 \\
		Hard Coal/new & 22 & 2600 & 6200 \\
		Lignite/old 1 & 32 & 5300 & 7300 \\ 
		Lignite/old 2 & 32 & 5300 & 7300 \\
		Lignite/new & 32 & 5300 & 7300 \\
		Gas/Conventional old 1 & 20 & 500 & 3000 \\
		Gas/Conventional old 2 & 20 & 500 & 3000 \\
		Gas/CCGT old 1 & 20 & 3000 & 8000 \\ 
		Gas/CCGT old 2 & 20 & 3000 & 8000 \\
		Gas/CCGT new & 20 & 3000 & 8000 \\ 
		Gas/OCGT old & 20 & 500 & 3000 \\ 
		Gas/OCGT new & 20 & 500 & 3000 \\
		Gas/CCGT present 1 & 20 & 3000 & 8000 \\
		Gas/CCGT present 2 & 20 & 3000 & 8000 \\
		Light oil/- & 6 & 895.83 & 895.83
	\end{tabular}
	\caption{FLH for Non-RES}
	\label{tab:fixed-opex-non-res}
\end{table}

As the FLH represent a lower bound for the actual operating hours, we assume that the upper bound describes the actual operating hours. Thus, we denote the upper bound of the FLH of a plant technology $k$ shown in Table \ref{tab:fixed-opex-non-res} as $\overline{FLH}_k$ and define the fixed cost of a non-renewable generation unit $s \in S$ as: 
\begin{align*}
	h_s = \frac{\underbar{P}_s \text{OPEXfix}_{k(s)}}{\overline{FLH}_{k(s)}}
\end{align*}
where $k(s)$ denotes the fuel/plant type of generation unit $s$.
\ref{app:mapping} describes what non-RES capacities we assume.

\color{black}

\subsection{Matching of Generators to Grid Locations} \label{app:mapping}
The BZR grid data specify the nominal capacity of thermal generating units at each node, but information on the specific plant type and its operating restrictions are not provided. A different data set provides the aggregate capacity of each specific plant type (e.g., Hard Coal old, Hard Coal new, Gas CCGT, etc.) as well as their operating characteristics (e.g., minimum uptime constraints). Unfortunately, no straightforward mapping exists between the two datasets.

As discussed in Section \ref{sec:data}, we first identify each thermal unit's broad plant type (e.g., hard coal, gas, etc.) by matching IDs with the \textit{Kraftwerksliste} (power plant list) provided by the Bundesnetzagentur. 
Next, we need to distribute these plants to the sub-categories provided in the seller data. For example, all hard coal units must be distributed between Hard Coal old1, Hard Coal old2, and Hard Coal new, with each sub-category having different operational characteristics.

To perform this mapping, we solve an integer program \citep{Schmitt.2023}. Let $K$ be the set of broad plant types (e.g., hard coal, gas) and $A_k$ be the set of sub-categories (e.g., Hard Coal old1, Hard Coal old2, Hard Coal new) for each type $k$. The total capacity and number of units of each specific plant type $a \in A_k$ is given as $P_a$ and $n_a$ from the seller data. From the grid data, we obtain a set of generators $I_k$ for each broad plant type $K$, each having a nominal capacity of $p_i, i \in I_k$. We denote as $x_{ia} \in \{0,1\}$ as the desired mapping of each generator to a specific plant type. 

With this notation, the mapping problem looks as follows:
\begin{align}
	\min_{x} \quad & \sum_{k \in K, a \in A_k} -n_a(P_a - \sum_{i \in I_k} p_i x_{ia}) + \sum_{k \in K, a \in A_k} -P_a(n_a - \sum_{i \in I_k} x_{ia})  & \label{model:gen_mapping} \\
	\text{subject to \quad} & \sum_{a \in A_k} x_{ia} = 1 \quad \forall i \in I_k, k \in K & \nonumber \\
	& 0 \leq P_a - \sum_{i \in I_k} p_i x_{ia} \geq 600 \quad \forall a \in A_k, k \in K & \nonumber \\
	& 0 \leq N_a - \sum_{i \in I_k} x_{ia} \geq 2 \quad \forall a \in A_k, k \in K & \nonumber
\end{align}

The objective function minimizes the aggregate differences in both capacities and number of units per plant type, with equal weight. The first constraint ensures that each unit of broad type $k$ (e.g., hard coal) is assigned to exactly one specific plant type $a \in A_k$ (e.g., Hard Coal old1). The last two constraints bound the maximum allowed deviations in capacities and the number of units for each specific plant type $a$. Note that a broad type $k$ could not be identified for all units, and such unidentified units could be matched freely to any specific type $a$. After solving the optimization problem, we obtain a specific type $a$ for each unit in the grid model and can use its respective operating characteristics as input to the clearing problem in (\ref{model:general}). 

The result of this mapping is presented in Table \ref{tab:gen_mapping}. Especially hard coal, lignite, and oil plants could be mapped almost perfectly. Concerning gas units, the grid and seller datasets exhibited more significant differences, which is reflected in the mapping.  

\begin{table}[!htp]
	\centering
	\begin{tabular}{l|cc|cc}
		& \multicolumn{2}{c}{Seller Data} & \multicolumn{2}{c}{Grid Data} \\
		Fuel / PlantType       & Capacity {[}MW{]} & Units & Capacity {[}MW{]} & Units \\
		\hline
		Hard Coal / old1        & 1,905.1           & 8     & 1,905             & 8     \\
		Hard Coal / old2        & 5,172.4           & 27    & 5,165.4           & 27    \\
		Hard Coal / new         & 5,178             & 8     & 5,178             & 7     \\
		Lignite / old1         & 4,989.3           & 13    & 4,989             & 13    \\
		Lignite / old2         & 3,111.1           & 11    & 3,111             & 11    \\
		Lignite / new          & 6,437.8           & 11    & 6,437             & 9     \\
		Gas Conventional / old1 & 776.01           & 26    & 775.97           & 26    \\
		Gas Conventional / old2 & 1215.4           & 23    & 1,215.33          & 23    \\
		Gas CCGT / old1         & 3,601.049         & 78    & 3,001.054         & 78    \\
		Gas CCGT / old2         & 5,356.22          & 40    & 5,055.723         & 38    \\
		Gas CCGT / new          & 7,389.1           & 45    & 6,789.321         & 43    \\
		Gas OCGT / old          & 760.1            & 15    & 760.025          & 15    \\
		Gas OCGT / new          & 2,613.168         & 66    & 2,013.224         & 66    \\
		Gas CCGT / present1     & 64               & 2     & 64               & 2     \\
		Gas CCGT / present2     & 282              & 2     & 282              & 2     \\
		Light oil           & 998.33           & 21    & 997.43           & 21   
	\end{tabular}
	\caption{Generator Mapping Results}
	\label{tab:gen_mapping}
\end{table}

\section{Analysis of Representative Days} \label{app:repr-days}
To corroborate our findings, we examine three selected days in greater detail. Specifically, we analyze November 23 for its low price levels, March 10 for medium price levels, and February 18 for high price levels. 

The generation costs for each of these days are presented in Table \ref{tab:costs_day}. On 2009/11/23, characterized by warm and windy conditions with low demand, generation costs were minimal, and no congestion occurred in the zonal models, resulting in identical allocations for the national, 2 Zones (k) and 3 Zones configurations. In contrast, February 18, as a cold and cloudy day, saw substantially higher generation costs, but cross-zonal flow constraints had little impact on the allocation. Across all days, the shift from zonal to nodal pricing led to a similar increase in generation costs.

Table \ref{tab:redis_day} indicates that minimum cost redispatch costs are not correlated with the generation costs presented in Table \ref{tab:costs_day}. This observation underscores that zonal models lack a representation of transmission bottlenecks, necessitating redispatch regardless of the zonal configuration or day.


\begin{table}[!htp]
	\centering
	\begin{tabular}{c|cccccc}
		in kEUR & National & 2 Zones (k) & 2 Zones (s) & 3 Zones & 4 Zones &  Nodal \\
		\hline
		2009/11/23 & 26,673.75 & 27,921.59 & 26,673.75 & 26,673.75 & 26,774.86 & 29,596.10 \\
		2009/03/10 & 31,501.65 & 31,512.26 & 31,491.45 & 31,667.54 & 31,667.18 & 35,200.69 \\
		2009/02/18 & 46,287.24 & 46,296.34 & 46,307.95 & 46,281.61 & 46,432.04 & 49,303.25
	\end{tabular}
	\caption{Generation Costs: Nov 23, Mar 10, Feb 18}
	\label{tab:costs_day}
\end{table}


\begin{table}[!htp]
	\centering
	\begin{tabular}{c|cccccc}
		in kEUR & National & 2 Zones (k) & 2 Zones (s) & 3 Zones & 4 Zones &  Nodal \\
		\hline
		2009/11/23 & 5,004.47 &	4,998.43 & 5,012.4 & 5,014.47 & 4,753.79 & 0 \\
		2009/03/10 & 6,295.56 &	6,262.94 & 6,305.71 & 6,129.62 & 6,129.98 & 0 \\
		2009/02/18 & 3,603.04 & 3,593.5 & 3,602.7 & 3,602.93 & 3,458.41 & 0
	\end{tabular}
	\caption{\textcolor{black}{RD-Min-Cost Redispatch Costs: Nov 23, Mar 10, Feb 18}}
	\label{tab:redis_day}
\end{table}

In summary, as outlined in Table \ref{tab:tot_costs_day}, the nodal allocation rule consistently yields lower total costs than any zonal configurations. Despite modest cost savings, nodal allocation enables locational price signals and incentivizes a long-run equilibrium. 

On 2009/11/23 and 2009/03/10, the allocation obtained from the Euphemia algorithm is identical to the national configuration, and no welfare loss occurs. On 2009/02/18, costs under the Euphemia algorithm are slightly higher by 0.36\%. This validates our assertion that differences between zonal models -- including Euphemia -- are minor and nuanced, while nodal pricing constitutes a more substantial change in generation and redispatch costs.


\begin{table}[!htp]
	\centering
	\begin{tabular}{c|cccccc}
		in kEUR & National & 2 Zones (k) & 2 Zones (s) & 3 Zones & 4 Zones &  Nodal \\
		\hline
		2009/11/23 & 31,678.22 & 32,920.02 & 31,686.15 & 31,688.22 & 31,528.65 & 29,596.10 \\
        2009/03/10 & 37,797.21 & 37,775.2 & 37,797.16 & 37,797.16 & 37,797.16 & 35,200.69 \\
        2009/02/18 & 49,890.28 & 49,889.84 & 49,910.65 & 49,884.54 & 49,890.45 & 49,303.25 \\
   	\end{tabular}
	\caption{\textcolor{black}{Total Costs: Nov 25, Mar 10, Feb 18}}
	\label{tab:tot_costs_day}
\end{table}

\begin{figure*}[!htp]
	\centering
	\begin{subfigure}[b]{0.32\textwidth}
		\centering
		\includegraphics[width=\textwidth]{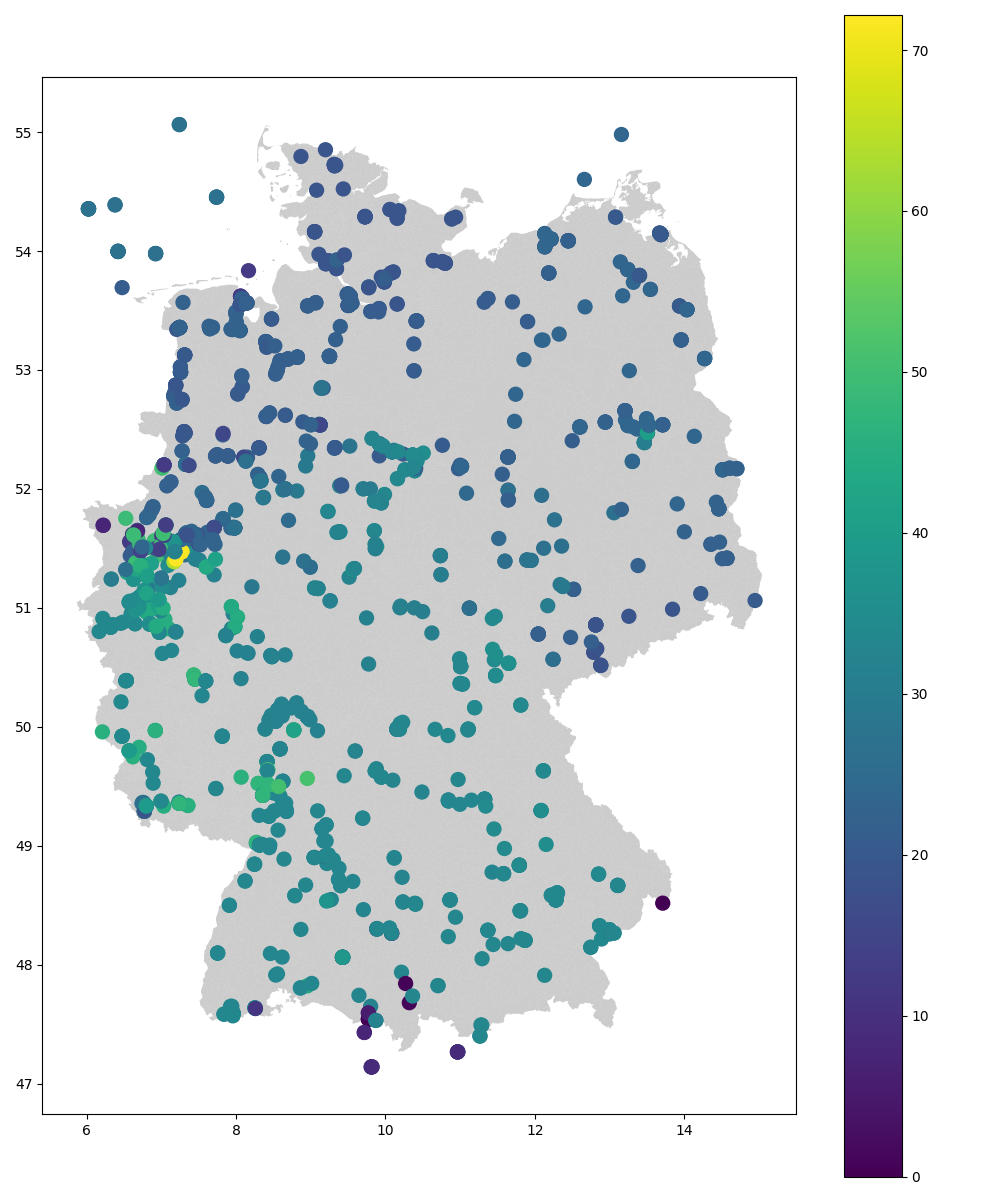}
		\caption{\small 2009/11/23 - Avg IP Prices}   
		\label{fig:nov_ip_map}
	\end{subfigure}
	\begin{subfigure}[b]{0.32\textwidth}  
		\centering 
		\includegraphics[width=\textwidth]{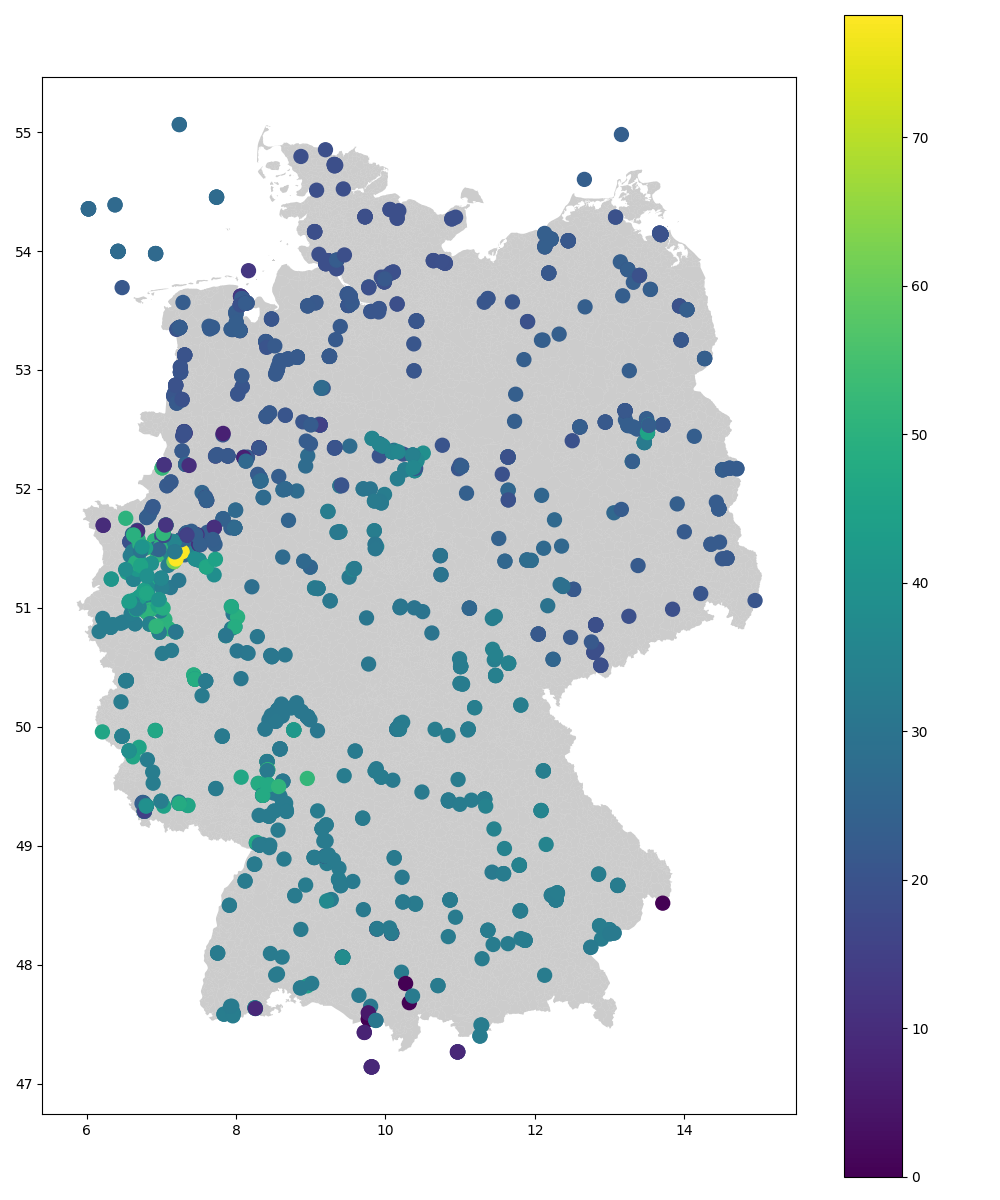}
		\caption{\small 2009/11/23 - Avg CH Prices}
		\label{fig:nov_ch_map}
	\end{subfigure}
	\begin{subfigure}[b]{0.32\textwidth}   
		\centering 
		\includegraphics[width=\textwidth]{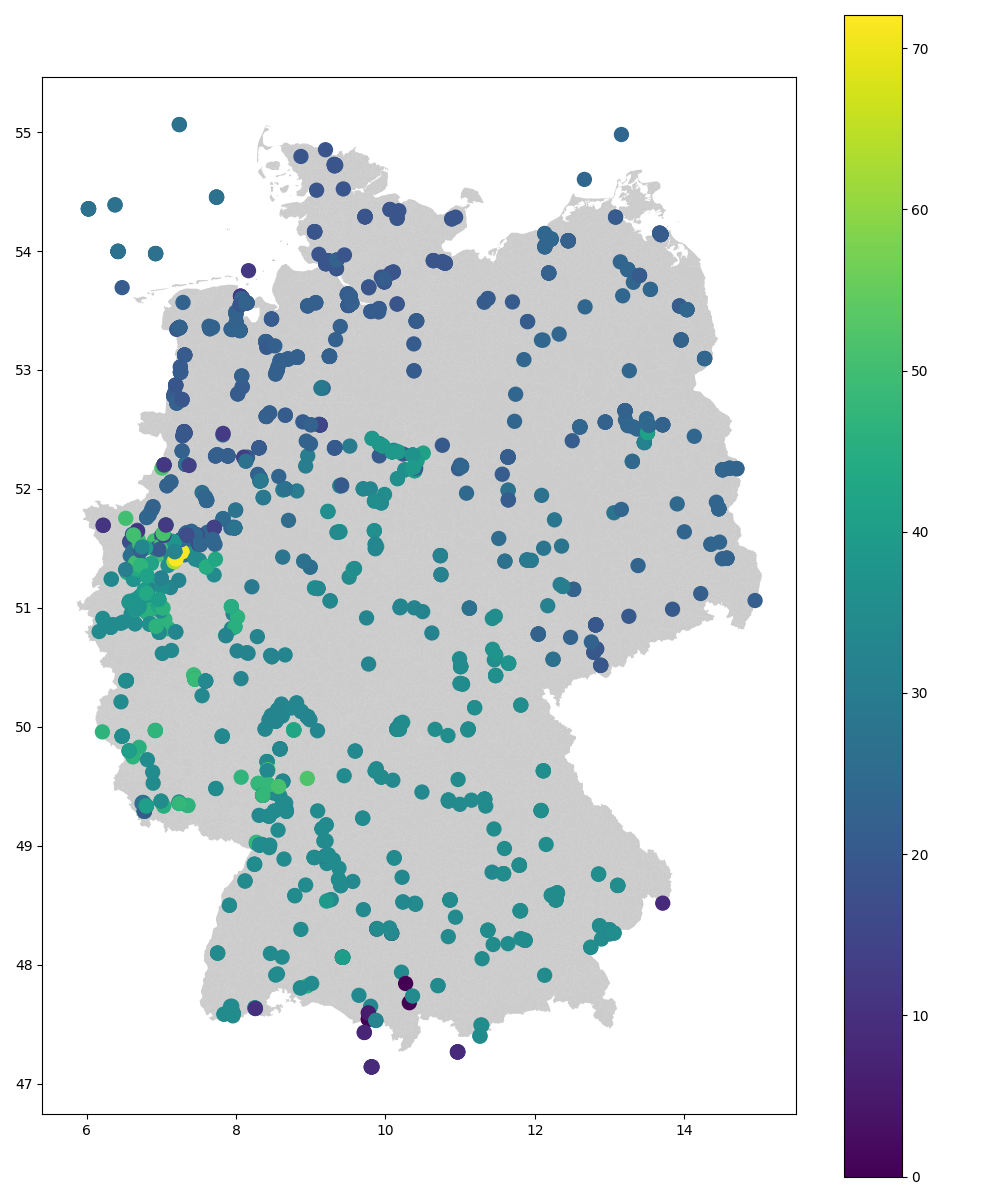}
		\caption{\small 2009/11/23 - Avg Join Prices}    
		\label{fig:nov_join_map}
	\end{subfigure}
	\vskip\baselineskip
	\begin{subfigure}[b]{0.32\textwidth}   
		\centering 
		\includegraphics[width=\textwidth]{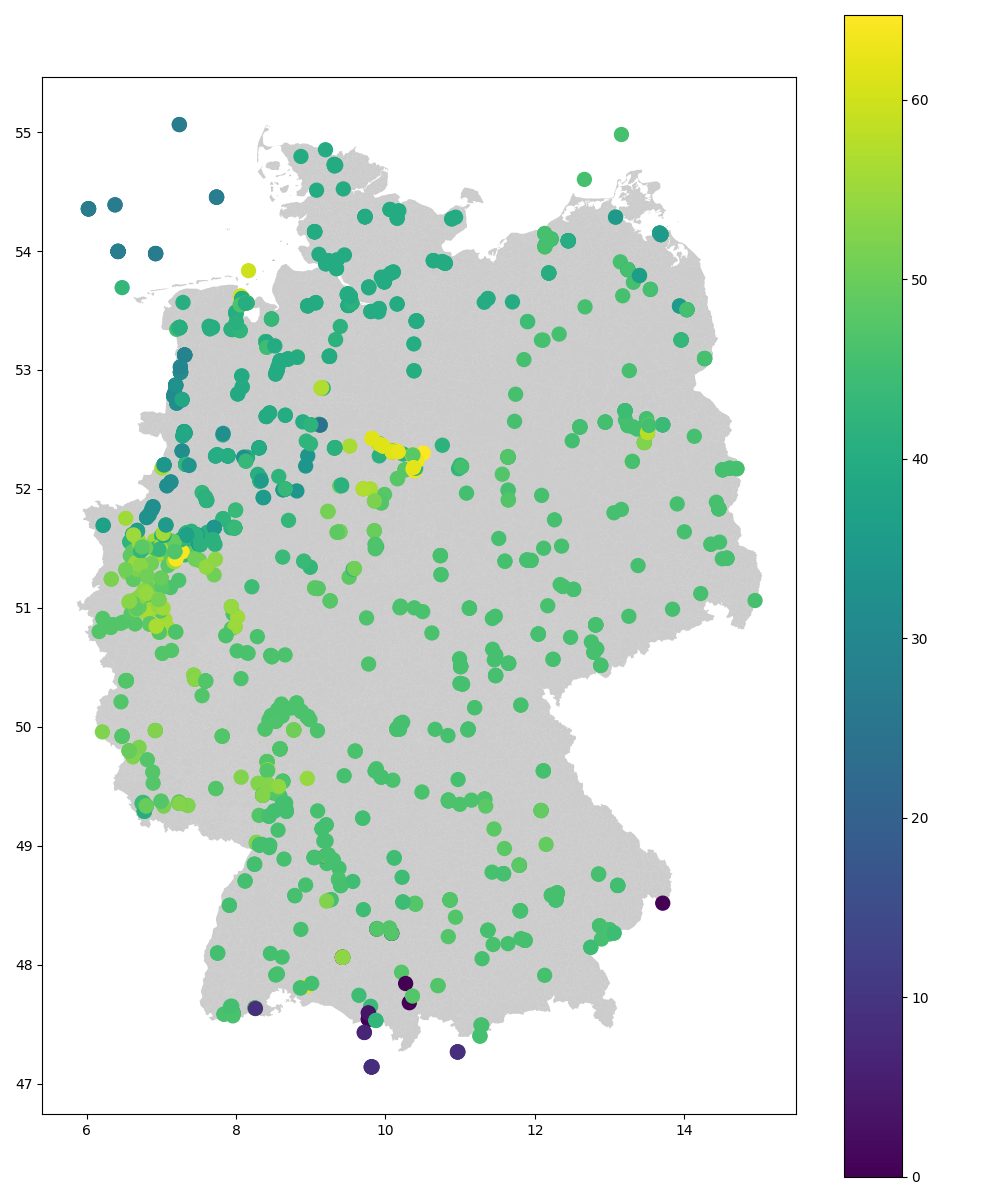}
		\caption{\small 2009/03/10 - Avg IP Prices}  
		\label{fig:mar_ip_map}
	\end{subfigure}
	\begin{subfigure}[b]{0.32\textwidth}   
		\centering 
		\includegraphics[width=\textwidth]{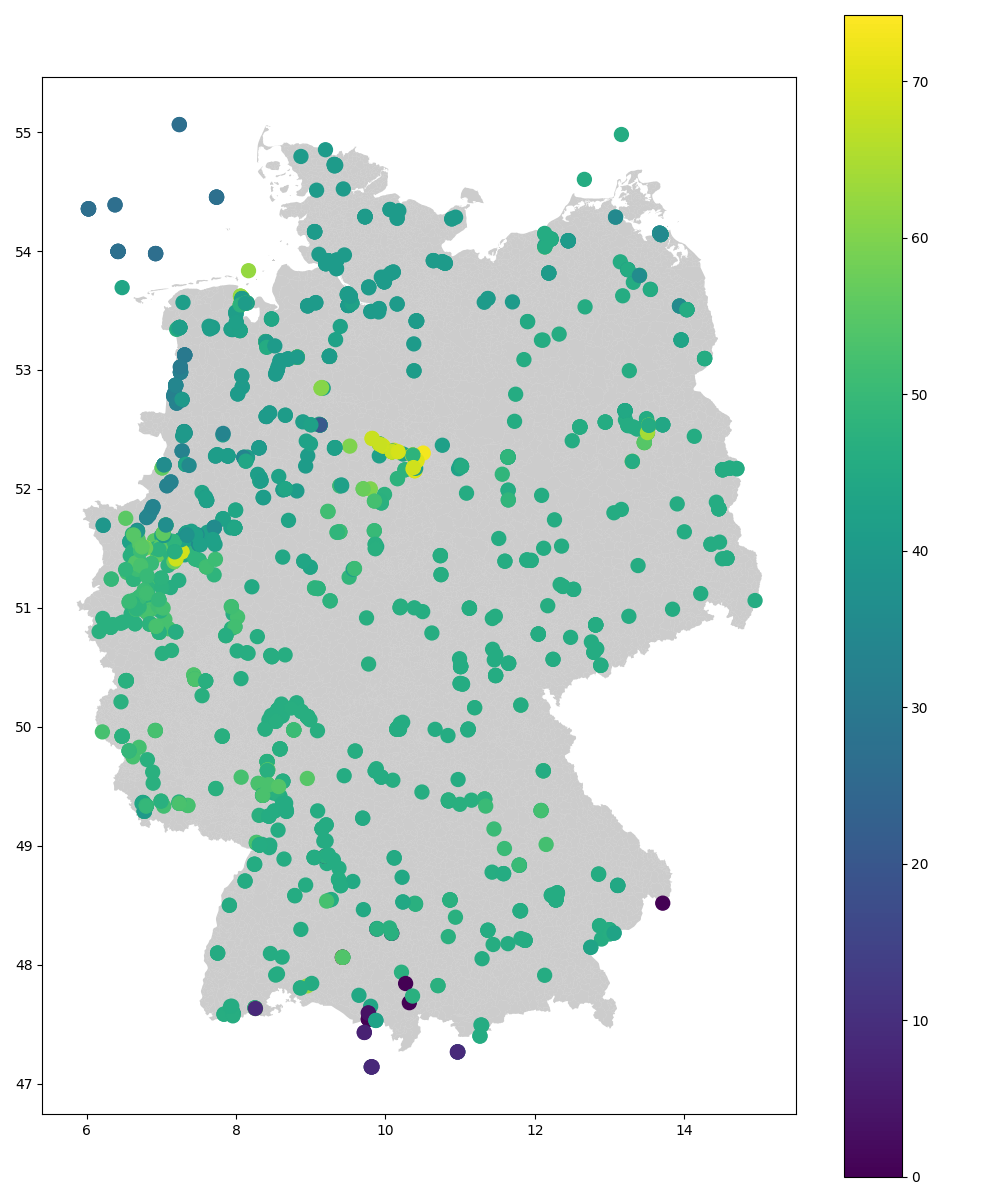}
		\caption{\small 2009/03/10 - Avg CH Prices}  
		\label{fig:mar_ch_map}
	\end{subfigure}
	\begin{subfigure}[b]{0.32\textwidth}   
		\centering 
		\includegraphics[width=\textwidth]{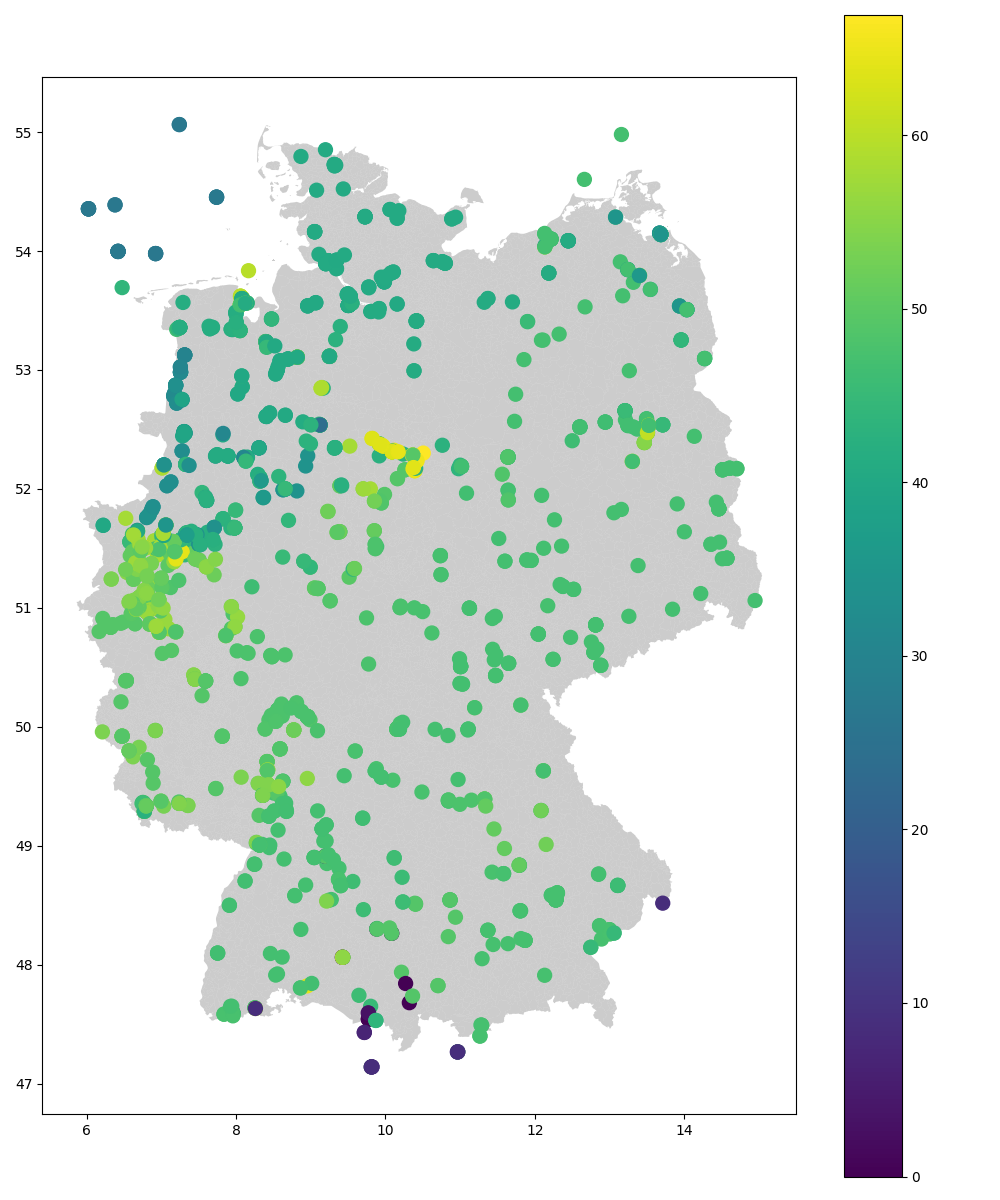}
		\caption{\small 2009/03/10 - Avg Join Prices}  
		\label{fig:mar_join_map}
	\end{subfigure}
	\vskip\baselineskip
	\begin{subfigure}[b]{0.32\textwidth}   
		\centering 
		\includegraphics[width=\textwidth]{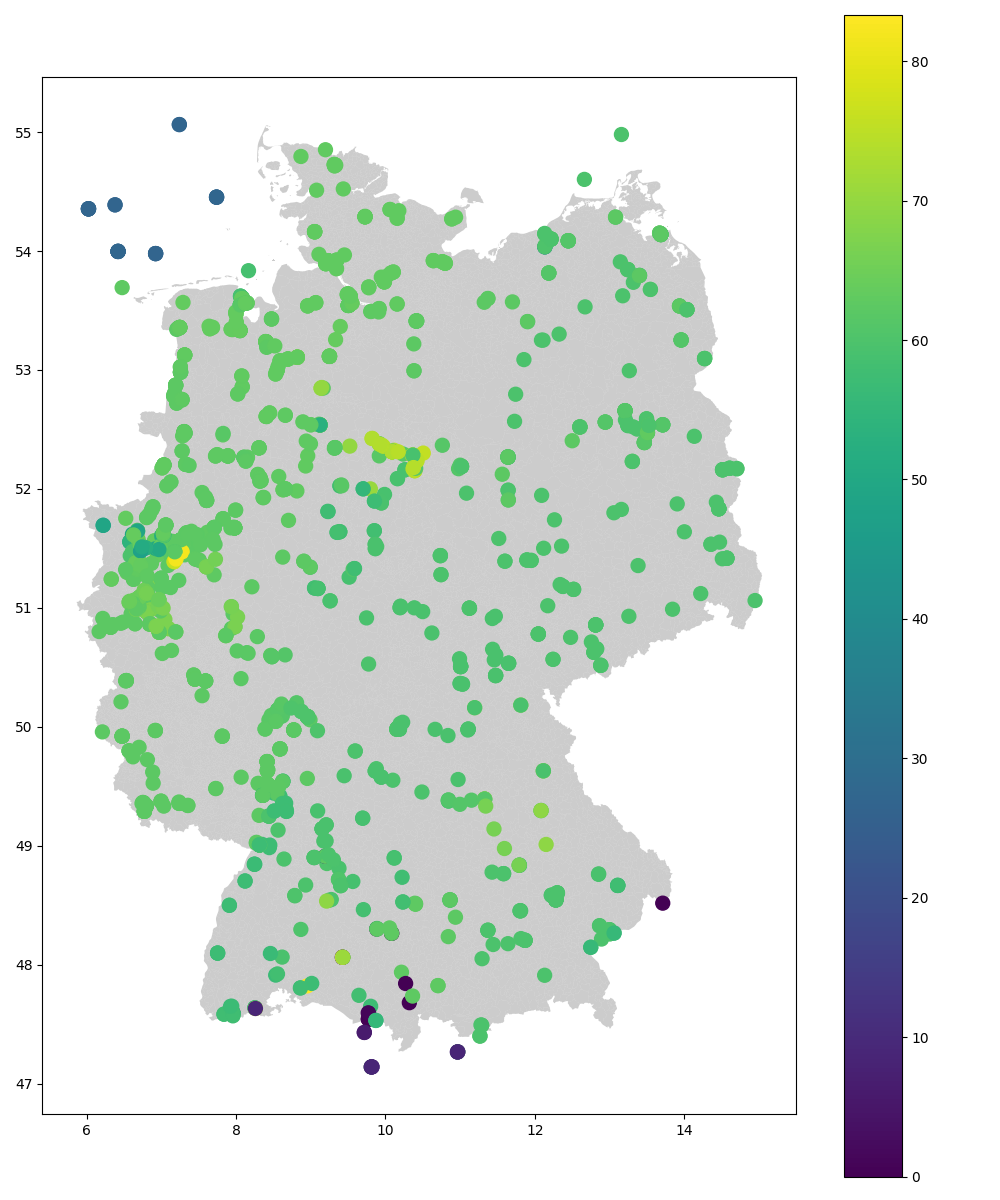}
		\caption{\small 2009/02/18 - Avg IP Prices}  
		\label{fig:feb_ip_map}
	\end{subfigure}
	\begin{subfigure}[b]{0.32\textwidth}   
		\centering 
		\includegraphics[width=\textwidth]{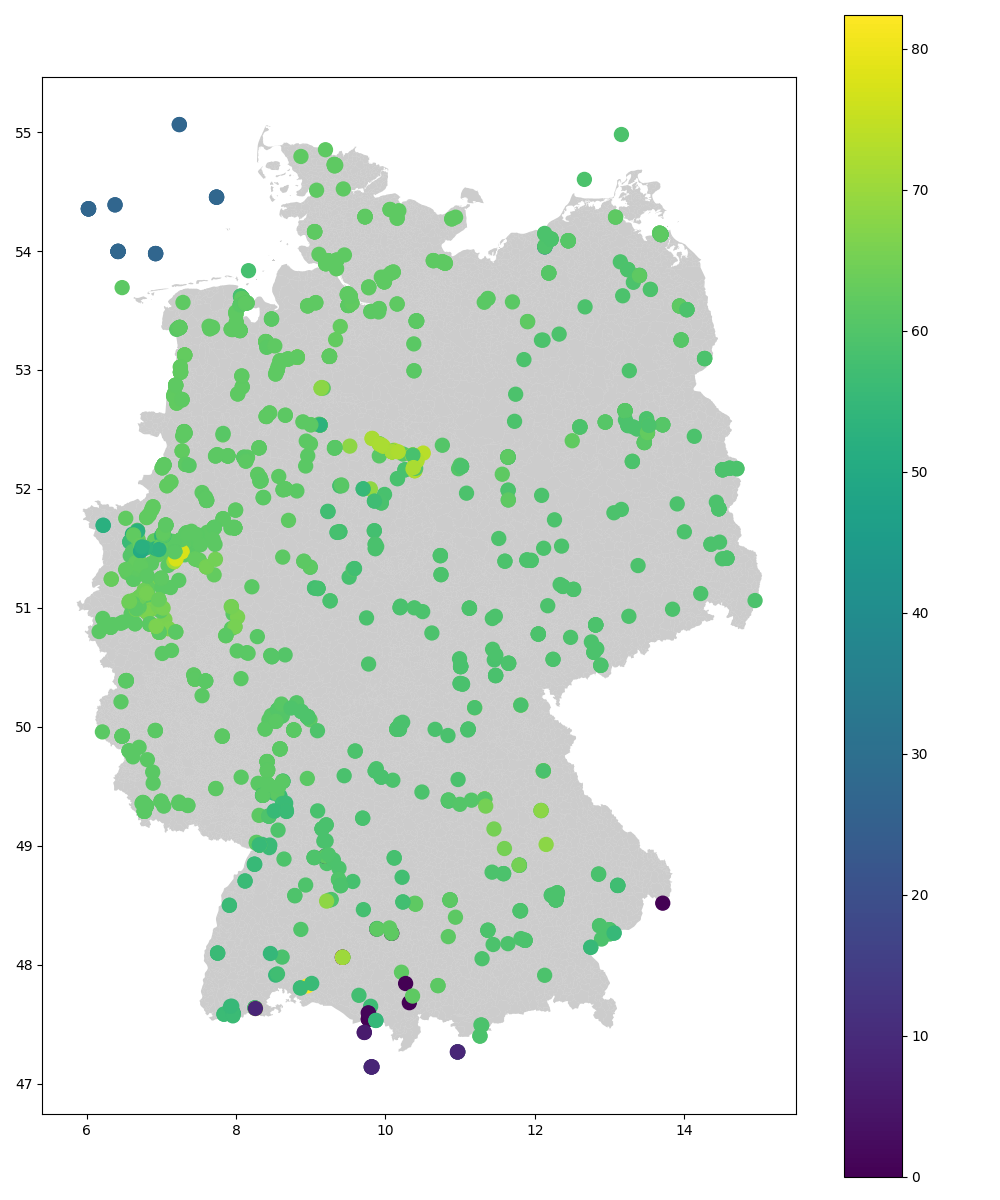}
		\caption{\small 2009/02/18 - Avg CH Prices}  
		\label{fig:feb_ch_map}
	\end{subfigure}
	\begin{subfigure}[b]{0.32\textwidth}   
		\centering 
		\includegraphics[width=\textwidth]{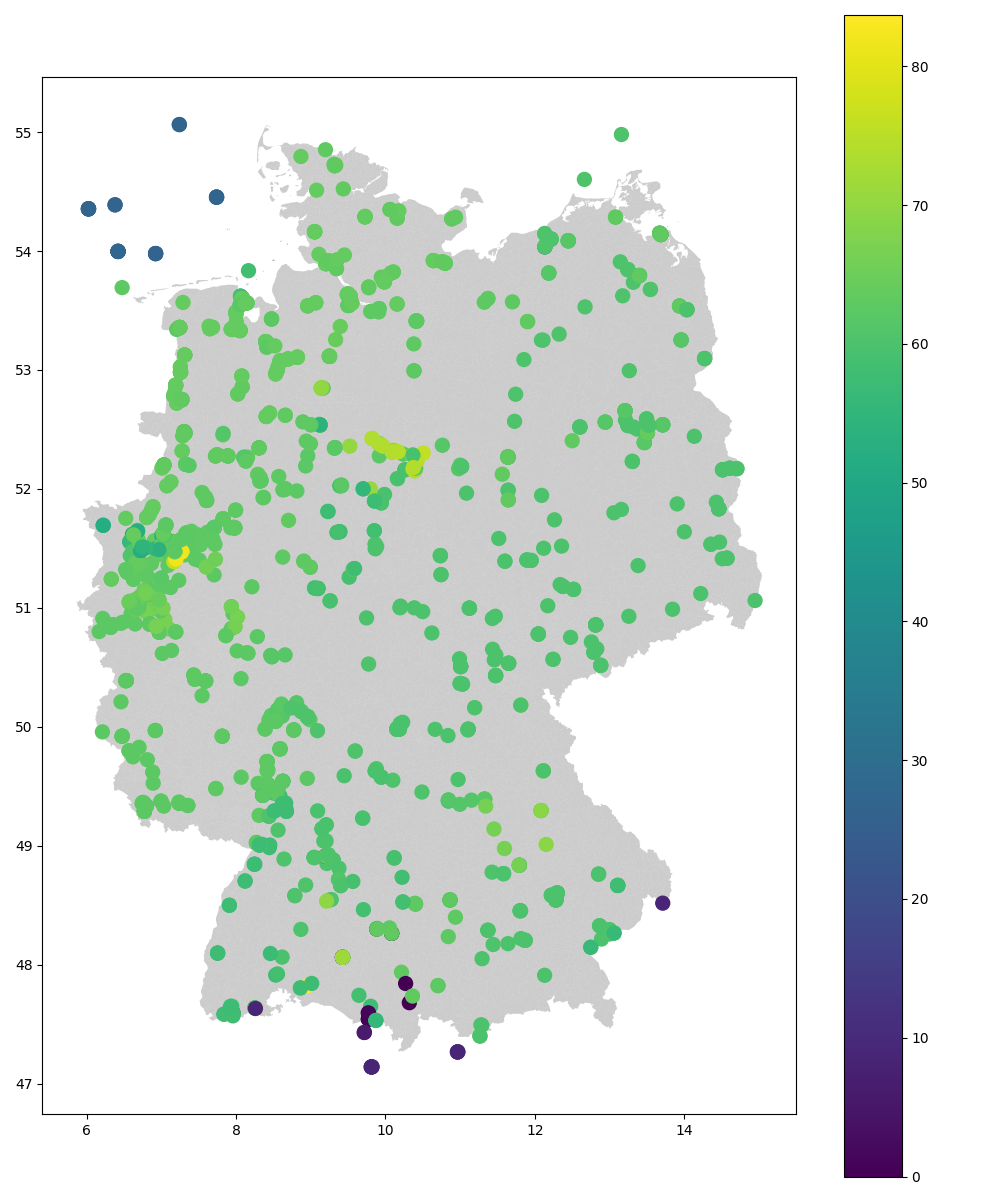}
		\caption{\small 2009/02/18 - Avg Join Prices}  
		\label{fig:feb_join_map}
	\end{subfigure}
	\caption{Average Nodal Prices} 
	\label{fig:map_local_prices}
\end{figure*}

Figure \ref{fig:map_local_prices} illustrates the locational IP, CH, and Join prices for the three selected days, averaged over 24 hours. As previously discussed, the prices on these days exhibit a positive correlation with generation costs, with the exception of Northern Central and Southern Germany with consistent price outliers due to local supply structures. Notably, even though at lower price levels, a more distinct north-south price gradient can be observed on 2009/11/23 compared to 2009/03/10. Specifically, on 2009/11/23, demand is low nationwide, yet inexpensive electricity from the north cannot be transmitted to the south, leading to uneven price levels. In contrast, on 2009/03/10, with higher demand but a balanced grid, prices are uniform across the country. This phenomenon holds for all three pricing rules. The simplified zonal models fail to detect any transmission bottlenecks on 2009/11/23, i.e., cross-zonal constraints are not tight and there are no price variations between zones. Consequently, if the zone configurations are suboptimal, zonal prices do not signal scarcity appropriately. 

Table \ref{tab:stats_days} summarizes the GLOCs, LLOCs, and MWPs for each selected day. On the low-demand day of 2009/11/23, a Walrasian equilibrium as in Definiton \ref{def:WE} is attainable for the national, 2 Zones (k) and 3 Zones models, resulting in zero GLOCs, LLOC, and MWPs. Crucially, all three pricing rules successfully obtain these equilibrium prices. The Walrasian equilibrium, however, can only persist before redispatch is considered and ceases to exist when nodal transmission constraints are taken into account. 

GLOCs, LLOCs, and MWPs do not necessarily increase with average price levels. For nodal pricing, they remain relatively consistent across all days. There is a substantial increase for zonal pricing on the medium-price day of 2009/03/10, compared to the high-price day of 2009/02/18. This indicates that incentives to deviate hinge more on the quality of the price signal than the actual price level. 

The simplification inherent in zonal models, which leads to redispatch, does not necessarily imply better economic properties of prices. For example, on 2009/02/18, zonal IP and Join prices result in higher GLOCs than nodal pricing. Specifically, a few zonal prices are insufficient to incentivize many market participants to adhere to the allocation. 

As discussed above, CH prices always minimize GLOCs, and IP prices always yield zero LLOCs. Although GLOCs are reduced by CH pricing compared to both IP and Join pricing, their LLOCs, and often MWPs, experience a slight increase. In practical terms, this implies that congestion signals are more distorted and higher side-payments are required. In contrast, the Join pricing rule achieves a favorable trade-off between LLOCs and MWPs, requiring only slightly higher penalties (corresponding to GLOCs) than IP pricing.

\begin{figure}[!htp]
	\begin{subfigure}{0.45\textwidth}
		\centering
		\resizebox{\textwidth}{!}{%
		\begin{tabular}{c|c|ccc}
			\multicolumn{1}{l}{in kEUR}   &      & IP           & CH         & Join         \\
			\hline
			\multirow{3}{*}{National}    & GLOC & 0 & 0 & 0      \\
			& LLOC & 0 & 0 & 0        \\
			& MWP  & 0 & 0 & 0       \\
			\hline
			\multirow{3}{*}{2 Zones (k)} & GLOC & 0 & 0 & 0   \\
			& LLOC & 0 & 0 & 0        \\
			& MWP  & 0 & 0 & 0        \\
			\hline
			\multirow{3}{*}{2 Zones (s)} & GLOC & 1,774.99 &  1,167.39 & 1,774.99  \\
			& LLOC & 0 & 515.51 & 0        \\
			& MWP  & 0 & 117.40 & 0       \\
			\hline
			\multirow{3}{*}{3 Zones}     & GLOC & 0 & 0 & 0    \\
			& LLOC & 0 & 0 & 0        \\
			& MWP  & 0 & 0 & 0        \\
			\hline
			\multirow{3}{*}{4 Zones}     & GLOC & 1,824.29 & 791.11 & 1,824.29  \\
			& LLOC & 0 & 292.25 & 0     \\
			& MWP  & 0 & 118.68 & 0      \\
			\hline
			\multirow{3}{*}{Nodal}       & GLOC & 1,290.22 & 273.99 & 1,507.04 \\
			& LLOC & 0 & 149.17 & 27.31 \\
			& MWP  & 167.98 & 96.80 & 34.41
		\end{tabular}}
		\caption{GLOC, LLOC, MWP for 2009/11/23}
		\label{tab:stats_nov}
	\end{subfigure}
	\begin{subfigure}{0.45\textwidth}
		\centering
		\resizebox{\textwidth}{!}{%
		\begin{tabular}{c|c|ccc}
			\multicolumn{1}{l}{in kEUR}   &      & IP           & CH         & Join         \\
			\hline
			\multirow{3}{*}{National}    & GLOC & 2,274.12 & 52.96 & 924.26 \\
			& LLOC & 0 & 16.61 & 1.93    \\
			& MWP  & 134.79 & 13.33 & 0     \\
			\hline
			\multirow{3}{*}{2 Zones (k)} & GLOC & 3,539.04 & 45.62 &  3,539.04 \\
			& LLOC & 0 & 8.64 & 0      \\
			& MWP  & 164.79 & 4.60 & 164.79      \\
			\hline
			\multirow{3}{*}{2 Zones (s)} & GLOC & 4,730.19 & 260.35 & 918.13 \\
			& LLOC & 0 & 27.39 & 0.98 \\
			& MWP  & 84.07 & 26.24 & 1.01      \\
			\hline
			\multirow{3}{*}{3 Zones}     & GLOC & 3,872.84 & 213.04 & 3,872.84 \\
			& LLOC & 0 & 172.13 & 0        \\
			& MWP  & 0 & 153.07 & 0     \\
			\hline
			\multirow{3}{*}{4 Zones}     & GLOC & 787.65 & 751.28 & 970.31 \\
			& LLOC & 0 & 101.38 & 5.95       \\
			& MWP  & 109.6 & 132.42 & 3.26     \\
			\hline
			\multirow{3}{*}{Nodal}       & GLOC & 1,906.47 & 256.65 & 1.942.01 \\
			& LLOC & 0 & 158.07 & 34.79    \\
			& MWP  & 191.19 & 84.14 & 20.73
		\end{tabular}}
		\caption{GLOC, LLOC, MWP for 2009/03/10}
		\label{tab:stats_mar}
	\end{subfigure}
	\vskip\baselineskip
	\centering
	\begin{subfigure}{0.45\textwidth}
		\centering
		\resizebox{\textwidth}{!}{%
		\begin{tabular}{c|c|ccc}
			\multicolumn{1}{l}{in kEUR}   &      & IP         & CH        & Join       \\
			\hline
			\multirow{3}{*}{National}    & GLOC & 4,442.86 & 318.73 & 2,618.64 \\
			& LLOC & 0 & 29.04 & 9.31   \\
			& MWP  & 41.66 & 27.1 & 8.59  \\
			\hline
			\multirow{3}{*}{2 Zones (k)} & GLOC & 2,258.24 & 502.83 & 2,600.59 \\
			& LLOC & 0 & 166.68 & 6.14  \\
			& MWP  & 35.23 & 155.67 & 6.02  \\
			\hline
			\multirow{3}{*}{2 Zones (s)} & GLOC & 1,727.29 & 319.07 & 1,956.49 \\
			& LLOC & 0 & 16.06 & 8.01   \\
			& MWP  & 26.55 & 23.60 & 13.90 \\
			\hline
			\multirow{3}{*}{3 Zones}     & GLOC & 3,284.92 & 318.84 & 3,284.92 \\
			& LLOC & 0 & 28.68 & 0    \\
			& MWP  & 46.26 & 26.01 & 46.26  \\
			\hline
			\multirow{3}{*}{4 Zones}     & GLOC & 1,523.92 & 464.18 & 2,642.54 \\
			& LLOC & 0 & 136.24 & 6.03   \\
			& MWP  & 35.19 & 102.14 & 5.92  \\
			\hline
			\multirow{3}{*}{Nodal}       & GLOC & 551.27 & 241.74 & 624.76 \\
			& LLOC & 0 & 51.26 & 25.56 \\
			& MWP  & 73.43 & 27.55 & 20.02
		\end{tabular}}
		\caption{GLOC, LLOC, MWP for 2009/02/18}
		\label{tab:stats_feb}
	\end{subfigure}
	\captionsetup{type=table}
	\caption{Daily GLOCs, LLOCs, MWPs}
	\label{tab:stats_days}
	\end{figure}

\section{Median prices} \label{app:median-prices}

Median prices are shown in Table \ref{tab:medians}.

\begin{table}[!htp]
	\centering
	\begin{tabular}{c|cccccc}
		in EUR/MWh & National & 2 Zones (k) & 2 Zones (s) & 3 Zones & 4 Zones &  Nodal \\
		\hline
		IP & 48.66 & 48.44 & 48.84 & 48.23 & 45.79 & 48.66 \\
		CH & 31.09 & 31.17 & 31.27 & 31.11 & 38.53 & 49.41 \\
		Join & 48.68 & 49.37 & 48.60 & 48.31 & 49.27 & 49.19 \\
		Euphemia & 48.84 & & & & &
	\end{tabular}
	\caption{Median Prices}
	\label{tab:medians}
\end{table}

\section{Modelling Differences Compared to the BZR Study} \label{app:modelling-differences}

In this section, we highlight the main modelling differences between our study and the BZR study. The BZR study includes (1) the LMP study \citep{ENTSOE.2022}, in which locational marginal prices (LMPs) were computed for each node in the CE grid, and (2) an assessment conducted at the CE level of the alternative bidding zone configurations identified based on the computed LMPs \citep{ENTSOE.2025MainReport}.

\subsection{Nodal Prices}
In the LMP study, nodal prices were derived as shadow prices by solving a linear unit commitment problem, formulated as a linear program comprising Continental Europe and Ireland. In contrast, we computed nodal prices based on a mixed-integer linear problem (see \ref{app:dcopf}). We note that the average nodal prices between the two studies are very close: for the climate year 2009, the average value of the prices in the LMP study is 46.83 EUR/MWh, while the average value of the prices computed in this study with IP pricing is 46.63 EUR/MWh. 

\subsection{Zonal Market Clearing Problem}
Several modelling differences relate to the zonal market clearing problem. To assess the impact of an alternative configuration at the CE level, the BZR study employs a modelling chain comprising five modules: (1) base case creation, (2) capacity calculation, (3) market coupling, (4) Operational Security Analysis (OSA) and Remedial Actions Optimisation (RAO), and (5) loop flow analysis. In module (1), a market simulation is performed for the entire EU with the goal of obtaining a market result forecast to be used for the capacity calculation (module (2)). In module (2), a flow-based market coupling and a coordinated net transfer capacity (cNTC) approach are implemented to compute the cross-zonal capacity. These approaches rely on multiple steps, such as calculating generation shift keys, zonal PTDFs and RAMs. Module (3) uses the FB parameters and cNTC values, as well as market data to perform a flow-based market coupling simulation for the CE region. The outcome of this module is the market dispatch, which is used, together with the available remedial actions (or, the available redispatch potential) as input in the OSA/RAO module. This module employs a DC load flow to identify congestions, which are subsequently alleviated in a cost-optimal way by solving a linear optimization problem. The last module receives the final dispatch and applies a power flow coloring approach to identify the share of flows not induced by cross-zonal trade (i.e., the loop flows).
The implementation details of each module are not publicly available.

Modules (3) and (4) are particularly relevant for our study. Module (3) returns the hourly dispatch for each power plant, the hourly market clearing (zonal) prices, the hourly net positions of each BZ and the total socio-economic welfare from the market dispatch. While the exact price computation methodology is not detailed, zonal prices and dispatch appear to be co-optimized.
This is different from the zonal market clearing problem that we consider. Our one-stage zonal market clearing problem (see Section \ref{sec:zonal-market-clearing}) sets the cross-zonal capacity using an NTC approach. The result of this problem is the hourly dispatch for each power plant, the flows on the interconnectors and the total welfare. In a subsequent step, zonal prices are computed as described in Section \ref{sec:pricing_on_non_convex_markets}.
We also emphasize that in the BZR study, the socio-economic welfare is defined as the sum of consumer surplus, producer surplus and congestion rent. Consumer surplus is calculated as the product of the BZ load multiplied by the difference between the value of loss load\footnote{The value of lost load is the estimated amount that customers receiving electricity with firm contracts would be willing to pay to avoid a disruption in their electricity service. This value is not published in \citep{ENTSOE.2025MainReport}.} and market clearing price. Producer surplus is calculated as the product of the generation of each plant in each BZ and the difference between the market clearing price and the marginal cost of that plant. Congestion rent is calculated as the product of price differences across each bidding zone border multiplied by the respective border flow.
In contrast, in our study, the welfare is defined as the negative generation costs because load is assumed to be inelastic. The generation costs are the objective value of the allocation problem.

\subsection{Redispatch Computation}
Several modelling differences are linked to module (4), which focuses on redispatch computation. This module calculates the redispatch volumes with an algorithm that aims to ``\textit{to remove grid congestions \textbf{under full coordination in CE} by minimising the necessary operational actions, aiming to reduce the resulting redispatching volumes}'' \citep{ENTSOE.2025MainReport}. Redispatch costs are computed based on the redispatch volumes considering the following components: (1) the short-term marginal costs for upward and downward redispatching using redispatching markups (see \citep{ENTSOE.2025.Annex2}), (2) the costs of ensuring the availability of units for redispatching purposes, (3) the costs of grid reserves, (4) the activation costs of explicit demand side response, and (5) the start-up costs of generation units.  
As described in Section \ref{subsec:redis}, we consider a minimum-cost redispatch approach formulated as a mixed-integer program that takes as input the zonal dispatch and modifies it such that it satisfies the DC power flow equations.\footnote{This means that, unlike module (4) of the BZR study, we do not explicitly identify congested intra-zonal lines.} The method assumes that all generators are redispatchable, and that redispatch is determined through a system-wide optimization, i.e., there is a cross-border cooperation between German TSOs. Moreover, we do not consider demand-side response.\footnote{As described in \citep{ENTSOE.2022}, demand-side response (DSR) is either explicit or implicit. Explicit DSR can participate in the redispatch process.}
Since our analysis is restricted to Germany, cheaper redispatch potential from other countries cannot be used to relieve a bottleneck. Our redispatch costs include the costs for upward and downward redispatch (without markups), and the start-up costs of generation units, corresponding primarily to components (1) and (5) of the BZR study. We also assume that the redispatch costs per MWh for each unit are identical to the individual costs used in the market clearing problem.

\section{Comparison with BZR Results} \label{sec:comparison}
In what follows, we compare our main results against the key findings from the BZR study \citep{ENTSOE.2025MainReport}. 
Importantly, the BZR study evaluates the alternative configurations at the CE level, whereas our study evaluates these configurations at the national German level. Also, our analysis is restricted to climate year 2009. The key modelling differences between our study and the BZR study are discussed in \ref{app:modelling-differences}.

First, the BZR report shows that all alternative configurations proposed for Germany have economic efficiency gains\footnote{Economic efficiency is defined as the change in socio-economic welfare, i.e., the average change (over all climate years) in socio-economic welfare coming from the market dispatch minus the average change (over all climate years) in total additional redispatch costs at the CE level.} ranging between EUR 251 - 339 million, which represent less than 1\% of the simulated costs in CE. The highest gain is associated with a five-zone split. Importantly, the economic efficiency gain is not monotonically increasing with the number of zones -- the configuration with three zones has an efficiency gain of EUR 251 million, while a two-zone configuration has an efficiency gain of EUR 264 million.
In contrast, we show that the decrease of nodal costs relative to the total costs associated with the alternative configurations is between 7 and 8\%.

Second, in the BZR results, all configurations are associated with average redispatch cost savings of $\approx$ 50\% compared to the status quo BZ layout. The lowest savings ($\approx$ 30\%) are observed for the climate year 2009. Our study reveals that the redispatch cost savings are below 10\%, regardless of the considered alternative configuration. This discrepancy can be explained by differences in the redispatch algorithms used in the two studies (see Sections \ref{subsec:redis} and \ref{app:modelling-differences}). While the exact redispatch methodology is not fully specified in \citep{ENTSOE.2025MainReport}, it assumes a fully coordinated process across Central Europe that includes demand response, battery storage, and pumped hydro storage -- features not captured in our model. Additionally, redispatch costs in the BZR study account for multiple components beyond the upward, downward, and curtailment costs considered in our analysis. Similar to the results in the BZR study, our results show that the redispatch savings do not necessary increase if we consider a configuration with more zones.  

Third, the BZR results show that all configurations induce two zones in Germany where average market clearing prices change in opposite directions. The northern zone(s) exhibit(s) lower prices compared to the southern one(s). For instance, in a two-zone configuration, the average price difference across all three climate years is 6.45 EUR/MWh. However, as illustrated in \citep{ENTSOE.2024}, the price difference is the lowest for climate year 2009 (4.77 EUR/MWh).
In contrast, our study finds that zonal price differences are below 5 EUR/MWh (see Table \ref{tab:zon_prices_drilldown}). This lower price spread can be attributed to the difference in price computation methodology and to the fact that Germany is modeled in isolation from the rest of the European bidding zones. As illustrated in \citep{ENTSOE.2025MainReport}, zonal prices across bidding zones are correlated -- flows and trades in one zone influence prices in others.

Finally, we calculated the generation and redispatch costs for the five-zone configuration proposed by the German TSOs following the publication of the ACER configurations.  Average generation and redispatch costs were 32,220.77 kEUR and 6,014.25 kEUR, respectively, totaling 38,235.02 kEUR -- this is comparable to the costs corresponding to the other zonal configurations shown in Table \ref{tab:costs}. Average zonal price differences were below 1 EUR/MWh.

\end{document}